\newif\ifOneCol

\OneColfalse

\ifOneCol
	
\documentclass[draftcls,onecolumn,12pt]{IEEEtran}	
\else
\documentclass[twocolumn,10pt]{IEEEtran}
\fi

\usepackage{cite}
\usepackage{citesort}
\usepackage{graphicx,import}
\usepackage{amsmath}
\usepackage{amsfonts}
\usepackage{epstopdf}
\usepackage{xcolor}
\usepackage{amssymb,bm,upgreek}
\usepackage{algorithm}
\usepackage{algorithmic}
\usepackage{multirow}
\usepackage{verbatim}
\usepackage{subfigure}
\usepackage{mleftright}

\usepackage{import}

\DeclareMathOperator\erf{erf}

\let\ss= \scriptscriptstyle
\newcommand{\RX}{\textnormal{RX}}
\newcommand{\R} {\textnormal{RX}}
\newcommand{\TX}{\textnormal{TX}}
\newcommand{\T}{\textnormal{TX}}
\newcommand{\FC} {\textnormal{FC}}
\newcommand{\trans}{\textrm{trans}}
\newcommand{\report}{\textrm{report}}
\newcommand{\metre}{\textnormal{m}}
\newcommand{\s}{\textnormal{s}}

\newcommand{\m}{\textnormal{m}}

\newcommand{\ob}{\textnormal{ob}}
\newcommand{\md}{\textnormal{md}}
\newcommand{\fa}{\textnormal{fa}}
\newcommand{\e}{\textnormal{e}}

\newcommand{\mdb}{\textnormal{mdb}}
\newcommand{\fab}{\textnormal{fab}}

\newcommand\ceil[1]{\lceil#1\rceil}
\newcommand{\ca}{${\textrm{Ca}^{2+}}$}

\newtheorem{theorem}{Theorem}

\begin{document}

\title{\LARGE Convex Optimization of Distributed Cooperative Detection in Multi-Receiver Molecular Communication}

\author{
\ifOneCol	
Yuting Fang, \IEEEmembership{Student Member, IEEE},
Adam Noel, \IEEEmembership{Member, IEEE},
Nan Yang, \IEEEmembership{Member, IEEE},
Andrew W. Eckford, \IEEEmembership{Senior Member, IEEE},
and Rodney A. Kennedy,	\IEEEmembership{Fellow, IEEE}\vspace{-6mm}

\else
Yuting Fang, \IEEEmembership{Student Member, IEEE},
Adam Noel, \IEEEmembership{Member, IEEE},
Nan Yang, \IEEEmembership{Member, IEEE},\\
Andrew W. Eckford, \IEEEmembership{Senior Member, IEEE},
and Rodney A. Kennedy,	\IEEEmembership{Fellow, IEEE}\vspace{-6mm}
\fi

\thanks{This work was presented in part at the IEEE GLOBECOM 2016~\cite{GC 2016}.}
\thanks{Y. Fang, N. Yang, and R. A. Kennedy are with the Research School of Engineering, The Australian National University, Canberra, ACT 2601, Australia (e-mail: \{yuting.fang, nan.yang, rodney.kennedy\}@anu.edu.au).}
\thanks{A. Noel is with the School of Engineering, University of Warwick, Coventry, CV4 7AL, UK (e-mail: anoel2@uottawa.ca).}
\thanks{A. W. Eckford is with the Department of Electrical Engineering and Computer Science, York University, Toronto, ON M3J 1P3, Canada (e-mail:aeckford@yorku.ca)}}

\markboth{Submitted to IEEE Transactions on Molecular, Biological And Multi-Scale Communications}{Fang \MakeLowercase{\textit{et al.}}: Convex Optimization of Distributed Cooperative Detection in Multi-Receiver Molecular Communication}

\maketitle

\begin{abstract}
In this paper, the error performance achieved by cooperative detection among $K$ distributed receivers in a diffusion-based molecular communication (MC) system is analyzed and optimized. In this system, the receivers first make local hard decisions on the transmitted symbol and then report these decisions to a fusion center (FC). The FC combines the local hard decisions to make a global decision using an $N$-out-of-$K$ fusion rule. Two reporting scenarios, namely, perfect reporting and noisy reporting, are considered. Closed-form expressions are derived for the expected global error probability of the system for both reporting scenarios. New approximated expressions are also derived for the expected error probability. Convex constraints are then found to make the approximated expressions jointly convex with respect to the decision thresholds at the receivers and the FC. Based on such constraints, suboptimal convex optimization problems are formulated and solved to determine the optimal decision thresholds which minimize the expected error probability of the system. Numerical and simulation results reveal that the system error performance is greatly improved by combining the detection information of distributed receivers. They also reveal that the solutions to the formulated suboptimal convex optimization problems achieve near-optimal global error performance.
\end{abstract}


\IEEEpeerreviewmaketitle

\section{Introduction}\label{sec:intro}

\IEEEPARstart{O}{ver} the past decades there have been considerable advancements in the fields of nanotechnology and biological science, where the design and manufacturing of nanoscale ($<0.1\mu\metre$) and microscale ($0.1-100\mu\metre$) devices, referred to as nanomachines, have begun to take shape~\cite{nanonetworks}. Since nanomachines 
are only capable of performing simple computing, data storing, sensing, and actuation tasks, it is envisioned that they can be interconnected to execute more elaborate and challenging tasks in a collaborative and distributed manner. The resulting network, i.e., nanonetwork, is anticipated to expand the capabilities of single nanomachines by allowing them to exchange information and interact with each other. Looking 10--20 years ahead, nanonetworks will advance a diverse number of potential applications, such as 
disease detection, targeted drug delivery, and pollution control~\cite{Opportunity and Challenge}.

Molecular communication (MC) has been acknowledged as one of the most promising nanoscale communication paradigms in bio-inspired nanonetworks, due to its unique potential benefits of bio-compatibility and low energy consumption~\cite{Andrew_Book}. In fact, MC is present in nature and used by biological entities and systems, such as molecules, cells, and microorganisms. In MC, a transmitter (TX) releases tiny particles such as molecules or lipid vesicles into a fluid medium, where the particles propagate until they arrive at a receiver (RX). The RX then detects the information encoded in these particles~\cite{Survey}. The simplest molecular propagation mechanism is free diffusion where the information-carrying particles propagate from the TX to the RX via Brownian motion. The TX does not need to expand any energy to use this mechanism.

One of the primary challenges posed by diffusion-based MC is that its reliability rapidly decreases when the TX-RX distance increases. One approach to enhancing its reliability is to use multiple RXs sharing common information to help transmission. In biological environments, some cells or organisms indeed share common information to achieve a specific task~\cite{single and multiple}, e.g., calcium (\ca) signaling~\cite{Ca+ signal}. In one process regulated by \ca\;signaling, named excitation-contraction coupling, the cells in skeletal muscle share \ca\;ions to induce the contraction of myofibrils~\cite{excitation-contraction coupling}.

The majority of the existing MC studies have focused on the modeling of a single-RX MC system. Recent studies, e.g.,~\cite{Multiple-access broadcast,M.J.Moore,Silico Experiment,Bacteria,Microfluidic MC,Tranmissiom Rate,Chun Tung,Molecular MIMO}, have considered {an} MC system which consists of multiple RXs.
{For example,~\cite{Multiple-access broadcast,M.J.Moore} analyzed the transmission rate of a molecular broadcast system where a single TX communicates with multiple {noncooperative} RXs. In~\cite{Silico Experiment}, simulations were performed to demonstrate the feasibility of a bio-nanosensor network where bacteria-based bio-nanomachines perform target detection and tracking. In~\cite{Bacteria}, communication between two populations of bacteria through a diffusion channel was studied where each population acts as a TX or a RX. In~\cite{Microfluidic MC}, a microfluidic channel with two TX and RX pairs was explored. Focusing on the communication between a group of TXs and a group of RXs,~\cite{Tranmissiom Rate} optimized the transmission rates that maximize the throughput and efficiency. A new stochastic model named reaction-diffusion master equation with exogenous input was proposed in~\cite{Chun Tung} to characterize a MC system with multiple TXs and RXs. Recently,~\cite{Molecular MIMO} designed a multiple-input multiple-output MC system and investigated the inter-symbol and inter-link interference therein, but did not consider cooperation between the links. While \cite{Multiple-access broadcast,M.J.Moore,Silico Experiment,Bacteria,Microfluidic MC,Tranmissiom Rate,Chun Tung,Molecular MIMO} stand on their own
merits, these studies did not consider any active cooperation among multiple RXs. This indicates that the role of the cooperation among multiple RXs in determining the TX's intended symbol sequence in a multi-RX MC system has not been established in the literature.}

We note that the cooperation among distributed detectors in wireless communications has been identified as an effective means of improving performance. For example, cooperative spectrum sensing is achieved by allowing multiple secondary users to share sensing data to improve the detection quality of a primary user~\cite{cooperative spectrum sensing}. Generally, in a distributed detection system the data of the individual detectors is shared at a fusion center (FC). This data may be hard (binary) decisions, soft (multi-level) decisions, or quantized observations. The FC then appropriately combines the received data to yield a global inference~\cite{Distributed Detection} using a fusion rule, such as the AND rule and OR rule for hard decisions. In fact, logic operations and corresponding computations required at the FC, e.g., AND, OR, and addition operations, can be implemented at the molecular level~\cite{Molecular logic,Molecular logic1}. Therefore, MC is a suitable domain to apply distributed detection to improve transmission reliability. We note that this application has not been previously studied.

In this paper, we for the first time quantify and maximize the benefits of multi-RX cooperation in a cooperative diffusion-based MC system. Our goal is to establish a fundamental understanding of the reliability improvement brought by combining the detection results of distributed RXs at an FC. In our considered system, for each symbol transmitted from the TX, the RXs first independently make local hard decisions on the transmitted symbol and then report their decisions to the FC. We note that the role of the RXs in our considered system appears similar to that of decode-and-forward (DF) relays in wireless systems~\cite{cooperative diversity}. However, the results for DF relaying cannot be used in the MC system, due to the fact that the characteristics of the propagation channel and the methods for recovering the received symbols in MC systems are completely different from those in wireless systems. After receiving the local hard decisions, the FC fuses all decisions to make a global decision on the transmitted symbol using an $N$-out-of-$K$ fusion rule. Here, we consider two different reporting scenarios from the RXs to the FC, namely, perfect reporting and noisy reporting. In this work, we assume that the FC does not feed back its global decision to RXs. We also assume that each RX transmits a unique type of molecule and the FC is able to simultaneously and independently detect the different types of molecules from the RXs\footnote{We note that releasing a unique type of molecule at each RX may not be realistic in some cases. However, this assumption gives a lower bound on the error performance of the cooperative MC system using a hard decision fusion rule.} (as in~\cite{single and multiple}). {We note that the use of a binary sequence is expected in molecular communications between nanomachines to exchange the amount of information required for executing complex collaborative tasks, such as disease detection and targeted drug delivery. Thus, in this work we consider the transmission of multiple binary symbols as a sequence and take the resultant inter-symbol interference (ISI) into account for the cooperative MC system.}

{To maximize the benefits of multi-RX cooperation in the system, we determine the jointly optimal decision thresholds at the RXs and FC such that the expected global error probability is minimized. We note that it is mathematically intractable to derive analytical expressions for such optimal thresholds. Therefore, we resort to convex optimization as an efficient and effective method to solve the joint optimization problem. Since the expected global error probability is not necessarily convex with respect to thresholds at the RXs and FC, we conduct new convex analysis of the error performance for the system having a symmetric topology. Based on this analysis, we formulate convex optimization problems and find the solution via an efficient convex optimization algorithm. We note that finding the optimal thresholds at the RXs and FC via exhaustive search is time-consuming and requires relatively high complexity, compared with the adopted convex optimization.} In the symmetric topology, the distances between the TX and the RXs are identical and the distances between the RXs and the FC are also identical. This results in independent and \emph{identically} distributed observations at the RXs. We clarify that the assumption of the symmetric topology is to improve the tractability of our convex analysis and this assumption will be relaxed in future work. {Also, for some practical applications, such as health monitoring, we may manually set the locations of the RXs and FC to ensure that the topology of the TX (e.g., monitored organism), the RXs (e.g., detectors), and the FC, is symmetric and reduce the complexity of system design and performance optimization. Also, the assumption of a symmetric topology is reasonable if the difference in distance between different $\TX-\RX-\FC$ links is negligible, compared to the distance between the TX and the RXs.} We note that the expected error probability of a point-to-point MC link is minimized in~\cite{multi hop}, by deriving a closed-form analytical expression for the optimal decision threshold at the RX. However, the derived optimal decision threshold in~\cite{multi hop} cannot be applied or extended to the cooperative MC system.




The primary contributions of this paper, especially relative to our previous work~\cite{GC 2016}, are summarized as follows:
\begin{enumerate}
\item
We derive closed-form expressions for the expected global error probabilities of the cooperative MC system in the perfect and noisy reporting scenarios. We clarify that a symbol-by-symbol detection with a constant decision threshold at all RXs and the FC is considered in this derivation.
\item
We derive new approximated expressions for
the expected global error probability of the cooperative MC system in both reporting scenarios. We also derive additional convex constraints under which the approximated expressions are jointly convex with respect to the decision thresholds at the RXs and the FC.
\item
Based on the derived convex approximations and constraints, we formulate suboptimal convex optimization problems for a given transmitted symbol sequence. For the sake of practicality, we then extend the formulated convex optimization problems such that a single optimal threshold is determined to minimize the average error performance over all realizations of transmitted symbol sequences.
\end{enumerate}

Using numerical and simulation results, we demonstrate that the error performance of the cooperative MC system is much better than that of the point-to-point MC link. We also demonstrate the effectiveness of our formulated suboptimal convex optimization problems by showing that near-optimal global error performance is achieved by using the solutions to our formulated problems, i.e., the optimal thresholds. In this work, the optimality of the performance refers to the accuracy of the optimization to find the optimal thresholds in the symbol-by-symbol detection.

The rest of this paper is organized as follows. In Section \ref{sec:system model}, we describe the system model. In Section \ref{sec:Error Performance Analysis}, we present the error performance analysis of the cooperative-RX MC system. In Section \ref{sec:Convex Optimization}, we formulate convex optimization problems of the cooperative-RX MC system. Numerical and simulation results are provided in Section \ref{sec:Numerical}. In Section \ref{sec:con}, we conclude and describe future directions for this work.

\section{System Model}\label{sec:system model}



In this paper we consider a cooperative MC system in a three-dimensional space, as depicted in Fig.~\ref{system model}, which consists of one TX, a ``cluster'' of $K$ RXs, and one device acting as an FC. {The} FC is not included in the set of RXs. {We assume that the RXs are generally closer to the FC than to the TX to ensure reliable reporting channels from the RXs to the FC.} We assume that all RXs and the FC are spherical observers. Accordingly, we denote $V_{\ss\R_k}$ and $r_{\ss\R_k}$ as the volume and the radius of the $k$th RX, $\RX_k$, respectively, where $k\in\{1,2,{\ldots},K\}$. We also denote $V_{\ss\FC}$ and $r_{\ss\FC}$ as the volume and the radius of the FC, respectively. We also assume that the RXs and the FC are independent passive observers such that molecules can diffuse through them without reacting{\footnote{Although we cannot guarantee perfect independence between different RXs, the dependence between observations made at different RXs is extremely small. This is due to the fact that the time between adjacent samples at RXs is sufficiently long to ensure that observations at RXs are independent and each RX observes a small fraction of the total number of released molecules. Moreover, the validity of assuming independence will be demonstrated by the excellent agreement between analytical and simulation results depicted in Section~\ref{sec:Numerical}.}}. We further assume that all individual observations are independent of each other. In addition, we assume that the RXs operate in the half-duplex mode such that they do not receive information and report their local decisions at the same time.

\ifOneCol	
\begin{figure}[!t]
\centering
\includegraphics[width=0.6\columnwidth]{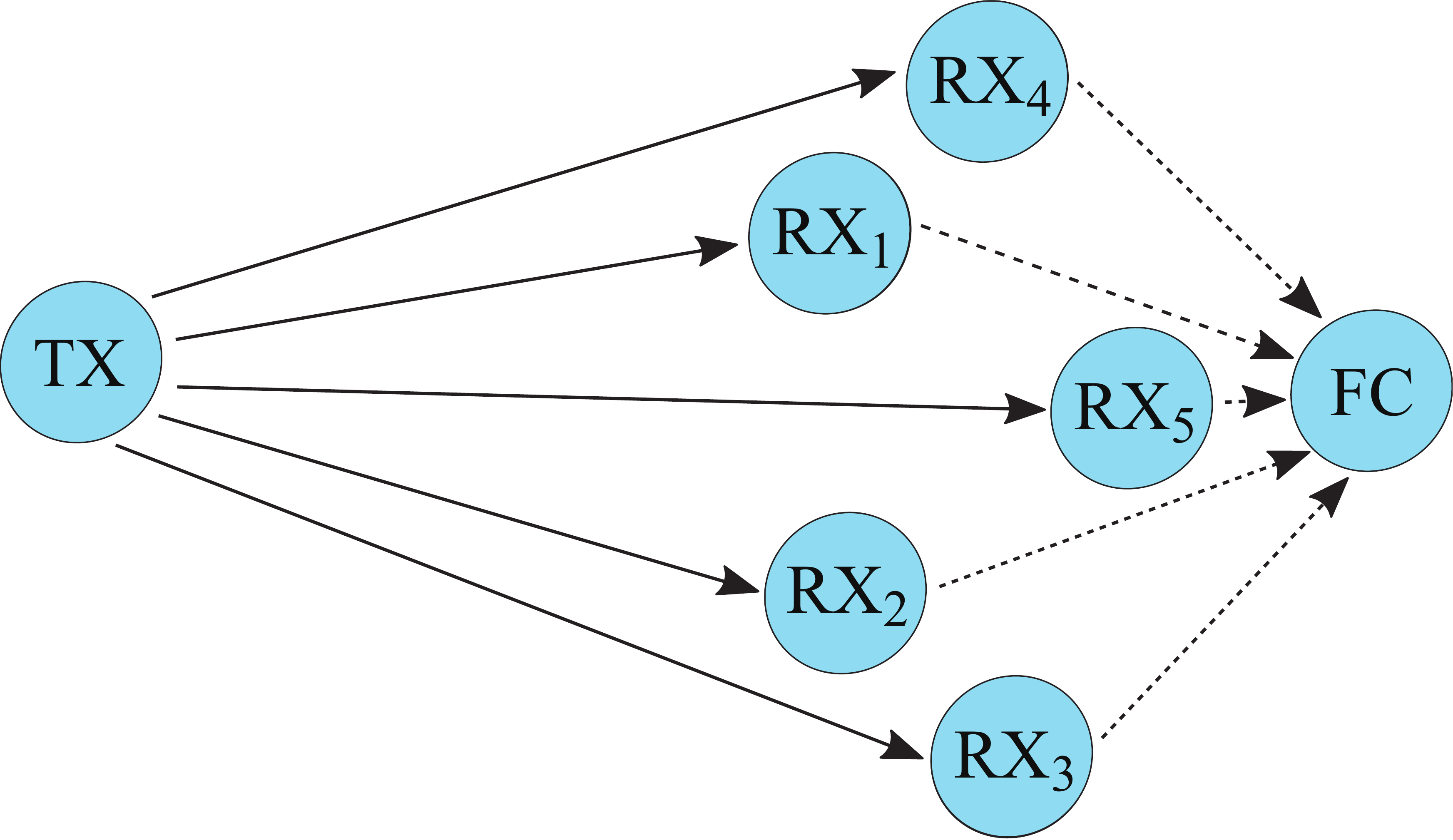}
\caption{An example of a cooperative MC system with $K=5$, where the transmission from the TX to the RXs is represented by solid arrows and the decision reporting from the RXs to the FC is represented by dashed arrows.}
\label{system model}
\end{figure}	
\else
\begin{figure}[!t]
\centering
\includegraphics[width=0.8\columnwidth]{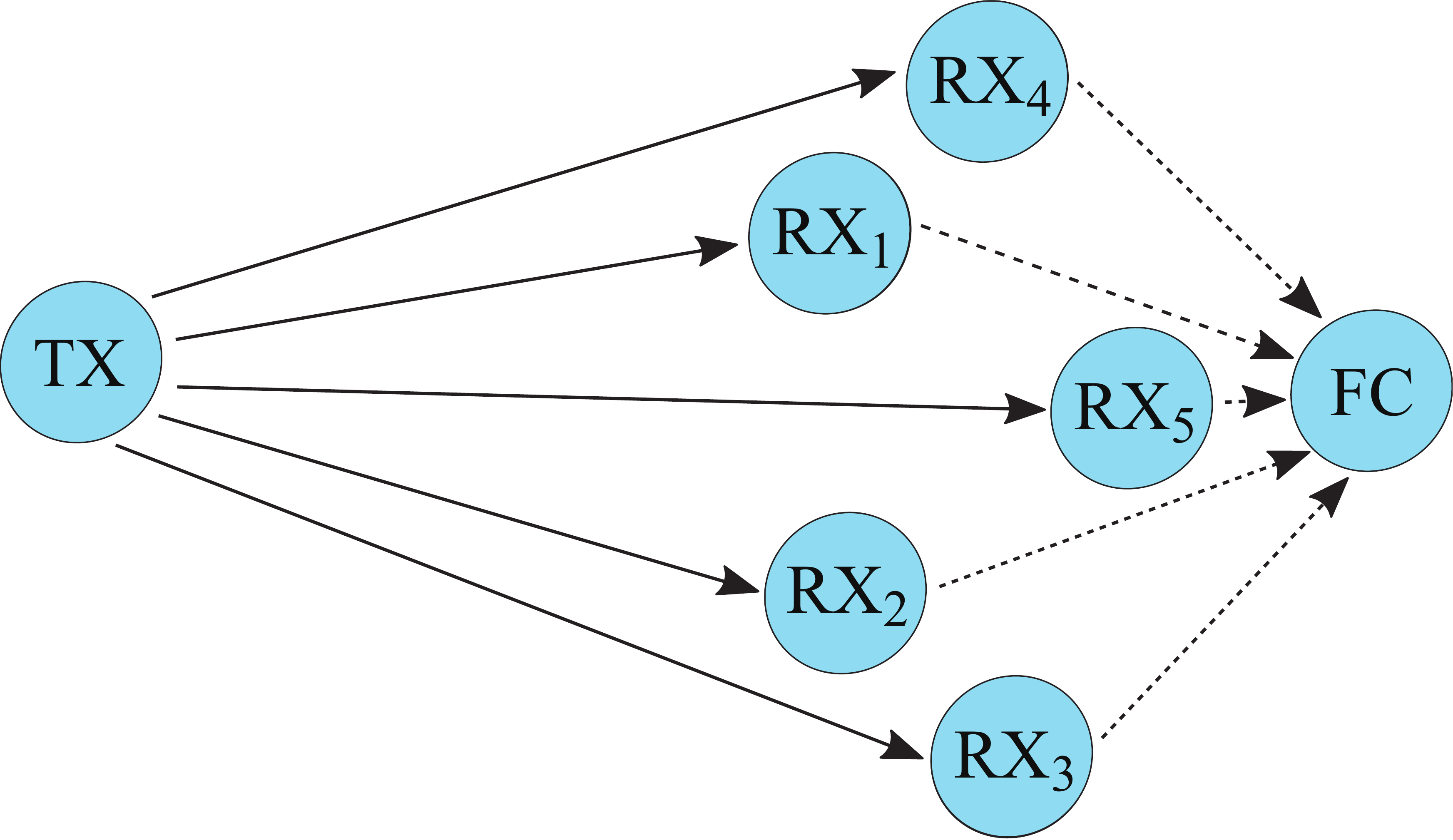}
\caption{An example of a cooperative MC system with $K=5$, where the transmission from the TX to the RXs is represented by solid arrows and the decision reporting from the RXs to the FC is represented by dashed arrows.}
\label{system model}
\end{figure}
\fi

In the considered system, the transmission of each information symbol from the TX to the FC via RXs is completed within three phases, detailed as follows:
\begin{itemize}
\item
In the first phase, the TX transmits one symbol of information via type $A_0$ molecules to the RXs through the diffusive channel. The number of the released type $A_0$ molecules is denoted by $S_0$. We assume that the diffusion of all individual molecules is independent. The type $A_0$ molecules transmitted by the TX are detected by all RXs. In this work we consider that the TX uses ON/OFF keying~\cite{modulation} to convey information. As per the rules of ON/OFF keying, the TX releases $S_{0}$ molecules of type $A_0$ to convey information symbol ``1'', and releases no molecules to convey information symbol ``0''. {To enable ON/OFF keying, the information transmitted by the TX {is represented by} an $L$-length binary sequence where each element is ``0'' or ``1''.} The sequence is denoted by $\textbf{W}_{\ss\T}=\{W_{\ss\T}[1],W_{\ss\T}[2],{\ldots},W_{\ss\T}[L]\}$, where $W_{\ss\T}[j]$, $j\in\{1,{\ldots},L\}$, is the $j$th symbol transmitted by the TX. We assume that the probability of transmitting ``1'' in the $j$th symbol is $P_{1}$ and the probability of transmitting ``0'' in the $j$th symbol is $1-P_{1}$, i.e., $\textrm{Pr}(W_{\ss\T}[j]=1)=P_1$ and $\textrm{Pr}(W_{\ss\T}[j]=0)=1-P_{1}$, where $\textrm{Pr}(\cdot)$ denotes probability.
\item
In the second phase, each RX makes a local hard decision on the current transmitted symbol. We denote $\hat{W}_{{\ss\R}_k}[j]$ as the local hard decision on the $j$th transmitted symbol at $\RX_k$. Then, the RXs simultaneously report their local $j$th hard decisions to the FC. We assume that $\RX_k$ transmits type $A_{k}$ molecules, which can be detected by the FC. The number of the released type $A_k$ molecules is denoted by $S_k$. We also assume that the channel between each RX and the FC is diffusion-based, and each RX uses ON/OFF keying to report its local hard decision.
\item
In the third phase, the FC obtains the decision at $\RX_k$ by receiving type $A_k$ molecules over the $\RX_{k}-\FC$ link. We assume that the $K$ $\RX_{k}-\FC$ links are independent. We denote $\hat{W}_{{\ss\FC}_{k}}[j]$ as the received local decision of $\RX_k$ on the $j$th transmitted symbol at the FC. {The FC combines all $\hat{W}_{{\ss\FC}_{k}}[j]$ using an $N$-out-of-$K$ fusion rule to make a global decision $\hat{W}_{\ss\FC}[j]$ on the $j$th symbol transmitted by the TX, where $N$ denotes the number of decisions of ``1'' received by the FC and $K$ denotes the number of RXs.} As per the $N$-out-of-$K$ fusion rule, the FC declares a global decision of ``1'' when it receives at least $N$ decisions of ``1''. There are several special cases of the $N$-out-of-$K$ fusion rule, such as 1) majority rule where $N=\ceil{K/2}$ and $\ceil{x}$ represents the smallest integer greater than or equal to $x$, 2) OR rule where $N=1$, and 3) AND rule where $N=K$.
\end{itemize}

We define $\textbf{W}_{{\ss\T}}^{l}=\{W_{\ss\T}[1],{\ldots},W_{\ss\T}[l]\}$ as an $l$-length subsequence of the information transmitted by the TX, where $l\leq{L}$. We also define $\hat{\textbf{W}}_{\ss\RX_k}^l=\{\hat{W}_{\ss\RX_k}[1],{\ldots},\hat{W}_{\ss\RX_k}[l]\}$ as an $l$-length subsequence of the local hard decisions at $\RX_k$. We then define $\hat{\textbf{W}}_{{\ss\FC_k}}^l=\{\hat{W}_{\ss\FC_k}[1],{\ldots},\hat{W}_{\ss\FC_k}[l]\}$ as an $l$-length subsequence of the received local decision of $\RX_k$ at the FC. We further define $\hat{\textbf{W}}_{\ss\FC}^{l}=\{\hat{W}_{\ss\FC}[1],{\ldots},\hat{W}_{\ss\FC}[l]\}$ as an $l$-length subsequence of the global decisions at the FC.

We denote $t_{\trans}$ as the transmission interval time from the TX to the RXs and $t_{\report}$ as the report interval time from the RXs to the FC. Thus, the symbol interval time from the TX to the FC is given by $T = t_{\trans}+t_{\report}$. At the beginning of the $j$th symbol interval, i.e., $(j-1)T$, the TX transmits $W_{\ss\T}[j]$. After this the TX keeps silent until the start of the $(j+1)$th symbol interval. We apply the weighted sum detector with equal weights\cite{Adam optimal} at the RXs and FC for detection. Thus, the RXs and FC each take multiple samples within their corresponding interval time, add the individual samples with equal weights, and compare the summation with a decision threshold. The decision thresholds at $\RX_k$ are denoted by $\xi_{\ss\R_k}$. {The decision thresholds at $\FC$ over the $\RX_{k}-\FC$ link denoted by $\xi_{\ss\FC_k}$.} Here, the assumption of equal weights for all samples is adopted to limit the computational complexity of the detector and facilitate its usage in MC.

We now describe the sampling schedules of the RXs and FC. {The FC or $\RX_k$ samples at {a certain} time $t$ by counting the number of the molecules observed.} All RXs sample at the same times{\footnote{We note that all RXs may not be synchronized perfectly in some cases. Thus, we make the assumption of same sampling times of all RXs to explore the best error performance achieved by the cooperative MC system, which serves as a performance bound for practical systems. We also note that various methods can be adopted to achieve time synchronization among nanomachines, e.g., \cite{blindsyn} and \cite{bio-inspiredsyn}. Therefore, the assumption of perfect synchronization is widely adopted in existing MC studies, e.g., \cite{multi hop}, \cite{Adam optimal}, and \cite{Molecular MIMO}.}} and take $M_{\ss\RX}$ samples per symbol interval. The time of the $m$th sample for each RX in the $j$th symbol interval is given by $t_{\ss\R}(j,m) = (j-1)T + m\Delta{t_{\ss\R}}$, where $\Delta{t_{\ss\R}}$ is the time step between two successive samples at each RX, $m\in\left\{1,2,{\ldots},M_{\ss\RX}\right\}$, and $M_{\ss\RX}\Delta{t_{\ss\R}}<t_\trans$. At the time $(j-1)T + t_{\trans}$, each RX reports its local decision for the $j$th interval via diffusion to the FC. We assume that the FC takes $M_{\ss\FC}$ samples of \emph{each} type of molecule in every reporting interval. The time of the $\tilde{m}$th sample of type $A_k$ molecules at the FC in the $j$th symbol interval is given by $t_{\ss\FC}(j,\tilde{m})=(j-1)T+t_{\trans}+\tilde{m}\Delta{t_{\ss\FC}}$, where $\Delta{t_{\ss\FC}}$ is the time step between two successive samples at the FC and $\tilde{m}\in\left\{1,2,{\ldots},M_{\ss\FC}\right\}$.

\section{Error Performance Analysis}\label{sec:Error Performance Analysis}

In this section, we analyze the expected global error probability\footnote{All the expected error probabilities throughout this paper are derived for given $\textbf{W}_{\ss\T}^{j-1}$, unless otherwise specified.} of the cooperative MC system. To this end, we denote $Q_{\ss\FC}[j]$ as the expected global error probability in the $j$th symbol interval for a \emph{given} transmitter sequence $\textbf{W}_{\ss\T}^{j-1}$. Under the assumption that there is no \emph{a priori} knowledge of $W_{\ss\T}[j]$, we express $Q_{\ss\FC}[j]$ as
\begin{align}\label{overall probability}
Q_{\ss\FC}[j] = P_1Q_{\md}[j] + \left(1-P_{1}\right)Q_{\fa}[j],
\end{align}
where $Q_{\md}[j]$ denotes the expected global missed detection probability (MDP) in the $j$th symbol interval and $Q_{\fa}[j]$ denotes the expected global false alarm probability (FAP) in the $j$th symbol interval. By averaging $Q_{\ss\FC}[j]$ over all possible realizations of $\textbf{W}_{\ss\T}^{j-1}$ and across all symbol intervals, the expected average error probability of the cooperative MC system, $\overline{Q}_{\ss\FC}$, can be obtained. In the analysis, we address two different reporting scenarios, namely, perfect reporting and noisy reporting. In the perfect reporting scenario, we assume that no error occurs when $\RX_k$ reports to the FC, i.e., $\hat{W}_{{\ss\FC}_{k}}[j]= \hat{W}_{{\ss\R}_k}[j]$. In the noisy reporting scenario, errors can occur when $\RX_k$ reports to the FC via diffusion\footnote{In this paper, the notations for the symbol interval time, the number of molecules for symbol ``1'' released by the TX, and the sampling schedules of the RXs in the perfect reporting scenario are the same as those in the noisy reporting scenario.}.

\subsection{Perfect Reporting}\label{Perfect Error Performance Analysis}

In this subsection, we start our analysis by examining the error performance of the $\TX-\RX_k$ link. This examination is based on the analysis in~\cite{multi hop}. We then use the results of this examination to analyze $Q_{\md}[j]$ and $Q_{\fa}[j]$ in the perfect reporting scenario to obtain $Q_{\ss\FC}[j]$.

\subsubsection{$\TX-\RX_k$ Link}

We first evaluate the probability of observing a given type $A_0$ molecule, emitted from the TX at $t=0$, inside $V_{\ss\R_k}$ at time $t$, $P_{\ob,0}^{({\ss{\T},{\ss\R_k}})}\left(t\right)$. Given independent molecular behavior and assuming that the RXs are sufficiently far from the TX, we use \cite[Eq.~(1)]{Opportunity and Challenge} to write $P_{\ob,0}^{({\ss{\T},{\ss\R_k}})}\left(t\right)$ as
\begin{align}\label{probability}
P_{\ob,0}^{({\ss{\T},{\ss\R_k}})}(t) = \frac{V_{\ss\R_k}}{(4\pi D_{0}t)^{3/2}}\exp\left(-\frac{d_{\ss\T_k}^{2}}{4D_{0}t}\right),
\end{align}
where $D_{0}$ is the diffusion coefficient of type $A_0$ molecules in $\frac{\m^{2}}{\s}$ and $d_{\ss\T_k}$ is the distance between the TX and $\RX_k$ in $\m$.

We denote $S_{\ob,0}^{({\ss{\T},{\ss\R_k}})}[j]$ as the sum of the number of molecules observed within $V_{\ss\R_k}$ in the $j$th symbol interval, due to the emission of molecules from the current and previous symbol intervals at the TX, $\textbf{W}_{\ss\T}^j$. As discussed in~\cite{multi hop}, $S_{\ob,0}^{({\ss{\T},{\ss\R_k}})}[j]$ can be accurately approximated by a Poisson random variable (RV) with the mean given by
\ifOneCol	
\begin{align}\label{observed molecular numbers R}
\bar{S}_{\ob,0}^{({\ss{\T},{\ss\R_k}})}[j]
=&\;S_{0}\sum\limits^{j}_{i=1}W_{\ss\T}[i]
\sum\limits^{M_{\ss\RX}}_{m=1}P_{\ob,0}^{({\ss{\T},{\ss\R_k}})}\left(\left(j-i\right)T + m\Delta{t_{\ss\R}}\right).
\end{align}
\else
\begin{align}\label{observed molecular numbers R}
\bar{S}_{\ob,0}^{({\ss{\T},{\ss\R_k}})}[j]
=&\;S_{0}\sum\limits^{j}_{i=1}W_{\ss\T}[i]\nonumber\\
&\times\sum\limits^{M_{\ss\RX}}_{m=1}P_{\ob,0}^{({\ss{\T},{\ss\R_k}})}\left(\left(j-i\right)T + m\Delta{t_{\ss\R}}\right).
\end{align}
\fi

We also denote $U_{z,k}[j]$, $z\in{\{0,1\}}$, as the conditional mean of $S_{\ob,0}^{({\ss{\T},{\ss\R_k}})}[j]$ when the most recent information symbol transmitted by the TX is $W_{\ss\T}[j] = z$. Then, the decision at $\RX_k$ in the $j$th symbol interval is given by
\begin{align}\label{detectorRX}
\hat{W}_{\ss\R_k}[j]=
\begin{cases}
1,&\mbox{if $S_{\ob,0}^{({\ss{\T},{\ss\R_k}})}[j]\geq\xi_{\ss\R_k}$,}\\
0,&\mbox{otherwise}.
\end{cases}
\end{align}

Moreover, based on \cite[Eq.~(9)]{multi hop}, the expected MDP of the $\TX-\RX_k$ link in the $j$th symbol interval for given $\textbf{W}_{\ss\T}^{j-1}$ is written as
\begin{align}\label{Pm1}
P_{\md,k}[j] = \textrm{Pr}\left(S_{\ob,0}^{({\ss{\T},{\ss\R_k}})}[j]<\xi_{\ss\R_k}{\Big{|}}W_{\ss\T}[j]=1,\textbf{W}_{\ss\T}^{j-1}\right),
\end{align}
and the corresponding expected FAP is written as
\begin{align}\label{Pf1}
P_{\fa,k}[j] = \textrm{Pr}\left(S_{\ob,0}^{({\ss{\T},{\ss\R_k}})}[j]\geq\xi_{\ss\R_k}{\Big{|}}W_{\ss\T}[j]=0,\textbf{W}_{\ss\T}^{j-1}\right).
\end{align}

{Based on \cite[Eq.~2.5]{probability}, the cumulative distribution function (CDF) of $S_{\ob,0}^{({\ss{\T},{\ss\R_k}})}[j]$ is given by
\ifOneCol	
\begin{align}\label{RX}
\textrm{Pr}\left(S_{\ob,0}^{({\ss{\T},{\ss\R_k}})}[j]<\xi_{\ss\R_k}{\Big{|}}\textbf{W}_{\ss\T}^{j}\right) =&\;\exp\left(-\bar{S}_{\ob,0}^{({\ss{\T},{\ss\R_k}})}[j]\right)
\sum_{\omega=0}^{\xi_{\ss\R_k}-1}\frac{\bar{S}_{\ob,0}^{({\ss{\T},{\ss\R_k}})}[j]^\omega}{\omega!}.
\end{align}}
\else
\begin{align}\label{RX}
\textrm{Pr}\left(S_{\ob,0}^{({\ss{\T},{\ss\R_k}})}[j]<\xi_{\ss\R_k}{\Big{|}}\textbf{W}_{\ss\T}^{j}\right) =&\;\exp\left(-\bar{S}_{\ob,0}^{({\ss{\T},{\ss\R_k}})}[j]\right)\nonumber\\
&\times\sum_{\omega=0}^{\xi_{\ss\R_k}-1}\frac{\bar{S}_{\ob,0}^{({\ss{\T},{\ss\R_k}})}[j]^\omega}{\omega!}.
\end{align}}
\fi

{Using \eqref{RX} and its complementary function, we can find the closed-form expressions for $P_{\md,k}[j]$ and $P_{\fa,k}[j]$.}

\subsubsection{Global Error Probability}\label{expected cooperative noisy}

We consider the cooperative MC system having a symmetric topology such that the RXs have independent and \emph{identically} distributed observations. Under this consideration, we have $U_{z,k}[j]=U_{z}[j]$. Accordingly, we assume that the decision thresholds at the RXs are the same, i.e., $\xi_{\ss\RX_k} = \xi_{\ss\RX}$. Thus, we have $P_{{\md},k}[j]= P_{\md}[j]$ and $P_{{\fa},k}[j]= P_{\fa}[j]$.

We first consider the $N$-out-of-$K$ fusion rule. Using~\cite[Eq.~(3.4.30)]{Distributed Detection} and~\cite[Eq.~(3.4.31)]{Distributed Detection} we evaluate
$Q_{\md}[j]$ as
\begin{align}\label{Qm_KN}
Q_{\md}[j]= 1-\sum\limits^{K}_{n=N}\binom{K}{n}\left(1-P_{{\md}}[j]\right)^{n}{P_{{\md}}[j]}^{K-n}
\end{align}
and evaluate $Q_{\fa}[j]$ as
\begin{align}\label{Qf_KN}
Q_{\fa}[j] = \sum\limits^{K}_{n=N}\binom{K}{n}{P_{\fa}[j]}^{n}\left(1-P_{{\fa}}[j]\right)^{K-n}.
\end{align}
For the OR rule, we obtain $Q_{\md}[j]$ and $Q_{\fa}[j]$ by substituting $N=1$ into \eqref{Qm_KN} and \eqref{Qf_KN}, leading to
\begin{align}\label{Qm_OR}
Q_{\md}[j] = P_{{\md}}[j]^K
\end{align}
and
\begin{align}\label{Qf_OR}
Q_{\fa}[j] = 1-(1-P_{\fa}[j])^K,
\end{align}
respectively. For the AND rule, we obtain $Q_{\md}[j]$ and $Q_{\fa}[j]$ by substituting $N=K$ into \eqref{Qm_KN} and \eqref{Qf_KN}, resulting in
\begin{align}\label{Qm_AND}
Q_{\md}[j] = 1-(1-P_{{\md}}[j])^K
\end{align}
and
\begin{align}\label{Qf_AND}
Q_{\fa}[j] = P_{{\fa}}[j]^K,
\end{align}
respectively.

We note that the single-RX MC system, which consists of one TX, one RX, and one FC, is a special case of the cooperative MC system. Therefore, the expected error probability of the single-RX MC system in the $j$th symbol interval for a \emph{given} transmitter sequence $\textbf{W}_{\ss\T}^{j-1}$ in the perfect reporting scenario, $P_{\e,1}[j]$, can be obtained by setting $K=1$ in \eqref{Qm_KN} and \eqref{Qf_KN}. 
Accordingly, the expected average error probability of the single-RX MC system, $\overline{P}_{\e,1}$, can be obtained by averaging $P_{\e,1}[j]$ over all possible realizations of $\textbf{W}_{\ss\T}^{j-1}$ and across all symbol intervals.

\subsection{Noisy Reporting}\label{Noisy Error Performance Analysis}

In this subsection, we first examine the error performance of the $\TX-\RX_k-\FC$ link,
based on the analysis in~\cite{multi hop,Adam dimen}. We then use the results of this examination to analyze $Q_{\md}[j]$ and $Q_{\fa}[j]$ in the noisy reporting scenario, enabling us to obtain $Q_{\ss\FC}[j]$.

\subsubsection{$\TX-\RX_k-\FC$ Link}

We denote $P_{\ob,{k}}^{({\ss\R_k,\ss\FC})}(t)$ as the probability of observing a given $A_k$ molecule, emitted from the $\RX_k$ at $t=0$, inside $V_{\ss\FC}$ at time $t$. {Due to the relatively close distance between the RXs and the FC}, we find that \eqref{probability} or \cite[Eq.~(1)]{Opportunity and Challenge} cannot be used to evaluate $P_{\ob,{k}}^{({\ss\R_k,\ss\FC})}(t)$. Thus, we resort to \cite[Eq. (27)]{Adam dimen} to evaluate $P_{\ob,{k}}^{(\ss\R_k,\ss\FC)}(t)$ as
\ifOneCol
\begin{align}\label{general prob}
P_{\ob,{k}}^{({\ss\R_k,\ss\FC})}(t)=&\;\frac{1}{2}\left[\erf\left(\tau_{1}\right)+\erf\left(\tau_{2}\right)\right]
-\frac{\sqrt{D_{{k}}t}}{d_{\ss\FC_k}\sqrt{\pi}}\left[\exp\left(-\tau_{1}^{2}\right)-\exp\left(-\tau_{2}^{2}\right)\right],
\end{align}
\else
\begin{align}\label{general prob}
P_{\ob,{k}}^{({\ss\R_k,\ss\FC})}(t)=&\;\frac{1}{2}\left[\erf\left(\tau_{1}\right)+\erf\left(\tau_{2}\right)\right]\nonumber\\
&-\frac{\sqrt{D_{{k}}t}}{d_{\ss\FC_k}\sqrt{\pi}}\left[\exp\left(-\tau_{1}^{2}\right)-\exp\left(-\tau_{2}^{2}\right)\right],
\end{align}
\fi
where $\tau_{1}=\frac{r_{\ss\FC}+d_{\ss\FC_k}}{2\sqrt{D_{{k}}t}}$,  $\tau_{2}=\frac{r_{\ss\FC}-d_{\ss\FC_k}}{2\sqrt{D_{{k}}t}}$, $D_{{k}}$ is the diffusion coefficient of type $A_{k}$ molecules in $\frac{\m^{2}}{\s}$, and $d_{\ss\FC_k}$ is the distance between $\RX_k$ and the $\FC$ in $\m$.

We denote ${S}_{\ob,{k}}^{({\ss\R_k,\ss\FC})}[j]$ as the number of molecules observed within $V_{\ss\FC}$ in the $j$th symbol interval, due to the emissions of molecules from the current and the previous symbol intervals at $\RX_k$, $\hat{\textbf{W}}_{\ss\RX_k}^j$. We note that the TX and $\RX_k$ use the same modulation method and the $\TX-\RX_k$ and $\RX_k-\FC$ links are both diffusion-based. Therefore, ${S}_{\ob,{k}}^{({\ss\R_k,\ss\FC})}[j]$ can be accurately approximated by a Poisson RV. We denote ${\bar{S}}_{\ob,{k}}^{({\ss\R_k,\ss\FC})}[j]$ as the mean of ${S}_{\ob,{k}}^{({\ss\R_k,\ss\FC})}[j]$ and obtain it by replacing $S_{0}$, $W_{\ss\T}[i]$, $P_{\ob,0}^{({\ss{\T},{\ss\R_k}})}$, $M_{\ss\RX}$, $m$, and $\Delta{t_{\ss\R}}$ with $S_{{k}}$, $\hat{W}_{\ss\R_k}[i]$, $P_{\ob,{k}}^{({\ss\R_k,\ss\FC})}$, $M_{\ss\FC}$, $\tilde{m}$, and $\Delta{t_{\ss\FC}}$ in \eqref{observed molecular numbers R}, respectively. 
{We define $V_{\tilde{z},k}\left[j\right]$, $\tilde{z}\in{\{0,1\}}$, as the conditional mean of ${S}_{\ob,{k}}^{({\ss\R_k,\ss\FC})}[j]$ when the most recent information symbol transmitted by $\RX_k$ is $\hat{W}_{{\ss\R_k}}[j] = \tilde{z}$, given previous decisions at $\RX_k$, $\hat{\textbf{W}}_{\ss\RX_k}^{j-1}$.} Furthermore, we note that $\hat{W}_{{\ss\FC}_{k}}[j]$ can be obtained by replacing $S_{\ob,0}^{({\ss{\T},{\ss\R_k}})}[j]$ and $\xi_{\ss\R_k}$ with ${S}_{\ob,{k}}^{({\ss\R_k,\ss\FC})}[j]$ and $\xi_{\ss\FC_k}$ in \eqref{detectorRX}, respectively.
{We now derive the expected MDP and FAP of the $\TX-\RX_k-\FC$ link in the $j$th symbol interval averaging over \emph{all possible realizations} of $\hat{\textbf{W}}_{\ss\R_k}^{j-1}$ for given $\textbf{W}_{\ss\T}^{j-1}$, denoted by $\tilde{P}_{\md,k}[j]$ and $\tilde{P}_{\fa,k}[j]$, respectively. For a given $\textbf{W}_{\ss\T}^{j-1}$, there are $2^{(j-1)}$ different possible realizations of $\hat{\textbf{W}}_{\ss\R_k}^{j-1}$. We define $\mathcal{W}_j$ as the set containing all realizations of $\hat{\textbf{W}}_{\ss\R_k}^{j-1}$. Considering all possible realizations of $\hat{\textbf{W}}_{\ss\R_k}^{j-1}$ and their likelihood of occurrence, we first derive $\tilde{P}_{\md,k}[j]$ and $\tilde{P}_{\fa,k}[j]$ as
\ifOneCol	
\begin{align}\label{Pem2,origin}
\hspace{-2.5mm}\tilde{P}_{\md,k}[j]=&\;
\sum_{\hat{\textbf{W}}_{\ss\R_k}^{j-1}\in\mathcal{W}_j}\left[\textrm{Pr}\left(\hat{\textbf{W}}_{\ss\R_k}^{j-1}\Big{|}\textbf{W}_{\ss\T}^{j-1}\right)\right.\nonumber\\
&\times\left(\textrm{Pr}\left(S_{\ob,{0}}^{({\ss{\T},{\ss\R_k}})}[j]\geq\xi_{\ss\R}\Big{|}W_{\ss\T}[j]=1,\textbf{W}_{\ss\T}^{j-1}\right)\right.
\textrm{Pr}\left(S_{\ob,{k}}^{({\ss\R_k,\ss\FC})}[j]<\xi_{\ss\FC_k}\Big{|}\hat{W}_{\ss\R_k}[j]=1,\hat{\textbf{W}}_{\ss\R_k}^{j-1}\right)\nonumber\\
&+\textrm{Pr}\left(S_{\ob,{0}}^{({\ss{\T},{\ss\R_k}})}[j]<\xi_{\ss\R}\Big{|}W_{\ss\T}[j]=1,\textbf{W}_{\ss\T}^{j-1}\right)
\left.\left.\textrm{Pr}\left(S_{\ob,{k}}^{({\ss\R_k,\ss\FC})}[j]<\xi_{\ss\FC_k}\Big{|}\hat{W}_{\ss\R_k}[j]=0,\hat{\textbf{W}}_{\ss\R_k}^{j-1}\right)\right)\right]
\end{align}
\else
\begin{align}\label{Pem2,origin}
\hspace{-2.5mm}\tilde{P}_{\md,k}[j]=&\;
\sum_{\hat{\textbf{W}}_{\ss\R_k}^{j-1}\in\mathcal{W}_j}\left[\textrm{Pr}\left(\hat{\textbf{W}}_{\ss\R_k}^{j-1}\Big{|}\textbf{W}_{\ss\T}^{j-1}\right)\right.\nonumber\\
&\times\left(\textrm{Pr}\left(S_{\ob,{0}}^{({\ss{\T},{\ss\R_k}})}[j]\geq\xi_{\ss\R}\Big{|}W_{\ss\T}[j]=1,\textbf{W}_{\ss\T}^{j-1}\right)\right.\nonumber\\
&\times\textrm{Pr}\left(S_{\ob,{k}}^{({\ss\R_k,\ss\FC})}[j]<\xi_{\ss\FC_k}\Big{|}\hat{W}_{\ss\R_k}[j]=1,\hat{\textbf{W}}_{\ss\R_k}^{j-1}\right)\nonumber\\
&+\textrm{Pr}\left(S_{\ob,{0}}^{({\ss{\T},{\ss\R_k}})}[j]<\xi_{\ss\R}\Big{|}W_{\ss\T}[j]=1,\textbf{W}_{\ss\T}^{j-1}\right)\nonumber\\
&\times\left.\left.\textrm{Pr}\left(S_{\ob,{k}}^{({\ss\R_k,\ss\FC})}[j]<\xi_{\ss\FC_k}\Big{|}\hat{W}_{\ss\R_k}[j]=0,\hat{\textbf{W}}_{\ss\R_k}^{j-1}\right)\right)\right]
\end{align}
\fi
and
\ifOneCol	
\begin{align}\label{Pef2,origin}
\hspace{-2.5mm}\tilde{P}_{\fa,k}[j]=&\;
\sum_{\hat{\textbf{W}}_{\ss\R_k}^{j-1}\in\mathcal{W}_j}\left[\textrm{Pr}\left(\hat{\textbf{W}}_{\ss\R_k}^{j-1}\Big{|}\textbf{W}_{\ss\T}^{j-1}\right)\right.\nonumber\\
&\times\left(\textrm{Pr}\left(S_{\ob,{0}}^{({\ss{\T},{\ss\R_k}})}[j]\geq\xi_{\ss\R}\Big{|}W_{\ss\T}[j]=0,\textbf{W}_{\ss\T}^{j-1}\right)\right.
\textrm{Pr}\left(S_{\ob,{k}}^{({\ss\R_k,\ss\FC})}[j]\geq\xi_{\ss\FC_k}\Big{|}\hat{W}_{\ss\R_k}[j]=1,\hat{\textbf{W}}_{\ss\R_k}^{j-1}\right)\nonumber\\
&+\textrm{Pr}\left(S_{\ob,{0}}^{({\ss{\T},{\ss\R_k}})}[j]<\xi_{\ss\R}\Big{|}W_{\ss\T}[j]=0,\textbf{W}_{\ss\T}^{j-1}\right)\left.\left.\textrm{Pr}\left(S_{\ob,{k}}^{({\ss\R_k,\ss\FC})}[j]\geq\xi_{\ss\FC_k}\Big{|}\hat{W}_{\ss\R_k}[j]=0,\hat{\textbf{W}}_{\ss\R_k}^{j-1}\right)\right)\right],
\end{align}
\else
\begin{align}\label{Pef2,origin}
\hspace{-2.5mm}\tilde{P}_{\fa,k}[j]=&\;
\sum_{\hat{\textbf{W}}_{\ss\R_k}^{j-1}\in\mathcal{W}_j}\left[\textrm{Pr}\left(\hat{\textbf{W}}_{\ss\R_k}^{j-1}\Big{|}\textbf{W}_{\ss\T}^{j-1}\right)\right.\nonumber\\
&\times\left(\textrm{Pr}\left(S_{\ob,{0}}^{({\ss{\T},{\ss\R_k}})}[j]\geq\xi_{\ss\R}\Big{|}W_{\ss\T}[j]=0,\textbf{W}_{\ss\T}^{j-1}\right)\right.\nonumber\\
&\times\textrm{Pr}\left(S_{\ob,{k}}^{({\ss\R_k,\ss\FC})}[j]\geq\xi_{\ss\FC_k}\Big{|}\hat{W}_{\ss\R_k}[j]=1,\hat{\textbf{W}}_{\ss\R_k}^{j-1}\right)\nonumber\\
&+\textrm{Pr}\left(S_{\ob,{0}}^{({\ss{\T},{\ss\R_k}})}[j]<\xi_{\ss\R}\Big{|}W_{\ss\T}[j]=0,\textbf{W}_{\ss\T}^{j-1}\right)\nonumber\\
&\times\left.\left.\textrm{Pr}\left(S_{\ob,{k}}^{({\ss\R_k,\ss\FC})}[j]\geq\xi_{\ss\FC_k}\Big{|}\hat{W}_{\ss\R_k}[j]=0,\hat{\textbf{W}}_{\ss\R_k}^{j-1}\right)\right)\right],
\end{align}
\fi
respectively. Considering that the cooperative MC system having a symmetric topology, each $\RX_k$ has independent and identically distributed (i.i.d.) observations over each $\TX-\RX_k-\FC$ link. Under this consideration, we have $V_{\tilde{z},k}[j]=V_{\tilde{z}}[j]$. Accordingly, we assume that the decision thresholds at the FC over RXs-FC links are the same, i.e., $\xi_{\ss\FC_k} = \xi_{\ss\FC}$. Thus, in \eqref{Pem2,origin} and \eqref{Pef2,origin}, the likelihood of occurrence of each realization of $\hat{\textbf{W}}_{\ss\R_k}^{j-1}$ is the same for each $\RX_k$. Also, the conditional MDP and FAP for the given realization is the same for each $\RX_k$. This indicates that $\tilde{P}_{\md,k}[j]$ and $\tilde{P}_{\fa,k}[j]$ are the same for all RXs, i.e., $\tilde{P}_{{\md},k}[j]= \tilde{P}_{\md}[j]$ and $\tilde{P}_{{\fa},k}[j]= \tilde{P}_{\fa}[j]$. We note that the high complexity caused by considering $2^{(j-1)}$ possible realizations of $\hat{\textbf{W}}_{\ss\R_k}^{j-1}$ and their likelihood of occurrence make the evaluation of \eqref{Pem2,origin} and \eqref{Pef2,origin} cumbersome. To facilitate the calculation of \eqref{Pem2,origin} and \eqref{Pef2,origin}, we consider only one possible realization of $\hat{\textbf{W}}_{\ss\R_k}^{j-1}$ and refer to this considered realization as the candidate. By only considering the candidate of $\hat{\textbf{W}}_{\ss\R_k}^{j-1}$, we then approximate \eqref{Pem2,origin} and \eqref{Pef2,origin} as
\ifOneCol	
\begin{align}\label{Pem2}
\hspace{-2.5mm}\tilde{P}_{\md}[j]\approx&\;
\textrm{Pr}\left(S_{\ob,{0}}^{({\ss{\T},{\ss\R_k}})}[j]\geq\xi_{\ss\R}\Big{|}W_{\ss\T}[j]=1,\textbf{W}_{\ss\T}^{j-1}\right)
\textrm{Pr}\left(S_{\ob,{k}}^{({\ss\R_k,\ss\FC})}[j]<\xi_{\ss\FC}\Big{|}\hat{W}_{\ss\R_k}[j]=1,\hat{\textbf{W}}_{\ss\R_k}^{j-1}\right)\nonumber\\
&+\textrm{Pr}\left(S_{\ob,{0}}^{({\ss{\T},{\ss\R_k}})}[j]<\xi_{\ss\R}\Big{|}W_{\ss\T}[j]=1,\textbf{W}_{\ss\T}^{j-1}\right)
\textrm{Pr}\left(S_{\ob,{k}}^{({\ss\R_k,\ss\FC})}[j]<\xi_{\ss\FC}\Big{|}\hat{W}_{\ss\R_k}[j]=0,\hat{\textbf{W}}_{\ss\R_k}^{j-1}\right)
\end{align}
\else
\begin{align}\label{Pem2}
\hspace{-2.5mm}\tilde{P}_{\md}[j]\approx&\;
\textrm{Pr}\left(S_{\ob,{0}}^{({\ss{\T},{\ss\R_k}})}[j]\geq\xi_{\ss\R}\Big{|}W_{\ss\T}[j]=1,\textbf{W}_{\ss\T}^{j-1}\right)\nonumber\\
&\times\textrm{Pr}\left(S_{\ob,{k}}^{({\ss\R_k,\ss\FC})}[j]<\xi_{\ss\FC}\Big{|}\hat{W}_{\ss\R_k}[j]=1,\hat{\textbf{W}}_{\ss\R_k}^{j-1}\right)\nonumber\\
&+\textrm{Pr}\left(S_{\ob,{0}}^{({\ss{\T},{\ss\R_k}})}[j]<\xi_{\ss\R}\Big{|}W_{\ss\T}[j]=1,\textbf{W}_{\ss\T}^{j-1}\right)\nonumber\\
&\times\textrm{Pr}\left(S_{\ob,{k}}^{({\ss\R_k,\ss\FC})}[j]<\xi_{\ss\FC}\Big{|}\hat{W}_{\ss\R_k}[j]=0,\hat{\textbf{W}}_{\ss\R_k}^{j-1}\right)
\end{align}
\fi
and
\ifOneCol	
\begin{align}\label{Pef2}
\hspace{-2mm}\tilde{P}_{\fa}[j]\approx&\;
\textrm{Pr}\left(S_{\ob,{0}}^{({\ss{\T},{\ss\R_k}})}[j]\geq\xi_{\ss\R}\Big{|}W_{\ss\T}[j]=0,\textbf{W}_{\ss\T}^{j-1}\right)
\textrm{Pr}\left(S_{\ob,{k}}^{({\ss\R_k,\ss\FC})}[j]\geq\xi_{\ss\FC}\Big{|}\hat{W}_{\ss\R_k}[j]=1,\hat{\textbf{W}}_{\ss\R_k}^{j-1}\right)\nonumber\\
&+\textrm{Pr}\left(S_{\ob,{0}}^{({\ss{\T},{\ss\R_k}})}[j]<\xi_{\ss\R}\Big{|}W_{\ss\T}[j]=0,\textbf{W}_{\ss\T}^{j-1}\right)
\textrm{Pr}\left(S_{\ob,{k}}^{({\ss\R_k,\ss\FC})}[j]\geq\xi_{\ss\FC}\Big{|}\hat{W}_{\ss\R_k}[j]=0,\hat{\textbf{W}}_{\ss\R_k}^{j-1}\right),
\end{align}
\else
\begin{align}\label{Pef2}
\hspace{-2mm}\tilde{P}_{\fa}[j]\approx&\;
\textrm{Pr}\left(S_{\ob,{0}}^{({\ss{\T},{\ss\R_k}})}[j]\geq\xi_{\ss\R}\Big{|}W_{\ss\T}[j]=0,\textbf{W}_{\ss\T}^{j-1}\right)\nonumber\\
&\times\textrm{Pr}\left(S_{\ob,{k}}^{({\ss\R_k,\ss\FC})}[j]\geq\xi_{\ss\FC}\Big{|}\hat{W}_{\ss\R_k}[j]=1,\hat{\textbf{W}}_{\ss\R_k}^{j-1}\right)\nonumber\\
&+\textrm{Pr}\left(S_{\ob,{0}}^{({\ss{\T},{\ss\R_k}})}[j]<\xi_{\ss\R}\Big{|}W_{\ss\T}[j]=0,\textbf{W}_{\ss\T}^{j-1}\right)\nonumber\\
&\times\textrm{Pr}\left(S_{\ob,{k}}^{({\ss\R_k,\ss\FC})}[j]\geq\xi_{\ss\FC}\Big{|}\hat{W}_{\ss\R_k}[j]=0,\hat{\textbf{W}}_{\ss\R_k}^{j-1}\right),
\end{align}
\fi
where the candidate of $\hat{\textbf{W}}_{\ss\R_k}^{j-1}$ can be obtained using a biased coin toss method. Particularly, we model the $i$th decision at $\RX_k$, $\hat{W}_{\ss\R_k}[i]$, as $\hat{W}_{\ss\R_k}[i] = |\lambda-W_{\ss\T}[i]|$, where $i\in\{1,\cdots,j-1\}$ and $\lambda\in\{0,1\}$ is the outcome of the coin toss with $\textrm{Pr}(\lambda=1)=P_{\md,k}[i]$ if $W_{\ss\T}[i]=1$ and $\textrm{Pr}(\lambda=1)=P_{\fa,k}[i]$ if $W_{\ss\T}[i]=0$. We assume that the candidate of $\hat{\textbf{W}}_{\ss\R_k}^{j-1}$ in \eqref{Pem2} and \eqref{Pef2} is the same at all RXs to ensure that $\tilde{P}_{{\md},k}[j]= \tilde{P}_{\md}[j]$ and $\tilde{P}_{{\fa},k}[j]= \tilde{P}_{\fa}[j]$ are still valid after adopting the approximations of $\tilde{P}_{\md}[j]$and $\tilde{P}_{\fa}[j]$ in \eqref{Pem2} and \eqref{Pef2}, respectively. We clarify that the candidate is only considered for the \emph{theoretical} evaluation of system error performance, i.e., the calculation of \eqref{Pem2} and \eqref{Pef2}. We emphasize that we do not consider the same realizations of $\hat{\textbf{W}}_{\ss\R_k}^{j-1}$ at all RXs in our system model. Furthermore, in our simulations, each RX makes decisions independently and the realizations of $\hat{\textbf{W}}_{\ss\R_k}^{j-1}$ at all RXs are not necessarily identical. Our simulation results in Section V demonstrate the accuracy of \eqref{Pem2} and \eqref{Pef2}.}

{In addition, the CDF of ${S}_{\ob,{k}}^{({\ss\R_k,\ss\FC})}[j]$ is obtained by replacing $S_{\ob,0}^{({\ss{\T},{\ss\R_k}})}[j]$, $\xi_{\ss\R_k}$, $\textbf{W}_{\ss\T}^{j}$ and $\bar{S}_{\ob,0}^{({\ss{\T},{\ss\R_k}})}[j]$ with $S_{\ob,{k}}^{({\ss\R_k,\ss\FC})}[j]$, $\xi_{\ss\FC}$, $\textbf{W}_{\ss\R_k}^{j}$, ${\bar{S}}_{\ob,{k}}^{({\ss\R_k,\ss\FC})}[j]$ in \eqref{RX}, respectively. Using the CDFs of $S_{\ob,0}^{({\ss{\T},{\ss\R_k}})}[j]$ and ${S}_{\ob,{k}}^{({\ss\R_k,\ss\FC})}[j]$ and their complementary functions, we can find the closed-form expressions for $\tilde{P}_{\md,k}[j]$ and $\tilde{P}_{\fa,k}[j]$.}

\subsubsection{Global Error Probability}
In the noisy reporting scenario, we obtain $Q_{\md}[j]$ and $Q_{\fa}[j]$ for the $N$-out-of-$K$ rule, OR rule, and AND rule by replacing $P_{\md}[j]$ and $P_{\fa}[j]$ with $\tilde{P}_{\md}[j]$ and $\tilde{P}_{\fa}[j]$, respectively, in \eqref{Qm_KN}--\eqref{Qf_AND}. We also note that the expected error probability of the single-RX MC system in the $j$th symbol interval for a \emph{given} transmitter sequence $\textbf{W}_{\ss\T}^{j-1}$ in the noisy reporting scenario 
can be obtained by setting $K=1$. 

{We note that the expected error probabilities of the cooperative MC system are also tractable if weighted sum detectors with different weights are considered at the RXs and FC for detection. Under this consideration, the $m$th sample at $\RX_k$ or $\FC$ in the $j$th symbol interval can be accurately approximated by a Poisson RV. Although the weighted sums of Poisson RVs, $S_{\ob,0}^{({\ss{\T},{\ss\R_k}})}[j]$ and ${S}_{\ob,{k}}^{({\ss\R_k,\ss\FC})}[j]$, are not Poisson RVs, the weighted sums of Gaussian approximations of the individual variables are Gaussian RVs. Thus, we can write the CDF of the Gaussian RVs and complementary functions to derive $P_{\md,k}[j]$, $P_{\fa,k}[j]$, $\tilde{P}_{\md,k}[j]$, and $\tilde{P}_{\fa,k}[j]$. Then we can derive the global error probabilities using \eqref{Qm_KN}--\eqref{Qf_AND} for the cooperative MC system in the perfect and noisy reporting scenarios.}


\section{Error Performance Optimization}\label{sec:Convex Optimization}

In this section, we present {a} novel analysis to determine the joint optimal $\xi_{\ss\R}$ and $\xi_{\ss\FC}$ that minimize the global error probability of the cooperative MC system. To this end, we first derive the convex upper bounds on $Q_{\ss\FC}[j]$ for the OR rule, AND rule, and $N$-out-of-$K$ rule\footnote{We clarify that the convex upper bounds for the OR rule, AND rule, and $N$-out-of-$K$ rule are derived separately. This is due to the fact that the derived convex upper bounds for the $N$-out-of-$K$ rule with $N=1$ and $N=K$ are not as tight as those derived for the OR rule and AND rule, respectively.} in the perfect and noisy reporting scenarios, allowing us to formulate the corresponding convex optimization problems for given $\textbf{W}_{\ss\T}^{j-1}$. We then extend the formulated convex optimization problems for given $\textbf{W}_{\ss\T}^{j-1}$ to the convex optimization problems for the \emph{average} error performance over all possible realizations of $\textbf{W}_{\ss\T}^{j-1}$ and across all symbol intervals. This extension is due to two reasons. First, optimizing the instantaneous error performance for given $\textbf{W}_{\ss\T}^{j-1}$ may not be feasible in practice. This optimization mandates the precise knowledge of $\textbf{W}_{\ss\T}^{j-1}$ at $\RX_k$, which cannot be realized in practice.
Second, the repeated optimization of the detection threshold for each realization of $\textbf{W}_{\ss\T}^{j-1}$ would demand a high computational overhead for $\RX_k$.



\subsection{Perfect Reporting}\label{sec:Perfect Reporting Optimization1}

In this subsection, we formulate the convex optimization problems with respect to $\xi_{\ss\R}$ for the OR rule, AND rule, and $N$-out-of-$K$ rule in the perfect reporting scenario. To achieve this, we first analyze the convexity of $P_{\md}[j]^K$ and $P_{\fa}[j]^K$ with respect to $\xi_{\ss\R}$. Since $S_{\ob,0}^{({\ss{\T},{\ss\R}})}[j]$ is a Poisson RV with a discrete distribution, its convexity analysis with respect to $\xi_{\ss\R}$ is cumbersome. To overcome this cumbersomeness, we approximate the CDF of a Poisson RV $X$ with mean $\lambda$ by the CDF of a continuous Gaussian RV. {We find that the accuracy of this approximation becomes higher when $\lambda$ increases. Thus, the tightness of the approximation can be ensured by any method achieving large $\lambda$, such as increasing the number of molecules released, increasing the volume (radius) of RXs and FC, and choosing the optimal sampling period.} Including a continuity correction, the CDF of the Gaussian RV is given by
\begin{equation}\label{Gaussian_Approximation}
\textrm{Pr}\left(X<x\right)=\frac{1}{2}\left[1+{\Lambda}\left(x,\lambda\right)\right],
\end{equation}
where ${\Lambda}\left(x,\lambda\right)=\erf\left((x-0.5-\lambda)/\sqrt{2\lambda}\right)$.
Applying \eqref{Gaussian_Approximation} into \eqref{Pm1} and \eqref{Pf1}, $P_{\md}[j]$ and $P_{\md}[j]$ are approximated as
\begin{align}\label{Pm1 approx}
P_{\md}[j] \approx \frac{1}{2}\left[1 +{\Lambda}\left(\xi_{\ss\R},U_{1}[j]\right)\right]
\end{align}
and
\begin{align}\label{Pf1 approx}
P_{\fa}[j] \approx 1-\frac{1}{2}\left[1 +{\Lambda}\left(\xi_{\ss\R},U_{0}[j]\right) \right],
\end{align}
respectively. We now present the constraints making $P_{\md}[j]^K$ and $P_{\fa}[j]^K$ convex in the following theorem.
\begin{theorem}\label{theorem 5}
$P_{\md}[j]^K$ and $P_{\fa}[j]^K$ are convex with respect to $\xi_{\ss\R}$, if we impose the following convex constraints:
\begin{align}\label{Constraint 1}
-0.5-U_{1}[j]+\xi_{\ss\R}\leq0
\end{align}	
and 	
\begin{align}\label{Constraint 2}
0.5+U_{0}[j]-\xi_{\ss\R}\leq0,
\end{align}
respectively.
\end{theorem}
\begin{IEEEproof}
The convexity of $P_{\md}[j]^K$ can be proven by showing that its second derivative with respect to $\xi_{\ss\R}$ is {nonnegative}~\cite{Convex Optimization}. We derive the second derivative of $P_{{\md}}[j]^K$ as
\ifOneCol
\begin{align}\label{QmdDe1 perfect}
\hspace{-2mm}\frac{\partial^2{P_{{\md}}[j]^K}}{\partial{\xi_{\ss\R}}^2}=&\;
\frac{1}{2^K}\left(\frac{2(-1+K)K}{\pi}\Xi\left(-2+K,2,1\right)\right.
\left.+\sqrt{\frac{2}{\pi}}K\left(0.5+U_{1}[j]-\xi_{\ss\R}\right)
\Xi\left(-1+K,1,\frac{3}{2}\right)\right),
\end{align}
\else
\begin{align}\label{QmdDe1 perfect}
\hspace{-2mm}\frac{\partial^2{P_{{\md}}[j]^K}}{\partial{\xi_{\ss\R}}^2}=&\;
\frac{1}{2^K}\left(\frac{2(-1+K)K}{\pi}\Xi\left(-2+K,2,1\right)\right.\nonumber\\
&\hspace{-15mm}\left.+\sqrt{\frac{2}{\pi}}K\left(0.5+U_{1}[j]-\xi_{\ss\R}\right)
\Xi\left(-1+K,1,\frac{3}{2}\right)\right),
\end{align}
\fi
where
\begin{align}
\Xi\left(\alpha,\beta,\gamma\right)=\frac{\left(1+{\Lambda}\left(\xi_{\ss\R},U_{1}[j]\right)\right)^{\alpha}{{\Theta}\left(\xi_{\ss\R},U_{1}[j]\right)}^{\beta}}{U_{1}[j]^{\gamma}}
\end{align}
and ${\Theta}\left(x,\lambda\right)\triangleq\exp{\big(}-\left(0.5+\lambda-x\right)^{2}/{2\lambda}{\big)}$. Due to the fact that the value of ${\Lambda}\left(x,\lambda\right)$ is between $-1$ and $1$ and the value of ${\Theta}\left(x,\lambda\right)$ is always greater than zero, \eqref{QmdDe1 perfect} is always {nonnegative} if we impose the constraint \eqref{Constraint 1}. Following a similar procedure, we prove that $P_{\fa}[j]^K$ is also convex with respect to $\xi_{\ss\R}$, if we impose the constraint \eqref{Constraint 2}.
\end{IEEEproof}

We now analyze the convexity of $Q_{\fa}[j]$ and $Q_{\md}[j]$ for the three rules. For the OR rule, an upper bound on $Q_{\fa}[j]$ is given by
\begin{align}\label{Upper Q_fa}
Q_{\fa}[j] \leq KP_{\fa}[j],
\end{align}
which is obtained by applying the first degree Taylor series approximation of $1-(1-P_{\fa}[j])^K$ into \eqref{Qf_OR} at $P_{\fa}[j]=0$. We find that this upper bound is tight when $P_{\fa}[j]$ is small. We note that $P_{\fa}[j]$ is convex with respect to $\xi_{\ss\R}$, if we impose the constraint \eqref{Constraint 2}, which can be proven by considering $K=1$ in {Theorem \ref{theorem 5}}. Thus, the upper bound in \eqref{Upper Q_fa} is also convex with respect to $\xi_{\ss\R}$ under the same constraint, since it scales a convex function with a {nonnegative} constant. Also based on {Theorem \ref{theorem 5}}, $Q_{\md}[j]$ for the OR rule, $P_{\md}[j]^K$, is convex with respect to $\xi_{\ss\R}$, if we impose the constraint \eqref{Constraint 1}. Therefore, the convex optimization problem for the cooperative MC system with the OR rule in the perfect reporting scenario is formulated as
\begin{equation}\label{perfect OR}
\begin{aligned}
& \underset{\xi_{\ss\R}}{\text{min}}
& & P_1P_{{\md}}[j]^K + \left(1-P_1\right)KP_{\fa}[j] \\
& \text{s.t.}
& & \eqref{Constraint 1}~\text{and}~\eqref{Constraint 2}. \\
\end{aligned}
\end{equation}

Due to the convexity of the objective function and the constraints, \eqref{perfect OR} can be quickly solved by efficient algorithms, e.g., the interior-point method~\cite{Convex Optimization}. Throughout this paper, we refer to the optimal threshold, i.e., the threshold in the feasible set that minimizes the objective function, as the solution to the convex optimization problem,
where the feasible set is the set containing all of the thresholds that satisfy all constraints.

Next, we focus on the AND rule. Using a similar method as in \eqref{Upper Q_fa}, $Q_{\md}[j]$ is upper-bounded by
\begin{align}\label{Upper Q_md}
Q_{\md}[j] \leq KP_{\md}[j].
\end{align}

We note that $P_{\md}[j]$ is convex with respect to $\xi_{\ss\R}$ under the constraint \eqref{Constraint 1}, which can be proven by considering $K=1$ in {Theorem \ref{theorem 5}}. Thus, \eqref{Upper Q_md} is also convex with respect to $\xi_{\ss\R}$ under the same constraint. Based on {Theorem \ref{theorem 5}}, $Q_{\fa}[j]$ for the AND rule, $P_{\fa}[j]^K$, is convex respect to $\xi_{\ss\R}$, if we impose the constraint \eqref{Constraint 2}. Therefore, the convex optimization problem for the cooperative MC system with the AND rule in the perfect reporting scenario can be formulated as
\begin{equation}\label{perfect AND}
\begin{aligned}
& \underset{\xi_{\ss\R}}{\text{min}}
& & P_1KP_{{\md}}[j]+ \left(1-P_1\right)P_{\fa}[j]^K \\
& \text{s.t.}
& & \eqref{Constraint 1}~\text{and}~\eqref{Constraint 2}. \\
\end{aligned}
\end{equation}

Finally, we consider the $N$-out-of-$K$ rule. We rewrite \eqref{Qm_KN} as
\begin{align}\label{Qm_KN re}
Q_{\md}[j]=\sum\limits^{K}_{\tilde{n}=K-N+1}\binom{K}{\tilde{n}}{P_{\md}[j]}^{\tilde{n}}\left(1-P_{{\md}}[j]\right)^{K-\tilde{n}}.
\end{align}

Based on \eqref{Qm_KN re} and \eqref{Qf_KN}, we verify that
\begin{align}\label{Qm_KN upp}
Q_{\md}[j]\leq\sum\limits^{K}_{\tilde{n}=K-N+1}\binom{K}{\tilde{n}}{P_{\md}[j]}^{\tilde{n}}\triangleq{}Q_{\md}^{+}[j]
\end{align}
and
\begin{align}\label{Qf_KN upp}
Q_{\fa}[j]\leq\sum\limits^{K}_{n=N}\binom{K}{n}{P_{\fa}[j]}^{n}\triangleq{}Q_{\fa}^{+}[j].
\end{align}

In {Theorem \ref{theorem 5}}, we showed that $P_{{\md}}[j]^K$ and $P_{{\fa}}[j]^K$ are convex with respect to $\xi_{\ss\R}$, if we impose the convex constraints \eqref{Constraint 1} and \eqref{Constraint 2}, respectively.
{We note that $P_{{\md}}[j]^{\tilde{n}}$ and $P_{{\fa}}[j]^n$, where $\tilde{n}\in\{K-N+1,{\ldots},K\}$ and $n\in\{N,{\ldots},K\}$, are also convex with respect to $\xi_{\ss\R}$, if we impose the convex constraints \eqref{Constraint 1} and \eqref{Constraint 2}, respectively. The convexity of $P_{{\md}}[j]^{\tilde{n}}$ and $P_{{\fa}}[j]^n$ with respect to $\xi_{\ss\R}$ can be proven by replacing $K$ with $\tilde{n}$ and $n$ in the proof to {Theorem \ref{theorem 5}}, respectively.} Since \eqref{Qm_KN upp} and \eqref{Qf_KN upp} are {nonnegative} weighted sums of convex functions, i.e., ${P_{\md}[j]}^{\tilde{n}}$ and ${P_{\fa}[j]}^{n}$, they are also convex with respect to $\xi_{\ss\R}$ under the same constraints. Therefore, the convex optimization problem for the cooperative MC system with the $N$-out-of-$K$ rule in the perfect reporting scenario is formulated as
\begin{equation}\label{perfect NK}
\begin{aligned}
& \underset{\xi_{\ss\R}}{\text{min}}
& &P_{1}Q_{\md}^{+}[j]+(1-P_1)Q_{\fa}^{+}[j]\\
& \text{s.t.}
& & \eqref{Constraint 1}~\text{and}~\eqref{Constraint 2}. \\
\end{aligned}
\end{equation}

We note that the convex optimization problem for the single-RX system in the perfect reporting scenario is a special case of problems \eqref{perfect OR}, \eqref{perfect AND}, and \eqref{perfect NK}, with $K=1$.

\subsection{Noisy Reporting}\label{sec:Noisy Reporting Optimization}

In this subsection, we first extend the formulated convex optimization problems from the perfect reporting scenario to the noisy reporting scenario, assuming that $\xi_{\ss\FC}$ is fixed. We then formulate the joint convex optimization problems with respect to \emph{both} $\xi_{\ss\R}$ and $\xi_{\ss\FC}$ for the OR rule, AND rule, and $N$-out-of-$K$ rule.

\subsubsection{Optimal $\xi_{\ss\R}$}

We first analyze the convexity of $\tilde{P}_{\md}[j]^K$ and $\tilde{P}_{\fa}[j]^K$ with respect to $\xi_{\ss\R}$. To facilitate the convexity analysis of $\tilde{P}_{\md}[j]^K$ and $\tilde{P}_{\fa}[j]^K$ with respect to $\xi_{\ss\R}$, we approximate \eqref{Pem2} and \eqref{Pef2} using \eqref{Gaussian_Approximation}, which result in
\ifOneCol
\begin{align}\label{Pem2 approx}
\tilde{P}_{\md}[j]\approx&\;\frac{1}{4}\left(2+\left(1+{\Lambda}\left(\xi_{\ss\R},U_{1}[j]\right)\right){\Lambda}\left(\xi_{\ss\FC},V_{0}[j]\right)\right.
\left.+\left(1-{\Lambda}\left(\xi_{\ss\R},U_{1}[j]\right)\right){\Lambda}\left(\xi_{\ss\FC},V_{1}[j]\right)\right)
\end{align}	
\else
\begin{align}\label{Pem2 approx}
\tilde{P}_{\md}[j]\approx&\;\frac{1}{4}\left[2+\left(1+{\Lambda}\left(\xi_{\ss\R},U_{1}[j]\right)\right){\Lambda}\left(\xi_{\ss\FC},V_{0}[j]\right)\right.\nonumber\\
&\left.+\left(1-{\Lambda}\left(\xi_{\ss\R},U_{1}[j]\right)\right){\Lambda}\left(\xi_{\ss\FC},V_{1}[j]\right)\right]
\end{align}		
\fi
and
\ifOneCol
\begin{align}\label{Pef2 approx}
\tilde{P}_{\fa}[j]\approx&\;\frac{1}{4}\left(2-\left(1+{\Lambda}\left(\xi_{\ss\R},U_{0}[j]\right)\right){\Lambda}\left(\xi_{\ss\FC},V_{0}[j]\right)\right.
\left.+\left(-1+{\Lambda}\left(\xi_{\ss\R},U_{0}[j]\right)\right){\Lambda}\left(\xi_{\ss\FC},V_{1}[j]\right)\right),
\end{align}
\else
\begin{align}\label{Pef2 approx}
\tilde{P}_{\fa}[j]\approx&\;\frac{1}{4}\left[2-\left(1+{\Lambda}\left(\xi_{\ss\R},U_{0}[j]\right)\right){\Lambda}\left(\xi_{\ss\FC},V_{0}[j]\right)\right.\nonumber\\
&\left.+\left(-1+{\Lambda}\left(\xi_{\ss\R},U_{0}[j]\right)\right){\Lambda}\left(\xi_{\ss\FC},V_{1}[j]\right)\right],
\end{align}
\fi
respectively. Recall that $V_{\tilde{z},k}[j]$, $\tilde{z}\in{\{0,1\}}$, denotes the conditional mean of ${S}_{\ob,{k}}^{({\ss\R_k,\ss\FC})}[j]$ when the most recent information symbol transmitted by the $\RX_k$ is $\tilde{z}$. We find that $V_{\tilde{z}}[j]$ depends on $\hat{\textbf{W}}_{\ss\R_k}^{j-1}$ and $\hat{\textbf{W}}_{\ss\R_k}^{j-1}$ depends on $\xi_{\ss\R}$. Thus, $V_{\tilde{z}}[j]$ depends on $\xi_{\ss\R}$, which complicates the convexity analysis of $\tilde{P}_{\md}[j]^K$ and $\tilde{P}_{\fa}[j]^K$ with respect to $\xi_{\ss\R}$. To avoid this complication, we consider a constant $V_{\tilde{z}}[j]$ in the $j$th symbol interval, denoted by $\overline{V}_{\tilde{z}}[j]$, which is averaged over all the realizations of $\hat{\textbf{W}}_{\ss\R_k}^{j-1}$, to approximate $V_{\tilde{z}}[j]$ in \eqref{Pem2 approx} and \eqref{Pef2 approx}{\footnote{{We note that the occurrence likelihood of each realization of $\hat{\textbf{W}}_{\ss\R_k}^{j-1}$ may not be the same in practice, since it depends on the value of $\xi_{\ss\R}$. For example, when $\xi_{\ss\RX}$ is very high, $\hat{\textbf{W}}_{\ss\R_k}^{j-1}$ would be all ``0''s, while when $\xi_{\ss\RX}$ is very small, $\hat{\textbf{W}}_{\ss\R_k}^{j-1}$ would be all ``1''s. In this paper, we assume an equal occurrence likelihood to keep a low evaluation complexity, but this does not have a significant impact on the analytical results.}}}. {By doing so, we obtain $\overline{V}_{\tilde{z}}[j]$ as
\begin{align}\label{constant V}
\overline{V}_{\tilde{z}}[j] = \frac{1}{|\omega_j|}\sum_{\hat{\textbf{W}}_{\ss\R_k}^{j-1}\in\omega_j} V_{\tilde{z}}[j],
\end{align}
where $\omega_j$ is the set containing all realizations of $\hat{\textbf{W}}_{\ss\R_k}^{j-1}$ and $|\omega_j|$ denotes the cardinality of $\omega_j$. }Using $\overline{V}_{\tilde{z}}[j]$, we further approximate $\tilde{P}_{\md}[j]$ and $\tilde{P}_{\fa}[j]$ as \footnote{{We clarify that we approximate $V_{\tilde{z}}[j]$ by $\overline{V}_{\tilde{z}}[j]$ in \eqref{Pem2 approx} and \eqref{Pef2 approx} to obtain \eqref{Pem2 approx averISI} and \eqref{Pef2 approx averISI} for the \emph{convexity analysis}. This approximation is not for deriving the expected MDP and FAP of each $\TX-\RX_k-\FC$ link in the $j$th symbol interval, i.e., $\tilde{P}_{\md,k}[j]$ and $\tilde{P}_{\fa,k}[j]$, since \eqref{Pem2 approx} and \eqref{Pef2 approx} are already the approximations of $\tilde{P}_{\md,k}[j]$ and $\tilde{P}_{\fa,k}[j]$, respectively.
}}
\ifOneCol
\begin{align}\label{Pem2 approx averISI}
\tilde{P}_{\md}[j]\approx&\;\frac{1}{4}\left[2+\left(1+{\Lambda}\left(\xi_{\ss\R},U_{1}[j]\right)\right){\Lambda}\left(\xi_{\ss\FC},\overline{V}_{0}[j]\right)\right.
\left.+\left(1-{\Lambda}\left(\xi_{\ss\R},U_{1}[j]\right)\right){\Lambda}\left(\xi_{\ss\FC},\overline{V}_{1}[j]\right)\right]
\end{align}	
\else
\begin{align}\label{Pem2 approx averISI}
\tilde{P}_{\md}[j]\approx&\;\frac{1}{4}\left[2+\left(1+{\Lambda}\left(\xi_{\ss\R},U_{1}[j]\right)\right){\Lambda}\left(\xi_{\ss\FC},\overline{V}_{0}[j]\right)\right.\nonumber\\
&\left.+\left(1-{\Lambda}\left(\xi_{\ss\R},U_{1}[j]\right)\right){\Lambda}\left(\xi_{\ss\FC},\overline{V}_{1}[j]\right)\right]
\end{align}		
\fi
and
\ifOneCol
\begin{align}\label{Pef2 approx averISI}
\tilde{P}_{\fa}[j]\approx&\;\frac{1}{4}\left[2-\left(1+{\Lambda}\left(\xi_{\ss\R},U_{0}[j]\right)\right){\Lambda}\left(\xi_{\ss\FC},\overline{V}_{0}[j]\right)\right.
\left.+\left(-1+{\Lambda}\left(\xi_{\ss\R},U_{0}[j]\right)\right){\Lambda}\left(\xi_{\ss\FC},\overline{V}_{1}[j]\right)\right],
\end{align}
\else
\begin{align}\label{Pef2 approx averISI}
\hspace{-2mm}\tilde{P}_{\fa}[j]\approx&\;\frac{1}{4}\left[2-\left(1+{\Lambda}\left(\xi_{\ss\R},U_{0}[j]\right)\right){\Lambda}\left(\xi_{\ss\FC},\overline{V}_{0}[j]\right)\right.\nonumber\\
&\left.+\left(-1+{\Lambda}\left(\xi_{\ss\R},U_{0}[j]\right)\right){\Lambda}\left(\xi_{\ss\FC},\overline{V}_{1}[j]\right)\right],
\end{align}
\fi
respectively. We now present the conditions making $\tilde{P}_{\md}[j]^K$ and $\tilde{P}_{\fa}[j]^K$ convex in the following theorem.
\begin{theorem}\label{theorem 2}
$\tilde{P}_{\md}[j]^K$ and $\tilde{P}_{\fa}[j]^K$ are convex with respect to $\xi_{\ss\R}$ when $\xi_{\ss\FC}$ is fixed, if we impose the convex constraints \eqref{Constraint 1} and \eqref{Constraint 2}, respectively.
\end{theorem}
\begin{IEEEproof}
We derive the second derivative of $\tilde{P}_{{\md}}[j]^K$ as
\ifOneCol
\begin{align}\label{QmdDe1}
\frac{\partial^2 {\tilde{P}_{{\md}}[j]^K}}{\partial {\xi_{\ss\R}}^2}=&\;\left(-0.5-U_{1}[j]+\xi_{\ss\R}\right)
\Upsilon\left(K-1,1,3/2\right)
+(K-1)
\Upsilon\left(K-2,2,1\right)
\end{align}
\else
\begin{align}\label{QmdDe1}
\frac{\partial^2 {\tilde{P}_{{\md}}[j]^K}}{\partial {\xi_{\ss\R}}^2}=&\;\left(-0.5-U_{1}[j]+\xi_{\ss\R}\right)
\Upsilon\left(K-1,1,3/2\right)\nonumber\\
&+(K-1)
\Upsilon\left(K-2,2,1\right)
\end{align}
\fi
where
\ifOneCol
\begin{align}\label{Upsilon}
\Upsilon\left(\alpha,\beta,\gamma\right) = &\;\biggl[\left(1+{\Lambda}\left(\xi_{\ss\R},U_{1}[j]\right)\right)\left(1+{\Lambda}\left(\xi_{\ss\FC},\overline{V}_{0}[j]\right)\right)\nonumber\\
&+\left(1-{\Lambda}\left(\xi_{\ss\R},U_{1}[j]\right)\right)\left(1+{\Lambda}\left(\xi_{\ss\FC},\overline{V}_{1}[j]\right)\right)\biggr]^\alpha\nonumber\\
&\times\left({\Lambda}\left(\xi_{\ss\FC},\overline{V}_{1}[j]\right)-{\Lambda}\left(\xi_{\ss\FC},\overline{V}_{0}[j]\right)\right)^\beta
\frac{K\Theta\left(\xi_{\ss\R},U_{1}[j]\right)^\beta}{U_{1}[j]^\gamma4^\alpha(2\sqrt{2\pi})^\beta}.
\end{align}
\else
\begin{align}\label{Upsilon}
\Upsilon\left(\alpha,\beta,\gamma\right) = &\;\biggl[\left(1+{\Lambda}\left(\xi_{\ss\R},U_{1}[j]\right)\right)\left(1+{\Lambda}\left(\xi_{\ss\FC},\overline{V}_{0}[j]\right)\right)\nonumber\\
&+\left(1-{\Lambda}\left(\xi_{\ss\R},U_{1}[j]\right)\right)\left(1+{\Lambda}\left(\xi_{\ss\FC},\overline{V}_{1}[j]\right)\right)\biggr]^\alpha\nonumber\\
&\times\left({\Lambda}\left(\xi_{\ss\FC},\overline{V}_{1}[j]\right)-{\Lambda}\left(\xi_{\ss\FC},\overline{V}_{0}[j]\right)\right)^\beta\nonumber\\
&\times\frac{K\Theta\left(\xi_{\ss\R},U_{1}[j]\right)^\beta}{U_{1}[j]^\gamma4^\alpha(2\sqrt{2\pi})^\beta}.
\end{align}
\fi

We next examine the monotonicity of ${\Lambda}\left(\xi_{\ss\FC},V\right)$ with respect to $V$, $V\in\{\overline{V}_{1,k}[j],\overline{V}_{0,k}[j]\}$. We derive the first derivative of ${\Lambda}\left(\xi_{\ss\FC},\lambda\right)$ with respect to $\lambda$ as
\begin{align}\label{prove1}
\frac{\partial{\Lambda}\left(\xi_{\ss\FC},\lambda\right)}{\partial\lambda}=
\frac{2{\Theta}\left(\xi_{\ss\FC},\lambda\right)\left(-\xi_{\ss\FC}+0.5-\lambda\right)}{2\sqrt{2\pi}\lambda^\frac{3}{2}}.
\end{align}
Since ${\Theta}\left(\xi_{\ss\FC},\lambda\right)>0$ and $\left(-\xi_{\ss\FC}+0.5-\lambda\right)<0$, we find that ${\Lambda}\left(\xi_{\ss\FC},\lambda\right)$ is a monotonically decreasing function with respect to $\lambda$. Therefore, we have $\left({\Lambda}\left(\xi_{\ss\FC},\overline{V}_{1}[j]\right)-{\Lambda}\left(\xi_{\ss\FC},\overline{V}_{0}[j]\right)\right)\leq0$. It follows that \eqref{QmdDe1} is always {nonnegative} if we impose the constraint \eqref{Constraint 1}, and thus $\tilde{P}_{{\md}}[j]^{K}$ is convex with respect to $\xi_{\ss\R}$. Similarly, we prove that $\tilde{P}_{\fa}[j]^K$ is convex with respect to $\xi_{\ss\R}$, if we impose the convex constraint \eqref{Constraint 2}.
\end{IEEEproof}

Similar to \eqref{Upper Q_fa} and \eqref{Upper Q_md}, we upper-bound $Q_{\fa}[j]$ for the OR rule and $Q_{\md}[j]$ for the AND rule as
\begin{align}\label{Upper Q_fa noisy}
Q_{\fa}[j]\leq K\tilde{P}_{\fa}[j]
\end{align}
and
\begin{align}\label{Upper Q_md noisy}
Q_{\md}[j]\leq K\tilde{P}_{\md}[j],
\end{align}
respectively. We note that $\tilde{P}_{\fa}[j]$ and $\tilde{P}_{\md}[j]$ are convex with respect to $\xi_{\ss\R}$, if we impose the constraints \eqref{Constraint 1} and \eqref{Constraint 2}, respectively, which can be proven by considering $K=1$ in {Theorem \ref{theorem 2}}. Since \eqref{Upper Q_fa noisy} and \eqref{Upper Q_md noisy} scale a convex function with a {nonnegative} constant, they are also convex with respect to $\xi_{\ss\R}$ under the same constraints. Next, we focus on $Q_{\md}[j]$ for the OR rule and $Q_{\fa}[j]$ for the AND rule. Based on {Theorem \ref{theorem 2}}, we note that $\tilde{P}_{{\md}}[j]^K$ and $\tilde{P}_{\fa}[j]^K$ are convex with respect to $\xi_{\ss\R}$ when $\xi_{\ss\FC}$ is fixed, if we impose the convex constraints \eqref{Constraint 1} and \eqref{Constraint 2}, respectively. {Then, we focus on the $N$-out-of-$K$ rule. We note that $\tilde{P}_{{\md}}[j]^{\tilde{n}}$ and $\tilde{P}_{\fa}[j]^n$ are convex with respect to $\xi_{\ss\R}$ when $\xi_{\ss\FC}$ is fixed, if we impose the convex constraints \eqref{Constraint 1} and \eqref{Constraint 2}, respectively, which can be proven by replacing $K$ with $\tilde{n}$ and $n$ in {Theorem \ref{theorem 2}}, respectively.} For the $N$-out-of-$K$ rule, using a similar method to \eqref{Qm_KN re}--\eqref{Qf_KN upp}, we can derive the upper bounds on $Q_{\md}[j]$ and $Q_{\fa}[j]$ that are convex with respect to $\xi_{\ss\R}$, given that $\tilde{P}_{{\md}}[j]^{\tilde{n}}$ and $\tilde{P}_{\fa}[j]^n$ are convex with respect to $\xi_{\ss\R}$.

In the noisy reporting scenario, we formulate the convex optimization problems with respect to $\xi_{\ss\R}$ given fixed $\xi_{\ss\FC}$ for the OR rule, AND rule, and $N$-out-of-$K$ rule by replacing $P_{\md}[j]$ and $P_{\fa}[j]$ with $\tilde{P}_{\md}[j]$ and $\tilde{P}_{\fa}[j]$, respectively, in \eqref{perfect OR}, \eqref{perfect AND}, and \eqref{perfect NK}. We note that the convex optimization problem with respect to $\xi_{\ss\R_k}$ for the single-RX system in the noisy reporting scenario is a special case of the corresponding problem for a cooperative MC system with $K=1$.

\subsubsection{Joint Optimal $\xi_{\ss\R}$ and $\xi_{\ss\FC}$}

We first analyze the joint convexity of $\tilde{P}_{\md}[j]^K$ and $\tilde{P}_{\fa}[j]^K$ with respect to $\xi_{\ss\R}$ and $\xi_{\ss\FC}$. To facilitate the joint convexity analysis of $\tilde{P}_{\md}[j]^K$ and $\tilde{P}_{\fa}[j]^K$ with respect to both $\xi_{\ss\R}$ and $\xi_{\ss\FC}$, we consider the approximations given by
\begin{align}\label{upp1}
\hspace{-7mm}\textrm{Pr}\left(S_{\ob,{k}}^{({\ss\R_k,\ss\FC})}[j]<\xi_{\ss\FC}{\Big{|}}\hat{W}_{\ss\R_k}[j]=0,\hat{\textbf{W}}_{\ss\R_k}^{j-1}\right){\approx}\;1
\end{align}
and
\begin{align}\label{upp2}
\textrm{Pr}\left(S_{\ob,{k}}^{({\ss\R_k,\ss\FC})}[j]\geq\xi_{\ss\FC}{\Big{|}}\hat{W}_{\ss\R_k}[j]=1,\hat{\textbf{W}}_{\ss\R_k}^{j-1}\right){\approx}\;1,
\end{align}
which are tight when the error probability of the $\RX_k-\FC$ link is low. {We emphasize that we still keep
\begin{align}
\textrm{Pr}\left(S_{\ob,{k}}^{({\ss\R_k,\ss\FC})}[j]<\xi_{\ss\FC}{\Big{|}}\hat{W}_{\ss\R_k}[j]=1,\hat{\textbf{W}}_{\ss\R_k}^{j-1}\right)
\end{align}
and
\begin{align}
\textrm{Pr}\left(S_{\ob,{k}}^{({\ss\R_k,\ss\FC})}[j]\geq\xi_{\ss\FC}{\Big{|}}\hat{W}_{\ss\R_k}[j]=0,\hat{\textbf{W}}_{\ss\R_k}^{j-1}\right)
\end{align}
in \eqref{Pem2 approx averISI} and \eqref{Pef2 approx averISI}, respectively.} Employing \eqref{upp1} and \eqref{upp2} into \eqref{Pem2 approx averISI} and \eqref{Pef2 approx averISI}, respectively, we further upper-bound $\tilde{P}_{\md}[j]$ and $\tilde{P}_{\fa}[j]$ as
\ifOneCol
\begin{align}\label{Pem2 upp}
\tilde{P}_{\mdb}[j]=&\;\frac{1}{4}\left[3+{\Lambda}\left(\xi_{\ss\R},U_{1}[j]\right)\right.
\left.+\left(1-{\Lambda}\left(\xi_{\ss\R},U_{1}[j]\right)\right){\Lambda}\left(\xi_{\ss\FC},\overline{V}_{1}[j]\right)\right]
\end{align}
\else
\begin{align}\label{Pem2 upp}
\tilde{P}_{\mdb}[j]=&\;\frac{1}{4}\left[3+{\Lambda}\left(\xi_{\ss\R},U_{1}[j]\right)\right.\nonumber\\
&\left.+\left(1-{\Lambda}\left(\xi_{\ss\R},U_{1}[j]\right)\right){\Lambda}\left(\xi_{\ss\FC},\overline{V}_{1}[j]\right)\right]
\end{align}
\fi
and
\ifOneCol
\begin{align}\label{Pef2 upp}
\tilde{P}_{\fab}[j]=&\;\frac{1}{4}\left[3-{\Lambda}\left(\xi_{\ss\FC},\overline{V}_{0}[j]\right)\right.
\left.-\left(1+{\Lambda}\left(\xi_{\ss\FC},\overline{V}_{0}[j]\right)\right){\Lambda}\left(\xi_{\ss\R},U_{0}[j]\right)\right],
\end{align}
\else
\begin{align}\label{Pef2 upp}
\tilde{P}_{\fab}[j]=&\;\frac{1}{4}\left[3-{\Lambda}\left(\xi_{\ss\FC},\overline{V}_{0}[j]\right)\right.\nonumber\\
&\left.-\left(1+{\Lambda}\left(\xi_{\ss\FC},\overline{V}_{0}[j]\right)\right){\Lambda}\left(\xi_{\ss\R},U_{0}[j]\right)\right],
\end{align}
\fi
respectively, {where $\tilde{P}_{\mdb}[j]$ and $\tilde{P}_{\fab}[j]$ are the upper bounds on $\tilde{P}_{\md}[j]$ and $\tilde{P}_{\fa}[j]$, respectively}. We now present the constraints making $\tilde{P}_{{\mdb}}[j]^K$ and $\tilde{P}_{{\fab}}[j]^K$ convex in the following two theorems.
\begin{theorem}\label{theorem 3}
$\tilde{P}_{{\mdb}}[j]^K$ is jointly convex with respect to $\xi_{\ss\R}$ and $\xi_{\ss\FC}$, if we impose the convex constraints \eqref{Constraint 1}, and the following constraints:
\begin{align}\label{Constraint 3}
-0.5-\overline{V}_{1}[j]+\xi_{\ss\FC}\leq0,
\end{align}	
\begin{align}\label{cooperative joint miss}
\Phi\left(\xi_{\ss\R},\xi_{\ss\FC}^+,K\right)\leq0,~\text{and}~ \Phi\left(\xi_{\ss\R}^+,\xi_{\ss\FC},K\right)\leq0,
\end{align}
where $\xi_{\ss\R}^-$ and $\xi_{\ss\R}^+$ are bounds on $\xi_{\ss\R}$, and $\xi_{\ss\FC}^-$ and $\xi_{\ss\FC}^+$ are bounds on $\xi_{\ss\FC}$, and $\Phi\left(\mu,\nu,K\right)$ is given
\ifOneCol	
by
\else
in \eqref{cons7} at the top of page~\pageref{cons7}.
\fi
\ifOneCol
\begin{align}\label{cons7}
\Phi\left(\mu,\nu,K\right)=&\;4{\Theta}\left(\xi_{\ss\R}^+,U_{1}[j]\right)\left(-4+K+K{\Lambda}\left(\xi_{\ss\FC}^-,\overline{V}_{1}[j]\right)+K{\Lambda}\left(\xi_{\ss\R}^-,U_{1}[j]\right)\left(1+{\Lambda}\left(\xi_{\ss\FC}^-,\overline{V}_{1}[j]\right)\right)\right)^2\nonumber\\
&\hspace{-15mm}-\frac{\left(1+{\Lambda}\left(\xi_{\ss\R}^-,U_{1}[j]\right)\right)}{\sqrt{U_{1}[j]\overline{V}_{1}[j]}}
\left(1+{\Lambda}\left(\xi_{\ss\FC}^-,\overline{V}_{1}[j]\right)\right)\Bigl(2\left(-1+K\right)\sqrt{\overline{V}_{1}[j]}\left(1+{\Lambda}\left(\xi_{\ss\R}^+,U_{1}[j]\right)\right)\nonumber\\
&\hspace{-15mm}-\frac{\sqrt{2\pi}}{{\Theta}\left(\xi_{\ss\FC}^+,\overline{V}_{1}[j]\right)}\left(0.5+\overline{V}_{1}[j]-\nu\right)
\left(-3+{\Lambda}\left(\xi_{\ss\FC}^+,\overline{V}_{1}[j]\right)\right.+{\Lambda}\left(\xi_{\ss\R}^+,U_{1}[j]\right)\left.\left(1+{\Lambda}\left(\xi_{\ss\FC}^+,\overline{V}_{1}[j]\right)\right)\right)\Bigr)\nonumber\\
&\hspace{-15mm}\times\Bigl({\Theta}\left(\xi_{\ss\R}^+,U_{1}[j]\right)\left(1+{\Lambda}\left(\xi_{\ss\FC}^-,\overline{V}_{1}[j]\right)\right)
\left(-1+K\right)2\sqrt{U_{1}[j]}-\sqrt{2\pi}\left(0.5+U_{1}[j]-\mu\right)\nonumber\\
&\hspace{-15mm}\times\left(-3+{\Lambda}\left(\xi_{\ss\FC}^+,\overline{V}_{1}[j]\right)+{\Lambda}\left(\xi_{\ss\R}^+,U_{1}[j]\right)
\left(1+{\Lambda}\left(\xi_{\ss\FC}^+,\overline{V}_{1}[j]\right)\right)\right)\Bigr).
\end{align}	
\else
\begin{figure*}
\begin{align}\label{cons7}
\Phi\left(\mu,\nu,K\right)=&\;4{\Theta}\left(\xi_{\ss\R}^+,U_{1}[j]\right)\left(-4+K+K{\Lambda}\left(\xi_{\ss\FC}^-,\overline{V}_{1}[j]\right)+K{\Lambda}\left(\xi_{\ss\R}^-,U_{1}[j]\right)\left(1+{\Lambda}\left(\xi_{\ss\FC}^-,\overline{V}_{1}[j]\right)\right)\right)^2-\frac{\left(1+{\Lambda}\left(\xi_{\ss\R}^-,U_{1}[j]\right)\right)}{\sqrt{U_{1}[j]\overline{V}_{1}[j]}}\nonumber\\
&\hspace{-10mm}\times\left(1+{\Lambda}\left(\xi_{\ss\FC}^-,\overline{V}_{1}[j]\right)\right)\Bigl(2\left(-1+K\right)\sqrt{\overline{V}_{1}[j]}\left(1+{\Lambda}\left(\xi_{\ss\R}^+,U_{1}[j]\right)\right)-\frac{\sqrt{2\pi}}{{\Theta}\left(\xi_{\ss\FC}^+,\overline{V}_{1}[j]\right)}\left(0.5+\overline{V}_{1}[j]-\nu\right)\nonumber\\
&\hspace{-10mm}\times\left(-3+{\Lambda}\left(\xi_{\ss\FC}^+,\overline{V}_{1}[j]\right)\right.+{\Lambda}\left(\xi_{\ss\R}^+,U_{1}[j]\right)\left.\left(1+{\Lambda}\left(\xi_{\ss\FC}^+,\overline{V}_{1}[j]\right)\right)\right)\Bigr)\Bigl({\Theta}\left(\xi_{\ss\R}^+,U_{1}[j]\right)\left(1+{\Lambda}\left(\xi_{\ss\FC}^-,\overline{V}_{1}[j]\right)\right)\nonumber\\
&\hspace{-10mm}\times\left(-1+K\right)2\sqrt{U_{1}[j]}-\sqrt{2\pi}\left(0.5+U_{1}[j]-\mu\right)\left(-3+{\Lambda}\left(\xi_{\ss\FC}^+,\overline{V}_{1}[j]\right)+{\Lambda}\left(\xi_{\ss\R}^+,U_{1}[j]\right)
\left(1+{\Lambda}\left(\xi_{\ss\FC}^+,\overline{V}_{1}[j]\right)\right)\right)\Bigr)
\end{align}
\hrulefill
\end{figure*}
\fi
\end{theorem}
\begin{theorem}\label{theorem 3_1}
$\tilde{P}_{{\fab}}[j]^K$ is jointly convex with respect to $\xi_{\ss\R}$ and $\xi_{\ss\FC}$, if we impose the convex constraints \eqref{Constraint 2} and the following constraints:
\begin{align}\label{Constraint 4}
0.5+\overline{V}_{0}[j]-\xi_{\ss\FC}\leq0,
\end{align}
\begin{align}\label{cooperative joint false}
\Psi\left(\xi_{\ss\R},\xi_{\ss\FC}^-,K\right)\leq0,~\text{and}~
\Psi\left(\xi_{\ss\R}^-,\xi_{\ss\FC},K\right)\leq0,
\end{align}
where $\Psi\left(\mu,\nu,K\right)$ is given
\ifOneCol	
by
\else
in \eqref{cons8} at the top of page~\pageref{cons8}.
\fi
\ifOneCol
\begin{align}\label{cons8}
\Psi\left(\mu,\nu,K\right)=&\;4{\Theta}\left(\xi_{\ss\R}^-,U_{0}[j]\right)\left(-4+K-K{\Lambda}\left(\xi_{\ss\FC}^+,\overline{V}_{0}[j]\right)+K{\Lambda}\left(\xi_{\ss\R}^+,U_{0}[j]\right)\left(-1+{\Lambda}\left(\xi_{\ss\FC}^+,\overline{V}_{0}[j]\right)\right)\right)^2\nonumber\\
&\hspace{-15mm}-\frac{\left(1-{\Lambda}\left(\xi_{\ss\R}^+,U_{0}[j]\right)\right)}{\sqrt{U_{0}[j]\overline{V}_{0}[j]}}\left(-1+{\Lambda}\left(\xi_{\ss\FC}^-,\overline{V}_{0}[j]\right)\right)\Bigl(-2\left(-1+K\right)\sqrt{\overline{V}_{0}[j]}\left(-1+{\Lambda}\left(\xi_{\ss\R}^-,U_{0}[j]\right)\right)\nonumber\\
&\hspace{-15mm}+\frac{\sqrt{2\pi}}{{\Theta}\left(\xi_{\ss\FC}^-,\overline{V}_{0}[j]\right)}\left(0.5+\overline{V}_{0}[j]-\nu\right)\left(-3-{\Lambda}\left(\xi_{\ss\FC}^+,\overline{V}_{0}[j]\right)\right.+{\Lambda}\left(\xi_{\ss\R}^-,U_{0}[j]\right)\left.\left(-1+{\Lambda}\left(\xi_{\ss\FC}^-,\overline{V}_{0}[j]\right)\right)\right)\Bigr)\nonumber\\
&\hspace{-15mm}\times\Bigl({\Theta}\left(\xi_{\ss\R}^-,U_{0}[j]\right)\left(-1+{\Lambda}\left(\xi_{\ss\FC}^-,\overline{V}_{0}[j]\right)\right)\left(-1+K\right)2\sqrt{U_{0}[j]}-\sqrt{2\pi}\left(0.5+U_{0}[j]-\mu\right)\nonumber\\
&\hspace{-15mm}\times\left(-3-{\Lambda}\left(\xi_{\ss\FC}^-,\overline{V}_{0}[j]\right)+{\Lambda}\left(\xi_{\ss\R}^-,U_{0}[j]\right)\left(-1+{\Lambda}\left(\xi_{\ss\FC}^+,\overline{V}_{0}[j]\right)\right)\right)\Bigr).
\end{align}
\else
\begin{figure*}
\begin{align}\label{cons8}
\Psi\left(\mu,\nu,K\right)=&\;4{\Theta}\left(\xi_{\ss\R}^-,U_{0}[j]\right)\left(-4+K-K{\Lambda}\left(\xi_{\ss\FC}^+,\overline{V}_{0}[j]\right)+K{\Lambda}\left(\xi_{\ss\R}^+,U_{0}[j]\right)\left(-1+{\Lambda}\left(\xi_{\ss\FC}^+,\overline{V}_{0}[j]\right)\right)\right)^2-\frac{\left(1-{\Lambda}\left(\xi_{\ss\R}^+,U_{0}[j]\right)\right)}{\sqrt{U_{0}[j]\overline{V}_{0}[j]}}\nonumber\\
&\hspace{-10mm}\times\left(-1+{\Lambda}\left(\xi_{\ss\FC}^-,\overline{V}_{0}[j]\right)\right)\Bigl(-2\left(-1+K\right)\sqrt{\overline{V}_{0}[j]}\left(-1+{\Lambda}\left(\xi_{\ss\R}^-,U_{0}[j]\right)\right)+\frac{\sqrt{2\pi}}{{\Theta}\left(\xi_{\ss\FC}^-,\overline{V}_{0}[j]\right)}\left(0.5+\overline{V}_{0}[j]-\nu\right)\nonumber\\
&\hspace{-10mm}\times\left(-3-{\Lambda}\left(\xi_{\ss\FC}^+,\overline{V}_{0}[j]\right)\right.+{\Lambda}\left(\xi_{\ss\R}^-,U_{0}[j]\right)\left.\left(-1+{\Lambda}\left(\xi_{\ss\FC}^-,\overline{V}_{0}[j]\right)\right)\right)\Bigr)\Bigl({\Theta}\left(\xi_{\ss\R}^-,U_{0}[j]\right)\left(-1+{\Lambda}\left(\xi_{\ss\FC}^-,\overline{V}_{0}[j]\right)\right)\nonumber\\
&\hspace{-10mm}\times\left(-1+K\right)2\sqrt{U_{0}[j]}-\sqrt{2\pi}\left(0.5+U_{0}[j]-\mu\right)\left(-3-{\Lambda}\left(\xi_{\ss\FC}^-,\overline{V}_{0}[j]\right)+{\Lambda}\left(\xi_{\ss\R}^-,U_{0}[j]\right)
\left(-1+{\Lambda}\left(\xi_{\ss\FC}^+,\overline{V}_{0}[j]\right)\right)\right)\Bigr)
\end{align}
\hrulefill
\end{figure*}
\fi
\end{theorem}
\begin{IEEEproof}
The proof of {Theorem \ref{theorem 3}} and {Theorem \ref{theorem 3_1}} is given in {the Appendix}.
\end{IEEEproof}

Similar to \eqref{Upper Q_fa} and \eqref{Upper Q_md}, we upper-bound $Q_{\fa}[j]$ for the OR rule and $Q_{\md}[j]$ for the AND rule as
\begin{align}\label{Upper Q_fa noisy1}
Q_{\fa}[j] \leq K\tilde{P}_{\fab}[j]
\end{align}
and
\begin{align}\label{Upper Q_md noisy1}
Q_{\md}[j] \leq K\tilde{P}_{\mdb}[j],
\end{align}
respectively. We note that $\tilde{P}_{\fab}[j]$ is convex with respect to $\xi_{\ss\R}$ and $\xi_{\ss\FC}$, if we impose the constraints \eqref{Constraint 2}, \eqref{Constraint 4}, $\Psi\left(\xi_{\ss\R},\xi_{\ss\FC}^-,1\right)\leq0$, and $\Psi\left(\xi_{\ss\R}^-,\xi_{\ss\FC},1\right)\leq0$, which can be proven by considering $K=1$ in {Theorem \ref{theorem 3_1}}. We also note that $\tilde{P}_{\mdb}[j]$ is convex with respect to $\xi_{\ss\R}$ and $\xi_{\ss\FC}$, if we impose the constraints \eqref{Constraint 1}, \eqref{Constraint 3}, $\Phi\left(\xi_{\ss\R},\xi_{\ss\FC}^+,1\right)\leq0$, and $\Phi\left(\xi_{\ss\R}^+,\xi_{\ss\FC},1\right)\leq0$, which can be proven by considering $K=1$ in {Theorem \ref{theorem 3}}. Since \eqref{Upper Q_fa noisy1} and \eqref{Upper Q_md noisy1} scale a convex function with a {nonnegative} constant, they are also convex with respect to $\xi_{\ss\R}$ and $\xi_{\ss\FC}$ under the same constraints. We then focus on the joint convexity analysis of $Q_{\md}[j]$ for the OR rule and $Q_{\fa}[j]$ for the AND rule. Based on {Theorem \ref{theorem 3}} and {Theorem \ref{theorem 3_1}}, we note that $\tilde{P}_{{\mdb}}[j]^K$ and $\tilde{P}_{{\fab}}[j]^K$ are jointly convex with respect to $\xi_{\ss\R}$ and $\xi_{\ss\FC}$, respectively. {For the $N$-out-of-$K$ rule, we note that $\tilde{P}_{{\mdb}}[j]^{\tilde{n}}$ is jointly convex with respect to $\xi_{\ss\R}$ and $\xi_{\ss\FC}$ under the constraints \eqref{Constraint 1}, \eqref{Constraint 3}, $\Phi\left(\xi_{\ss\R},\xi_{\ss\FC}^+,{\tilde{n}}\right)\leq0$, and $\Phi\left(\xi_{\ss\R}^+,\xi_{\ss\FC},{\tilde{n}}\right)\leq0$, which can be proven by replacing $K$ with $\tilde{n}$ in {Theorem \ref{theorem 3}}. We also note that $\tilde{P}_{{\fab}}[j]^n$ is jointly convex with respect to $\xi_{\ss\R}$ and $\xi_{\ss\FC}$ under the constraints \eqref{Constraint 2}, \eqref{Constraint 4}, $\Psi\left(\xi_{\ss\R},\xi_{\ss\FC}^-,n\right)\leq0$, and $\Psi\left(\xi_{\ss\R}^-,\xi_{\ss\FC},n\right)\leq0$, which can be proven by replacing $K$ with $n$ in {Theorem \ref{theorem 3_1}}.} Given that $\tilde{P}_{{\mdb}}[j]^n$ and $\tilde{P}_{{\fab}}[j]^n$ are jointly convex with respect to $\xi_{\ss\R}$ and $\xi_{\ss\FC}$ and applying a similar method to \eqref{Qm_KN re}--\eqref{Qf_KN upp}, we can derive the upper bounds on $Q_{\md}[j]$ and $Q_{\fa}[j]$ which are jointly convex with respect to $\xi_{\ss\R}$ and $\xi_{\ss\FC}$.

In the noisy reporting scenario, we formulate the convex optimization problems with respect to $\xi_{\ss\R}$ and $\xi_{\ss\FC}$ for the OR rule, AND rule, and $N$-out-of-$K$ rule as
\ifOneCol	
\begin{equation}\label{noisy OR}
\begin{aligned}
& \underset{\xi_{\ss\R},\; \xi_{\ss\FC}}{\text{min}}
& & P_1\tilde{P}_{{\mdb}}[j]^K + \left(1-\tilde{P}_1\right)KP_{\fab}[j] \\
& \text{s.t.}
& & \eqref{Constraint 1},\;\eqref{Constraint 2}, \;\eqref{Constraint 3}-\eqref{Constraint 4}, \Psi\left(\xi_{\ss\R},\xi_{\ss\FC}^-,1\right)\leq0, \;\text{and} \; \Psi\left(\xi_{\ss\R}^-,\xi_{\ss\FC},1\right)\leq0,\\
\end{aligned}
\end{equation}
\else
\begin{equation}\label{noisy OR}
\begin{aligned}
& \underset{\xi_{\ss\R},\; \xi_{\ss\FC}}{\text{min}}
& & P_1\tilde{P}_{{\mdb}}[j]^K + \left(1-\tilde{P}_1\right)KP_{\fab}[j] \\
& \text{s.t.}
& & \eqref{Constraint 1},\;\eqref{Constraint 2}, \;\eqref{Constraint 3}-\eqref{Constraint 4},\\
& & &\Psi\left(\xi_{\ss\R},\xi_{\ss\FC}^-,1\right)\leq0, \;\text{and} \; \Psi\left(\xi_{\ss\R}^-,\xi_{\ss\FC},1\right)\leq0,\\
\end{aligned}
\end{equation}
\fi
\ifOneCol	
\begin{equation}\label{noisy AND}
\begin{aligned}
& \underset{\xi_{\ss\R},\; \xi_{\ss\FC}}{\text{min}}
& & P_1K\tilde{P}_{{\mdb}}[j]+ \left(1-P_1\right)\tilde{P}_{\fab}[j]^K \\
& \text{s.t.}
& & \eqref{Constraint 1},\; \eqref{Constraint 2},\;\eqref{Constraint 3},\;\eqref{Constraint 4},\;\eqref{cooperative joint false}, \Phi\left(\xi_{\ss\R}^+,\xi_{\ss\FC},1\right)\leq0, \;\text{and} \; \Phi\left(\xi_{\ss\R},\xi_{\ss\FC}^+,1\right)\leq0,\\
\end{aligned}
\end{equation}
\else
\begin{equation}\label{noisy AND}
\begin{aligned}
& \underset{\xi_{\ss\R},\; \xi_{\ss\FC}}{\text{min}}
& & P_1K\tilde{P}_{{\mdb}}[j]+ \left(1-P_1\right)\tilde{P}_{\fab}[j]^K \\
& \text{s.t.}
& & \eqref{Constraint 1},\; \eqref{Constraint 2},\;\eqref{Constraint 3},\;\eqref{Constraint 4},\;\eqref{cooperative joint false}, \\
& & &\Phi\left(\xi_{\ss\R}^+,\xi_{\ss\FC},1\right)\leq0, \;\text{and} \; \Phi\left(\xi_{\ss\R},\xi_{\ss\FC}^+,1\right)\leq0,\\
\end{aligned}
\end{equation}
\fi
and
\ifOneCol	
\begin{equation}\label{noisy NK}
\begin{aligned}
& \underset{\xi_{\ss\R}, \;\xi_{\ss\FC}}{\text{min}}
& &P_1\tilde{Q}_{\md}^{+}[j]+(1-P_1)\tilde{Q}_{\fa}^{+}[j] \\
& \text{s.t.}
& & \eqref{Constraint 1},\; \eqref{Constraint 2},\;\eqref{Constraint 3},\; \eqref{Constraint 4},\\
& & &\Phi\left(\xi_{\ss\R}^+,\xi_{\ss\FC},\tilde{n}\right)\leq0,\;\Phi\left(\xi_{\ss\R},\xi_{\ss\FC}^+,\tilde{n}\right)\leq0,
\Psi\left(\xi_{\ss\R},\xi_{\ss\FC}^-,n\right)\leq0,\;\text{and} \;\Psi\left(\xi_{\ss\R}^-,\xi_{\ss\FC},n\right)\leq0,
\end{aligned}
\end{equation}
\else
\begin{equation}\label{noisy NK}
\begin{aligned}
& \underset{\xi_{\ss\R}, \;\xi_{\ss\FC}}{\text{min}}
& &P_1\tilde{Q}_{\md}^{+}[j]+(1-P_1)\tilde{Q}_{\fa}^{+}[j] \\
& \text{s.t.}
& & \eqref{Constraint 1},\; \eqref{Constraint 2},\;\eqref{Constraint 3},\; \eqref{Constraint 4},\\
& & &\Phi\left(\xi_{\ss\R}^+,\xi_{\ss\FC},\tilde{n}\right)\leq0,\;\Phi\left(\xi_{\ss\R},\xi_{\ss\FC}^+,\tilde{n}\right)\leq0,\\
& & & \Psi\left(\xi_{\ss\R},\xi_{\ss\FC}^-,n\right)\leq0,\;\text{and} \;\Psi\left(\xi_{\ss\R}^-,\xi_{\ss\FC},n\right)\leq0,
\end{aligned}
\end{equation}
\fi
respectively, where $\tilde{Q}_{\md}^{+}[j]\triangleq{}\sum_{\tilde{n}=K-N+1}^{K}\binom{K}{\tilde{n}}{\tilde{P}_{\mdb}[j]}^{\tilde{n}}$ and $\tilde{Q}_{\fa}^{+}[j]\triangleq\sum_{n=N}^{K}\binom{K}{n}{\tilde{P}_{\fab}[j]}^{n}$. {We emphasize that the constraints $\Phi\left(\xi_{\ss\R}^+,\xi_{\ss\FC},\tilde{n}\right)\leq0$, $\Phi\left(\xi_{\ss\R},\xi_{\ss\FC}^+,\tilde{n}\right)\leq0$, $\Psi\left(\xi_{\ss\R},\xi_{\ss\FC}^-,n\right)\leq0$, $\Psi\left(\xi_{\ss\R}^-,\xi_{\ss\FC},n\right)\leq0$ for each $\tilde{n}$ and $n$ are applied in \eqref{noisy NK}, where $\tilde{n}\in\{K-N+1,{\ldots},K\}$ and $n\in\{N,{\ldots},K\}$, to ensure the convexity of $\tilde{Q}_{\md}^{+}[j]$ and $\tilde{Q}_{\fa}^{+}[j]$.}
We note that the jointly convex optimization problem for the single-RX system in the noisy reporting scenario is a special case of problems \eqref{noisy OR}, \eqref{noisy AND}, and \eqref{noisy NK}, with $K=1$.

\subsection{Average Error Performance Optimization}\label{sec:Optimization of average performance1}

We emphasize that the solutions to the formulated optimization problems in Sections~\ref{sec:Perfect Reporting Optimization1} and~\ref{sec:Noisy Reporting Optimization} are the \emph{instantaneous} suboptimal thresholds which minimize the instantaneous system error performance for given $\textbf{W}_{\ss\T}^{j-1}$. As previously explained, it may not be realistic for the RXs and FC to calculate the instantaneous suboptimal thresholds and such calculation incurs significant computational overhead. Therefore, in this subsection we aim to obtain a \emph{single} suboptimal threshold which optimizes the average system error performance over all possible realizations of $\textbf{W}_{\ss\T}^{j-1}$ and across all symbol intervals.

If we aim to optimize  $\overline{Q}_{\ss\FC}$ for the OR rule in the perfect reporting scenario, based on \eqref{perfect OR}, then we formulate the problem as
\ifOneCol	
\begin{equation}\label{average opti}
\begin{aligned}
& \underset{\xi_{\ss\R}}{\text{minimize}}
& & \frac{1}{L}\sum_{j=1}^{L}\left(\frac{1}{|\omega_j|}\sum_{\omega_j}{P}_{{\md}}[j]^K+K{{P}_{\fa}[j]}\right) \\
& \text{s.t.}
& & \textrm{all constraints for all considered realizations of $\textbf{W}_{\ss\T}^{j-1}$ in $\omega_j$ for each symbol interval}.  \\
\end{aligned}
\end{equation}
\else
\begin{equation}\label{average opti}
\begin{aligned}
& \underset{\xi_{\ss\R}}{\text{minimize}}
& & \frac{1}{L}\sum_{j=1}^{L}\left(\frac{1}{|\omega_j|}\sum_{\omega_j}{P}_{{\md}}[j]^K+K{{P}_{\fa}[j]}\right) \\
& \text{s.t.}
& & \textrm{all constraints for all considered realizations}  \\
&&& \textrm{of $\textbf{W}_{\ss\T}^{j-1}$ in $\omega_j$ for each symbol interval}.  \\
\end{aligned}
\end{equation}
\fi

{The empirical average error performance of the system is optimized in \eqref{average opti}, since we assume that the occurrence likelihoods of the realizations of $\textbf{W}_{\ss\T}^{j-1}$ are equal.} Using a formulation similar to \eqref{average opti}, we can extend all convex optimization problems for optimizing the instantaneous system error performance to those for optimizing the average system error performance. Also, since all the derived inequality constraint functions are affine, the constraints define the lower limits and upper limits on $\xi_{\ss\R}$ and/or $\xi_{\ss\FC}$. We clarify that it is reasonable to only consider the minimum upper limit and the maximum lower limit on $\xi_{\ss\R}$ and/or $\xi_{\ss\FC}$ among all the upper and lower limits.


\section{Numerical Results and Simulations}\label{sec:Numerical}

In this section, we present numerical and simulation results to examine the error performance of the cooperative MC system. The simulation results are generated by a particle-based stochastic simulator, where we track the precise locations of all individual molecules over discrete time steps. {We clarify that all the approximations in Sections III and IV are only considered for facilitating our theoretical analysis, i.e., the theoretical evaluation and optimization of error performance. We do not adopt these approximations in our simulations. In our simulations, we consider a cooperative system as described in Section II.} In this section, we also demonstrate the effectiveness of the solutions to our formulated convex optimization problems, referred to as suboptimal solutions, by comparing them with the actual optimal solutions that minimize the expected average error probability of the system. We use the fmincon solver in MATLAB with the interior-point algorithm to obtain the suboptimal solutions. {We clarify that the actual optimal solutions are obtained via the exhaustive search of the numerical results of the expected \emph{average} error probability. Such solutions \emph{do not} require the information of $\textbf{W}_{\ss\T}^{j-1}$.} We denote $\xi_{\ss\R}^{\circ}$ and $\xi_{\ss\FC}^{\circ}$ as suboptimal solutions and denote $\xi_{\ss\R}^{\ast}$  and $\xi_{\ss\FC}^{\ast}$ as actual optimal solutions. We refer to the minimum upper bounds achieved by $\xi_{\ss\R}^{\circ}$ and $\xi_{\ss\FC}^{\circ}$ as suboptimal error probabilities. We refer to the expected error probability achieved by $\xi_{\ss\R}^{\circ}$ and $\xi_{\ss\FC}^{\circ}$ as the approximated error probabilities.


We list all the fixed environmental parameters adopted in this section in Table~\ref{tab:table1}. The varying parameters adopted in this section are the decision threshold at RXs, $\xi_{\ss\R}$, the decision threshold at the FC, $\xi_{\ss\FC}$, the number of RXs, $K$, the radius of $\RX_k$, $r_{\ss\RX_k}$, and the radius of the FC, $r_{\ss\FC}$. In particular, $r_{\ss\RX_k}$ is set as $0.225\,{\mu}\metre$ in all the figures except for Fig.~\ref{Pe_k} and $r_{\ss\FC}$ is fixed at $0.2\,{\mu}\metre$ in all the figures except for Fig.~\ref{Pe_rFC}. In Fig.~\ref{Pe_k}, we set $r_{\ss\RX_k}$ as $0.2\,{\mu}\metre$. In the following, we assume that the TX releases $S_{0} = 8000$ molecules for information symbol ``1'' and the total number of molecules released by all RXs for symbol ``1'' is fixed at $2000$, i.e., each RX releases $S_{k}=2000/K$ molecules to report its decision of symbol ``1''.
The locations of the TX, RXs, and FC are listed in Table~\ref{tab:coordinates1}. For each realization of $\textbf{W}_{\ss\T}^{j-1}$, we set $\xi_{\ss\R}^- = U_{0}[j]+1$, $\xi_{\ss\R}^+ = U_{1}[j]$, $\xi_{\ss\FC}^- = \overline{V}_{0}[j]+1$, and $\xi_{\ss\FC}^+=\overline{V}_{1}[j]$, since the initial convex feasible sets of  $\xi_{\ss\R}$ and $\xi_{\ss\FC}$ are $0.5+U_{0}[j]\leq\xi_{\ss\R}\leq0.5+U_{1}[j]$ and $0.5+\overline{V}_{0}[j]\leq\xi_{\ss\FC}\leq0.5+\overline{V}_{1}[j]$, respectively.

\begin{table}[!t]
\renewcommand{\arraystretch}{1.2}
\centering
\caption{Fixed Environmental Parameters Used in Section~\ref{sec:Numerical}}\label{tab:table1}\vspace{-2mm}
\begin{tabular}{c||c|c}
\hline
\bfseries Parameter &  \bfseries Symbol&  \bfseries Value \\
\hline\hline
Radius of RXs& $r_{\ss\R_k}$ & $\{0.225\,{\mu}\metre, 0.2\,{\mu}\metre\}$\\\hline
Radius of FC & $r_{\ss\FC}$ & $0.2\,{\mu}\metre$ \\\hline
Time step at RXs & $\Delta{t_{\ss\R}}$ & $100\,{\mu}\s$\\\hline
Time step at FC & $\Delta{t_{\ss\FC}}$ & $30\,{\mu}\s$ \\\hline
Number of samples of RXs& $M_{\ss\RX}$ & 5 \\\hline
Number of samples of FC& $M_{\ss\FC}$ & 5 \\\hline
Transmission time interval & $t_{\trans}$ & $1\,{\m}\s$\\\hline
Report time interval & $t_{\report}$ & $0.3\,{\m}\s$\\\hline
Bit interval time& $T$ & $1.3\,{\m}\s$\\\hline
Diffusion coefficient & $D_0=D_{k}$ & $5\times10^{-9}{\m^{2}}/{\s}$\\\hline
Length of symbol sequence & $L$ & $10$ \\\hline
Probability of binary 1 & $P_1$ & $0.5$ \\
\hline
\end{tabular}\vspace{-4mm}
\end{table}

\begin{table}[!t]
\renewcommand{\arraystretch}{1.2}
\centering
\caption{Locations of TX, RXs, and FC}\label{tab:coordinates1}\vspace{-2mm}
\begin{tabular}{c||c|c|c}
\hline
\bfseries Devices & \bfseries X-axis [${\mu}\metre$] & \bfseries Y-axis [${\mu}\metre$] & \bfseries Z-axis [${\mu}\metre$]\\\hline\hline
$\TX$   & $0$ & $0$ & $0$\\\hline
$\RX_1$ & $2$ & $0.6$ & $0$\\\hline
$\RX_2$ & $2$ & $-0.6$ & $0$ \\\hline
$\RX_3$ & $2$ & $-0.3$ & $0.5196$ \\\hline
$\RX_4$ & $2$ & $-0.3$ & $-0.519$ \\\hline
$\RX_5$ & $2$ & $0.3$ & $0.5196$ \\\hline
$\RX_6$ & $2$ & $0.3$ & $-0.5196$ \\\hline
$\FC$   & $2$ & $0$ & $0$ \\
\hline
\end{tabular}\vspace{-4mm}
\end{table}

Throughout this section, $\overline{Q}_{\ss\FC}$ are calculated by averaging ${P}_{\e,k}[j]$ and ${Q}_{\ss\FC}[j]$, respectively, over all considered realizations of  $\textbf{W}_{\ss\T}^{j-1}$ and across all symbol intervals. Here, we consider all possible realizations of $\textbf{W}_{\ss\T}^{j-1}$ except for the realization of all ``0'' bits, i.e., when the MDP is zero and there is no optimal threshold. Since we consider the length of the symbol sequence from the TX is 10 bits, we consider $1023$ different symbol sequences in total. The simulated error probabilities are averaged over at least $5\times10^4$ independent transmissions of the considered symbol sequences.
In Figs.~\ref{cooperative_perfect}--\ref{cooperative_noisy}, we plot the simulation for the expected error probabilities, while in Fig.~\ref{Pe_rFC}, we plot the simulation for the approximated error probabilities. {Moreover, we clarify that $\xi_{\ss\R}^{\circ}$ and $\xi_{\ss\FC}^{\circ}$ for the expected average error probabilities are obtained using the optimization method in Section~\ref{sec:Optimization of average performance1} only \emph{once} for all considered realizations of $\textbf{W}_{\ss\T}^{j-1}$ and across all symbol intervals, unless otherwise noted. In other words, suboptimal solutions \emph{do not} require the information of $\textbf{W}_{\ss\T}^{j-1}$, unless otherwise noted.} Furthermore, we clarify that the {noninteger} optimization solutions are rounded to integers in Figs. \ref{Pe_k}, \ref{joint_noisy}, and \ref{Pe_rFC}. Specifically, the most two nearest integers around the solution are compared and the one achieving the lower error probability is chosen.

\subsection{Perfect Reporting}\label{Numerical_Perfect}

In this subsection we consider the \emph{perfect} reporting scenario. In Fig.~\ref{cooperative_perfect}, we consider a three-RX cooperative system and plot the average global error probability versus the decision threshold at the RXs for the OR rule, AND rule, and majority rule. The expected curves for the three rules are obtained from \eqref{Qm_KN}--\eqref{Qf_AND} with \eqref{Pm1} and \eqref{Pf1}. The Gaussian approximation curves for the three rules are obtained from \eqref{Qm_KN}--\eqref{Qf_AND} with \eqref{Pm1 approx} and \eqref{Pf1 approx}. The upper bound curves for the OR rule, AND rule, and majority rule are obtained from \eqref{Qm_OR} and \eqref{Upper Q_fa}, \eqref{Qf_AND} and \eqref{Upper Q_md}, and \eqref{Qm_KN upp} and \eqref{Qf_KN upp}, respectively, with \eqref{Pm1 approx} and \eqref{Pf1 approx}. The value of $\xi_{\ss\R}^{\circ}$ for the OR rule, AND rule, and majority rule is obtained by solving \eqref{perfect OR}, \eqref{perfect AND}, and \eqref{perfect NK}, respectively, with \eqref{Pm1 approx} and \eqref{Pf1 approx}.

\ifOneCol	
\begin{figure}[!t]
\centering
\includegraphics[height=3in]{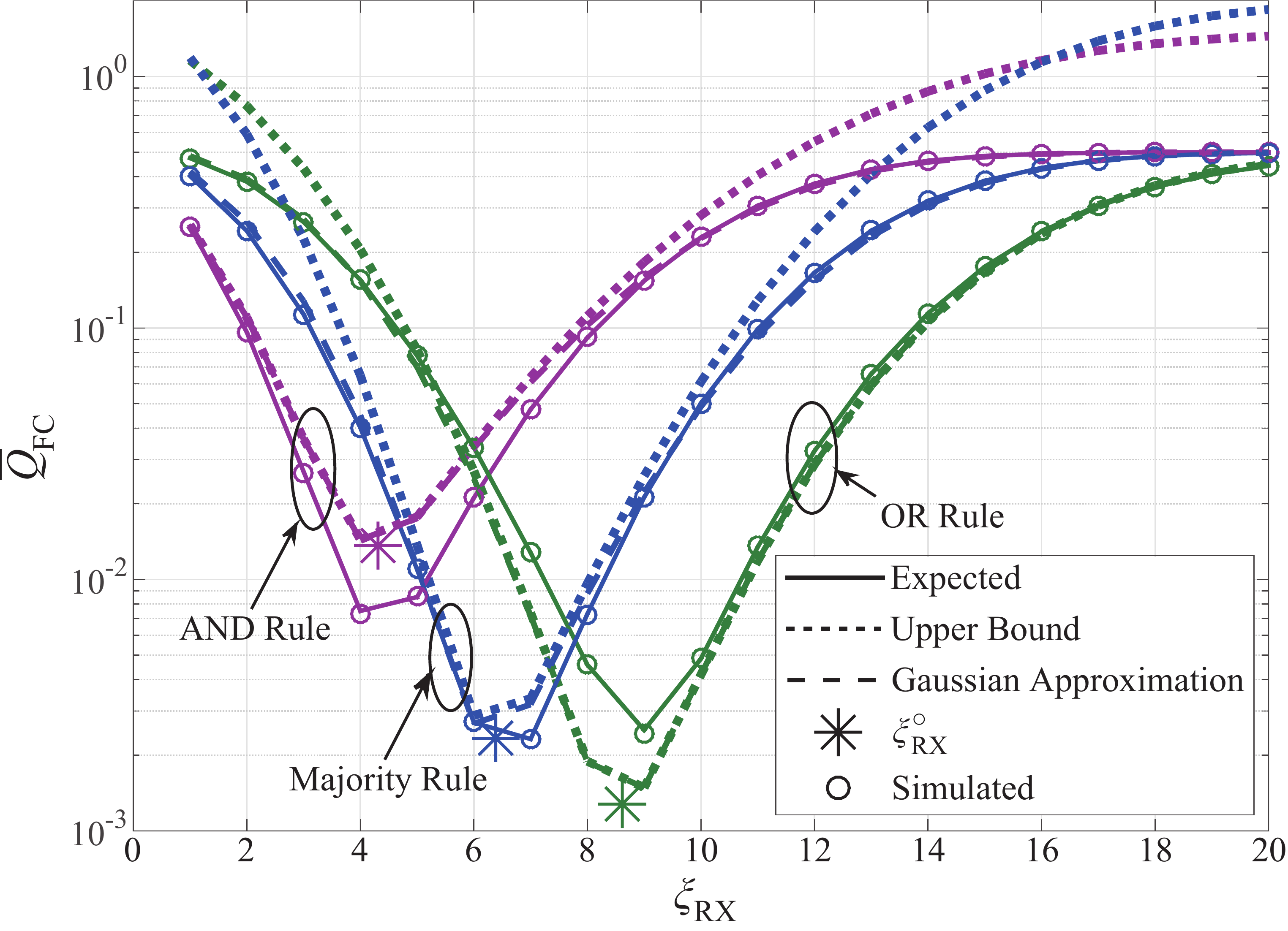}
\caption{Average global error probability $\overline{Q}_{\ss\FC}$ of different fusion rules versus the decision threshold at RXs $\xi_{\ss\R}$ with $K\!=\!3$ in the perfect reporting scenario.}
\label{cooperative_perfect}
\end{figure}
\else
\begin{figure}[!t]
\centering
\includegraphics[height=2.5in,width=3.25in]{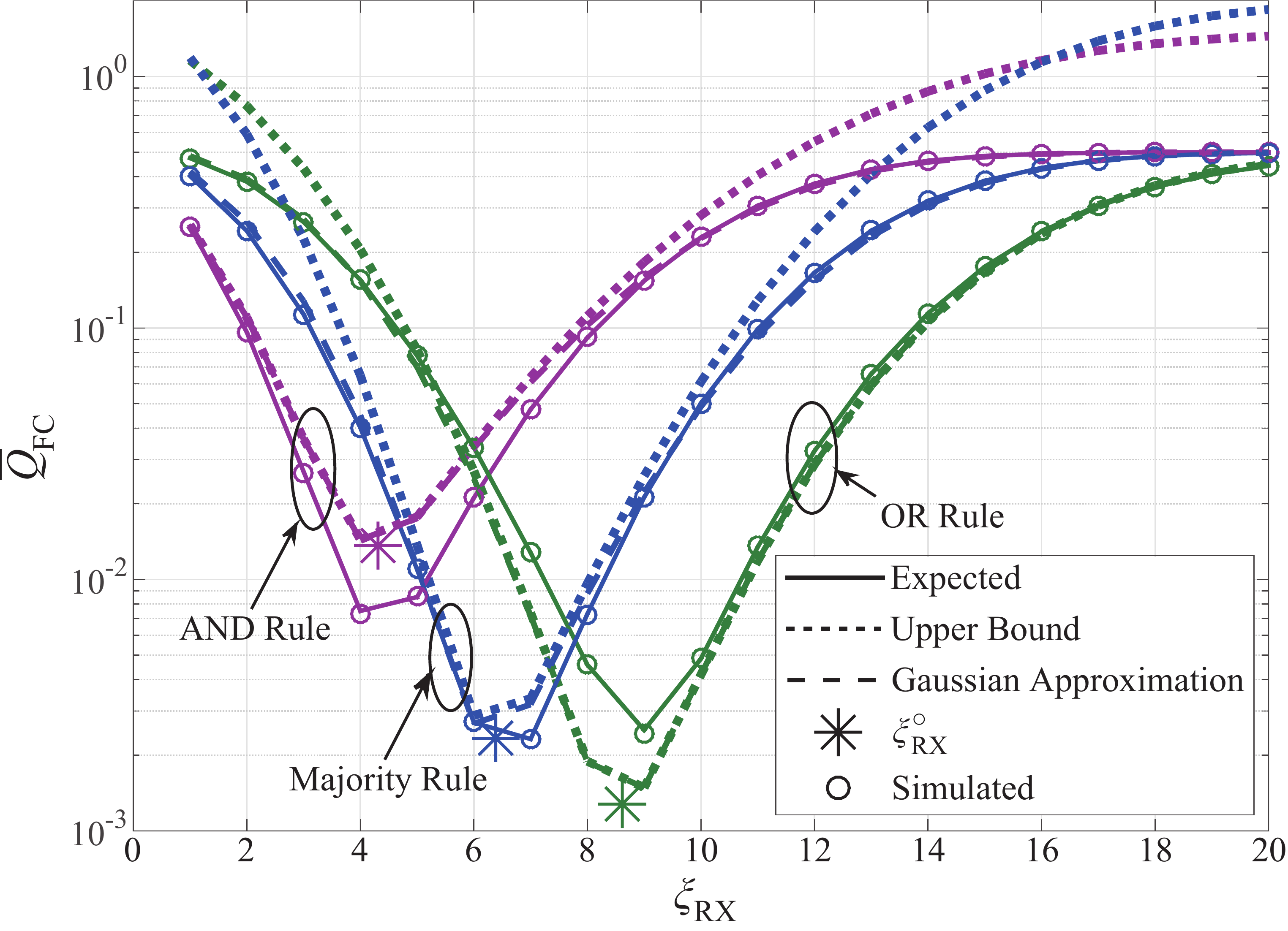}
\caption{Average global error probability $\overline{Q}_{\ss\FC}$ of different fusion rules versus the decision threshold at RXs $\xi_{\ss\R}$ with $K\!=\!3$ in the perfect reporting scenario.}
\label{cooperative_perfect}
\end{figure}
\fi

In Fig.~\ref{cooperative_perfect}, we first observe that the simulated points accurately match the expected curves, validating our analysis of the expected results. Second, we observe that $\xi_{\ss\R}^{\circ}$ is almost identical to $\xi_{\ss\R}^{\ast}$ for each fusion rule, confirming the accuracy of $\xi_{\ss\R}^{\circ}$. Third, we observe that the Gaussian approximation curves well approximate the expected curves. Fourth, we observe that the convex upper bound curve for the OR rule is lower than its expected curve. This can be explained as follows: In the single-RX system, the Gaussian approximations give an upper bound on $P_{\md}[j]$ and a lower bound on $P_{\fa}[j]$. For the OR rule, $Q_{\md}[j]$ is the product of $P_{\md}[j]$ and $Q_{\fa}[j]$ is the sum of $P_{\fa}[j]$. Since the Gaussian approximation of $Q_{\md}[j]$ is tighter than that of $Q_{\fa}[j]$, the Gaussian approximation of the global error probability for the OR rule is lower than the expected curve. 
Finally, observing the expected curves, we find that the majority rule outperforms the OR rule and the OR rule outperforms the AND rule at their corresponding optimal decision thresholds.

\ifOneCol	
\begin{figure}[!t]
\centering
\includegraphics[height=3in]{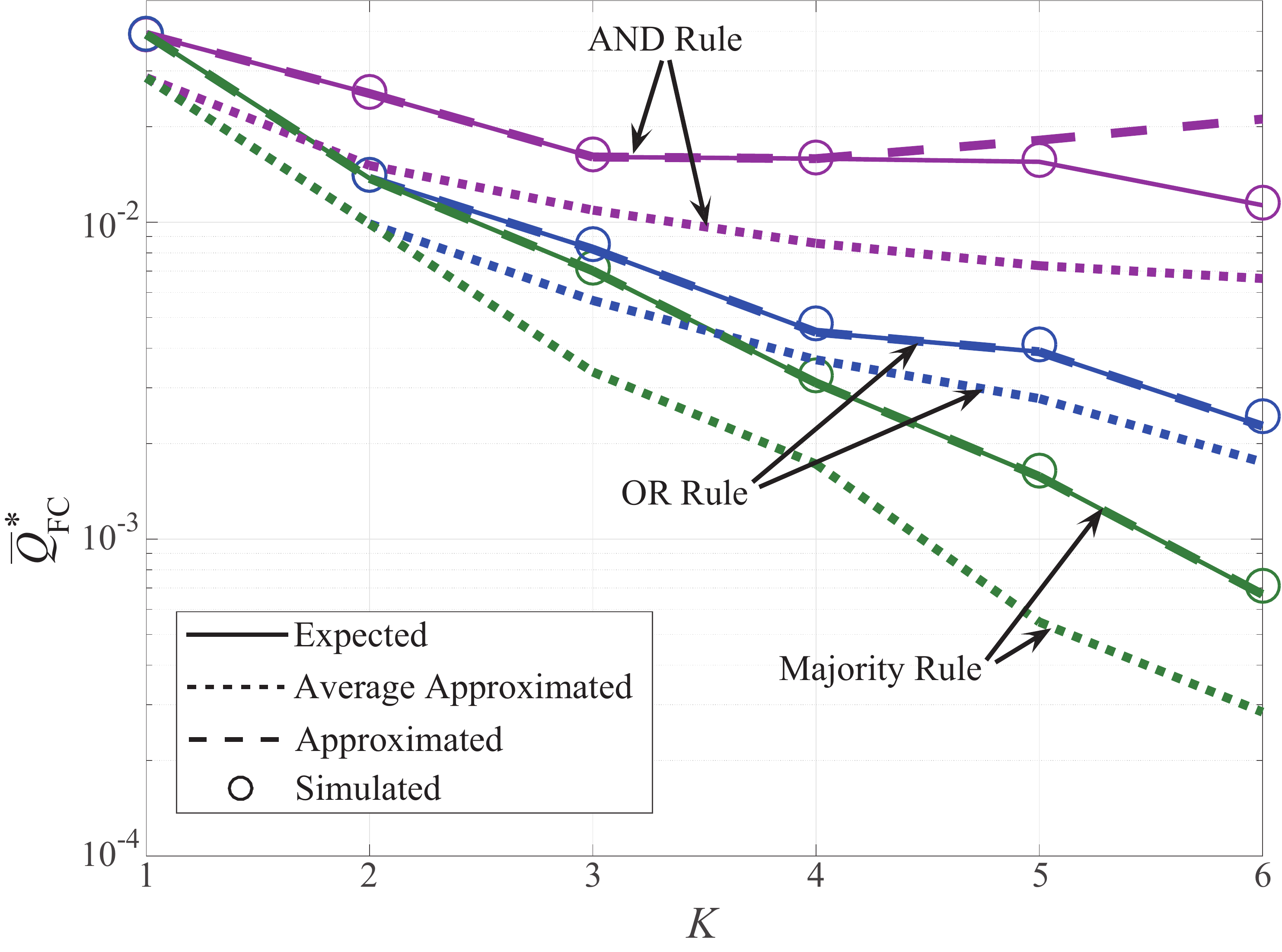}
\caption{Optimal average global error probability $\overline{Q}_{\ss\FC}^{\ast}$ of different fusion rules versus the number of cooperative RXs $K$ in the perfect reporting scenario.}
\label{Pe_k}
\end{figure}
\else
\begin{figure}[!t]
\centering
\includegraphics[height=2.5in,width=3.25in]{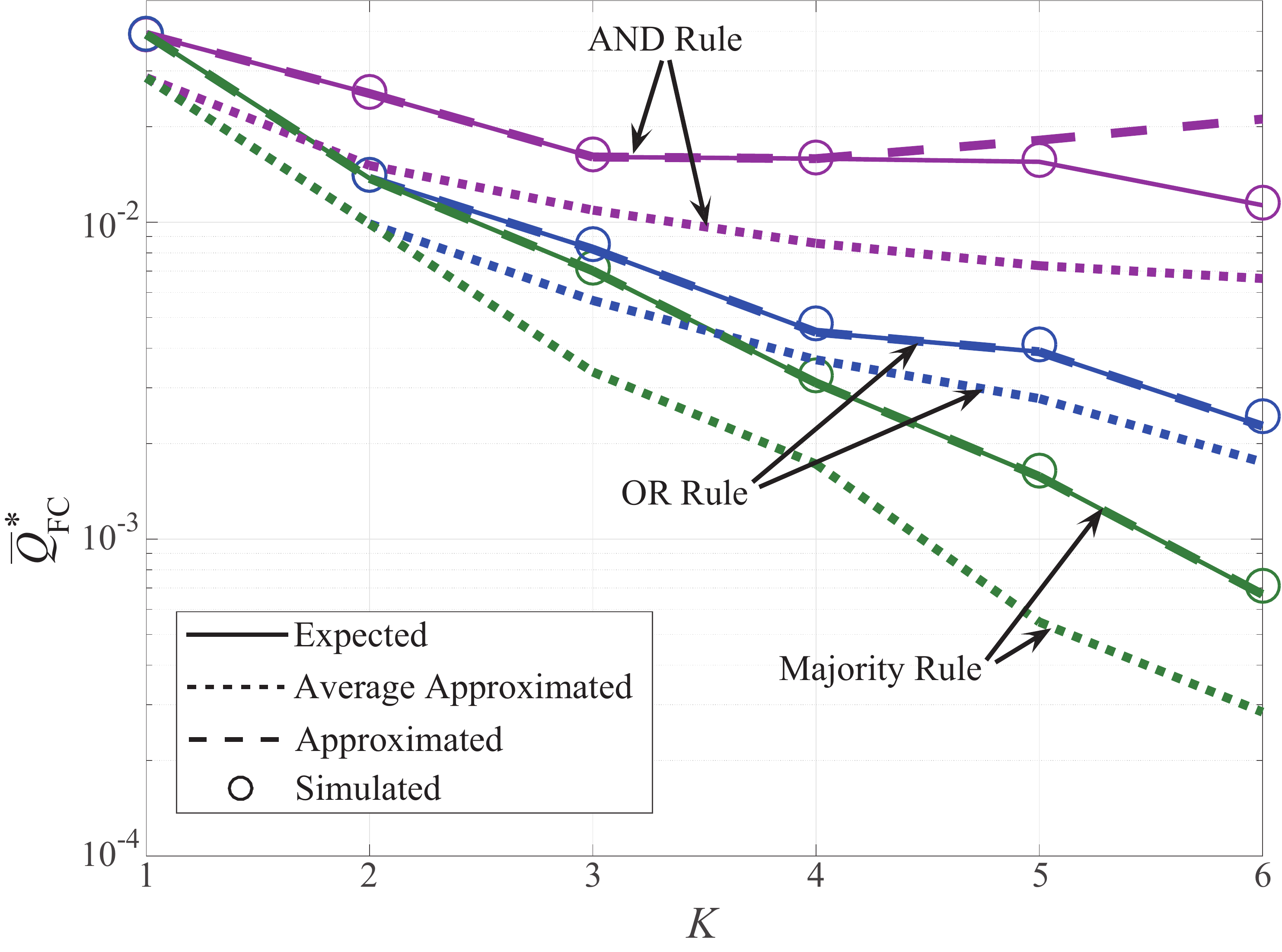}
\caption{Optimal average global error probability $\overline{Q}_{\ss\FC}^{\ast}$ of different fusion rules versus the number of cooperative RXs $K$ in the perfect reporting scenario.}
\label{Pe_k}
\end{figure}
\fi

In Fig.~\ref{Pe_k}, we plot the optimal average global error probability versus the number of cooperative RXs for the OR, AND, and majority rules. The baseline case is a single $\T-\R$ link with $K=1$, i.e., only one RX exists but no FC exists. In the baseline case, we assume that the RX is located at $(2\,{\mu}\metre,0.6\,{\mu}\metre,0)$, the TX releases $10000$ molecules, the time step between two successive samples is $100\,{\mu}\s$, and the symbol interval time is $T=1.3\,{\m}\s$, all of which ensure the fairness of the error performance comparison between the baseline case and the considered cooperative MC system. {We keep the total number of molecules released by all RXs fixed for the fairness of error performance comparison between the baseline case and the cooperative MC system with different $K$. Moreover, the fixed total number of molecules applies to realistic biological environments where the number of available molecules within the environment may be limited.} {Also, for a fair comparison of different $K$, we consider that all RXs sample at the same time and has the same number of samples for different $K$, since the sampling time $t_{\ss\R}(j,m)$ determines the mean number of molecules observed, based on \eqref{probability} and \eqref{general prob}.} The value of $\overline{Q}_{\ss\FC}^{\ast}$ for each $K$ in the expected curves for the three fusion rules is the minimum $\overline{Q}_{\ss\FC}$. For the expected curves, we consider that a single $\xi_{\ss\R}^{\ast}$ is applied to all considered realizations of $\textbf{W}_{\ss\T}^{j-1}$, which are obtained via exhaustive search of the expected expressions of \eqref{Qm_KN}--\eqref{Qf_AND} with \eqref{Pm1} and \eqref{Pf1}. On the other hand, the value of $\overline{Q}_{\ss\FC}^{\ast}$ for each $K$ in the approximated curves for the OR rule, AND rule, and majority rule are obtained by solving the corresponding average error performance optimization problems given by \eqref{perfect OR}, \eqref{perfect AND}, and \eqref{perfect NK}, respectively. To this end, we use a single $\xi_{\ss\R}^{\circ}$ for \emph{all} considered realizations of $\textbf{W}_{\ss\T}^{j-1}$, and then calculate the actual values of $\overline{Q}_{\ss\FC}$ achieved by $\xi_{\ss\R}^{\circ}$. The value of $\overline{Q}_{\ss\FC}^{\ast}$ for each $K$ in the \emph{average} approximated curves are obtained by solving \eqref{perfect OR}, \eqref{perfect AND}, and \eqref{perfect NK}, respectively, with \eqref{Pm1 approx} and \eqref{Pf1 approx} for all considered realizations. For this purpose, we consider a single $\xi_{\ss\R}^{\circ}$ for \emph{each} realization of $\textbf{W}_{\ss\T}^{j-1}$. {Hence, for average approximated curves, the information of $\textbf{W}_{\ss\T}^{j-1}$ is required for suboptimal solutions.} We then calculate the actual value of ${Q}_{\ss\FC}[j]$ achieved by $\xi_{\ss\R}^{\circ}$ for each realization of $\textbf{W}_{\ss\T}^{j-1}$, and refer to it as the instantaneous approximated error probabilities. Finally, we calculate the mean of all the instantaneous approximated error probabilities for all realizations of $\textbf{W}_{\ss\T}^{j-1}$.

In Fig.~\ref{Pe_k}, we first observe that for the OR rule and majority rule, the approximated curves match the expected curves, which confirms the accuracy of $\xi_{\ss\R}^{\circ}$. Second, we observe that for the AND rule, the approximated curve deviates from the expected curve when $K = 5$ and $K = 6$. This is due to the fact that $\xi_{\ss\R}^{\circ}$ is outside the feasible set restricted by all the constraints for all realizations of $\textbf{W}_{\ss\T}^{j-1}$. Third, we observe an accurate match between the simulated points and the expected curves. 
Fourth, we observe that for the three fusion rules, the error performance clearly improves when the optimization is performed for each realization of $\textbf{W}_{\ss\T}^{j-1}$. However, as previously explained, this performance gain may not be feasible in practice and thus, we consider the average approximated curves as the best performance bound of our considered system. Fifth, we observe from the expected curves that the majority rule outperforms the OR rule and AND rule, which is consistent with that in Fig.~\ref{cooperative_perfect}. Lastly, we observe that the cooperative MC system outperforms the baseline case for all fusion rules, even though the distance of the baseline case is shorter than that of the cooperative MC system. Importantly, we see that the system error performance significantly improves as $K$ increases. This is due to fact that an increasing number of cooperative RXs enables more independent observations of the transmitted information symbol. It follows that the probability that all RXs fail to detect the transmitted information symbol is reduced.

{We clarify that if we keep the \emph{total volume} of all RXs fixed, then the system error performance degrades as $K$ increases{\footnote{{For the numerical results to confirm this clarification, please refer to Addition Information in Page 17.}}}. We note that a single TX-RX link with one RX has the same error performance as our simple soft fusion rule proposed in \cite{GC 2016}, since both schemes have the same mean number of molecules observed under the assumption of uniform concentration. In simple soft fusion, the FC adds all RXs' observations in the $j$th symbol interval and then compares it with a decision threshold $\xi_{\ss\FC}$ to make a decision $\hat{W}_{\ss\FC}[j]$ (see \cite{GC 2016}). We then note that the simple soft fusion rule outperforms the majority rule, since local hard decisions are a quantization that decreases the granularity of the information available to the FC. Thus, for a fixed total volume of RXs, the single TX-RX link outperforms the majority rule.}

\subsection{Noisy Reporting}\label{Numerical_Noisy}

\ifOneCol	
\begin{figure}[!t]
\centering
\includegraphics[height=3in]{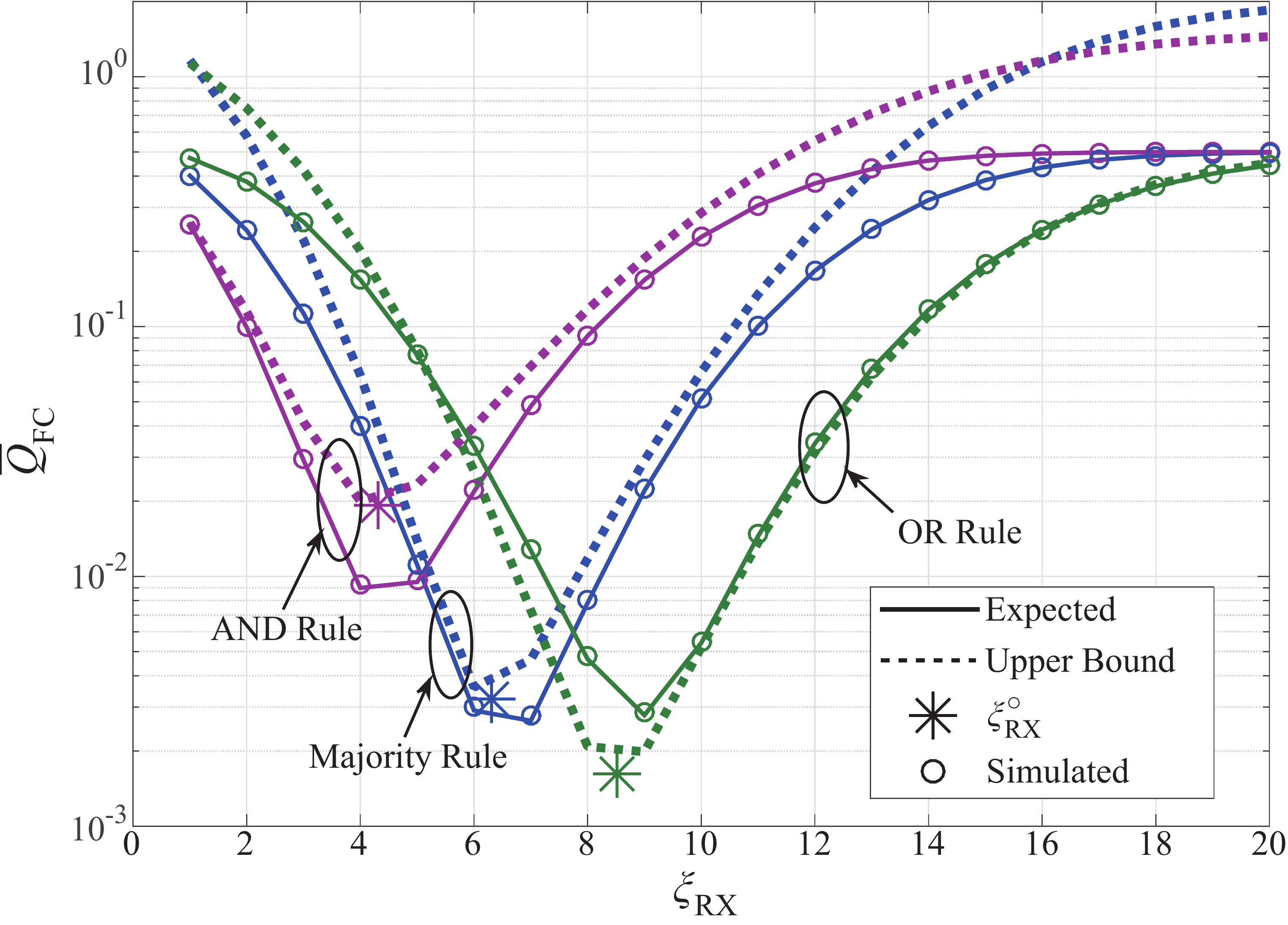}
\caption{Average global error probability $\overline{Q}_{\ss\FC}$ of different fusion rules versus the decision threshold  at RXs $\xi_{\ss\R}$ with $K=3$ in the noisy reporting scenario.}
\label{cooperative_noisy}
\end{figure}
\else
\begin{figure}[!t]
\centering
\includegraphics[height=2.5in,width=3.25in]{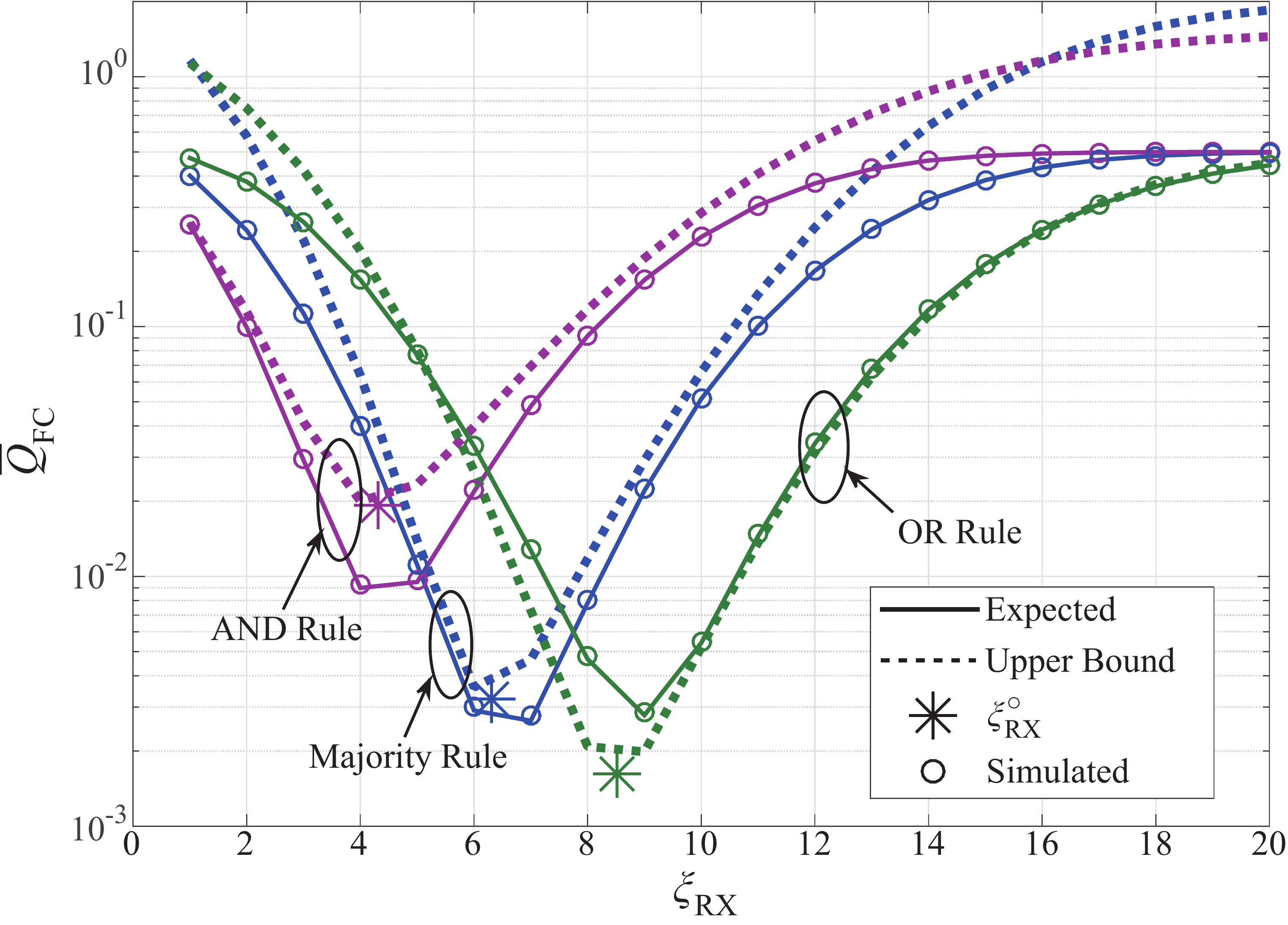}
\caption{Average global error probability $\overline{Q}_{\ss\FC}$ of different fusion rules versus the decision threshold  at RXs $\xi_{\ss\R}$ with $K=3$ in the noisy reporting scenario.}
\label{cooperative_noisy}
\end{figure}
\fi

In this subsection we focus on the \emph{noisy} reporting scenario. In Fig.~\ref{cooperative_noisy}, we consider a three-RX cooperative system and plot the average global error probability versus the decision threshold at the RXs for the AND rule, OR rule, and majority rule. {In this figure, we consider $\xi_{\ss\FC}=2$ for the AND rule, $\xi_{\ss\FC}=4$ for the OR rule, and $\xi_{\ss\FC}=3$ for the majority rule, since these thresholds are the values obtained when the thresholds at the RXs and FC are jointly optimized for the three fusion rules.} All curves in this figure are obtained from the same expressions and the same optimization problems as those in Fig.~\ref{cooperative_perfect}, except for replacing \eqref{Pm1}, \eqref{Pf1}, \eqref{Pm1 approx}, and \eqref{Pf1 approx} with \eqref{Pem2}, \eqref{Pef2}, \eqref{Pem2 approx averISI}, and \eqref{Pef2 approx averISI}, respectively. Similar to Fig.~\ref{cooperative_perfect}, we observe that $\xi_{\ss\R}^{\circ}$ is almost identical to $\xi_{\ss\R}^{\ast}$.
{By comparing Fig.~\ref{cooperative_perfect} with Fig.~\ref{cooperative_noisy}, we also observe that the expected error probabilities in Fig.~\ref{cooperative_perfect} are slightly lower than those in Fig.~\ref{cooperative_noisy}. We further observe that the optimal threshold at RXs is the same in Fig.~\ref{cooperative_perfect} and Fig.~\ref{cooperative_noisy}. This observation is not surprising, since the relatively short distance between $\RX_k$ and the FC, which leads to a relatively low error probability in the $\RX_k-\FC$ link. This low error probability does not significantly affect the error probability of the $\TX-\RX_k-\FC$ link.} In addition, we also confirmed that increasing $K$ significantly improves the system error performance in the noisy reporting scenario (figure omitted for brevity).


\begin{figure}[!t]
\centering
\subfigure[OR Rule]{\label{fig:a}\includegraphics[height=2in]{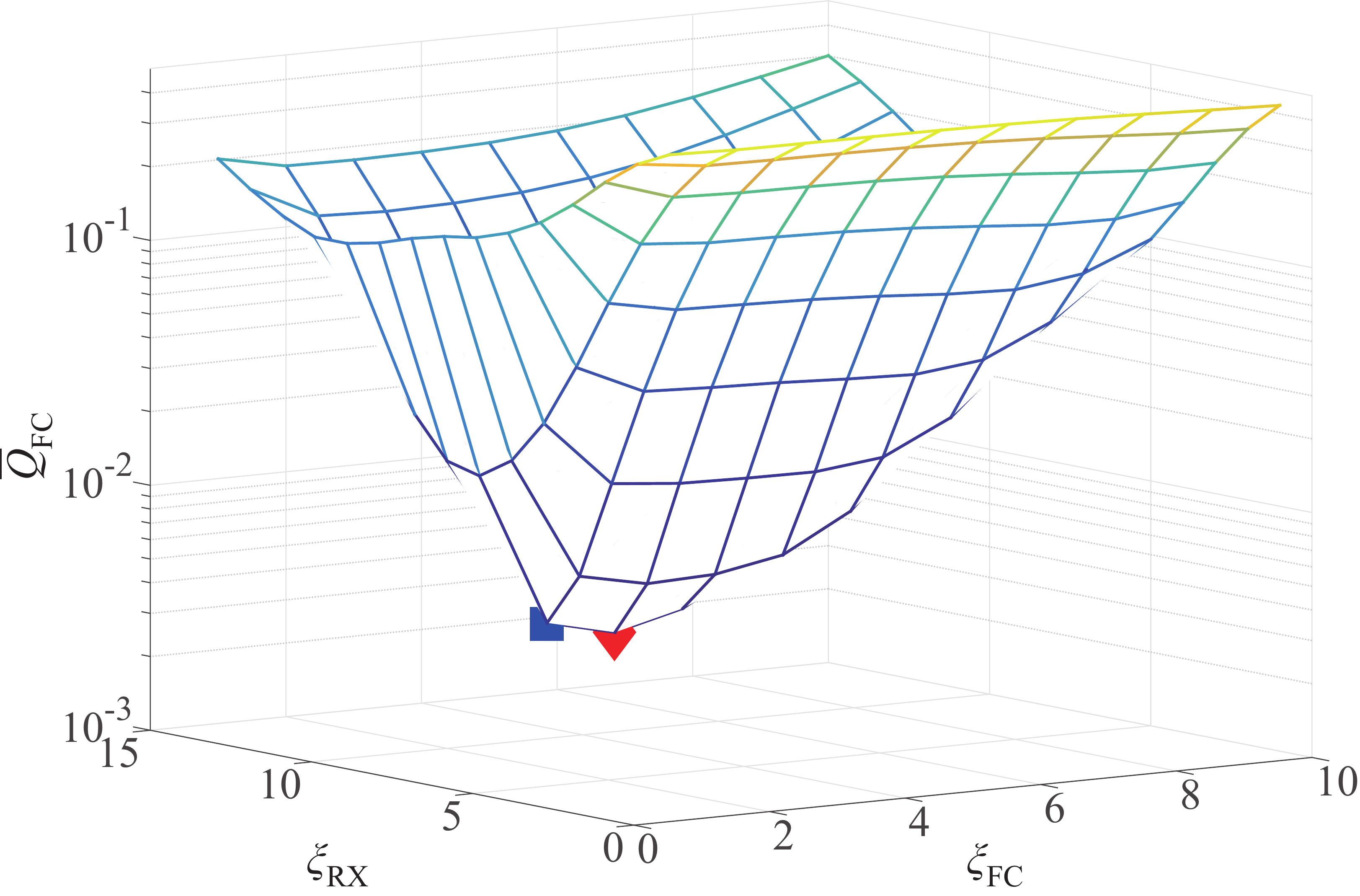}}
\subfigure[AND Rule]{\label{fig:b}\includegraphics[height=2in]{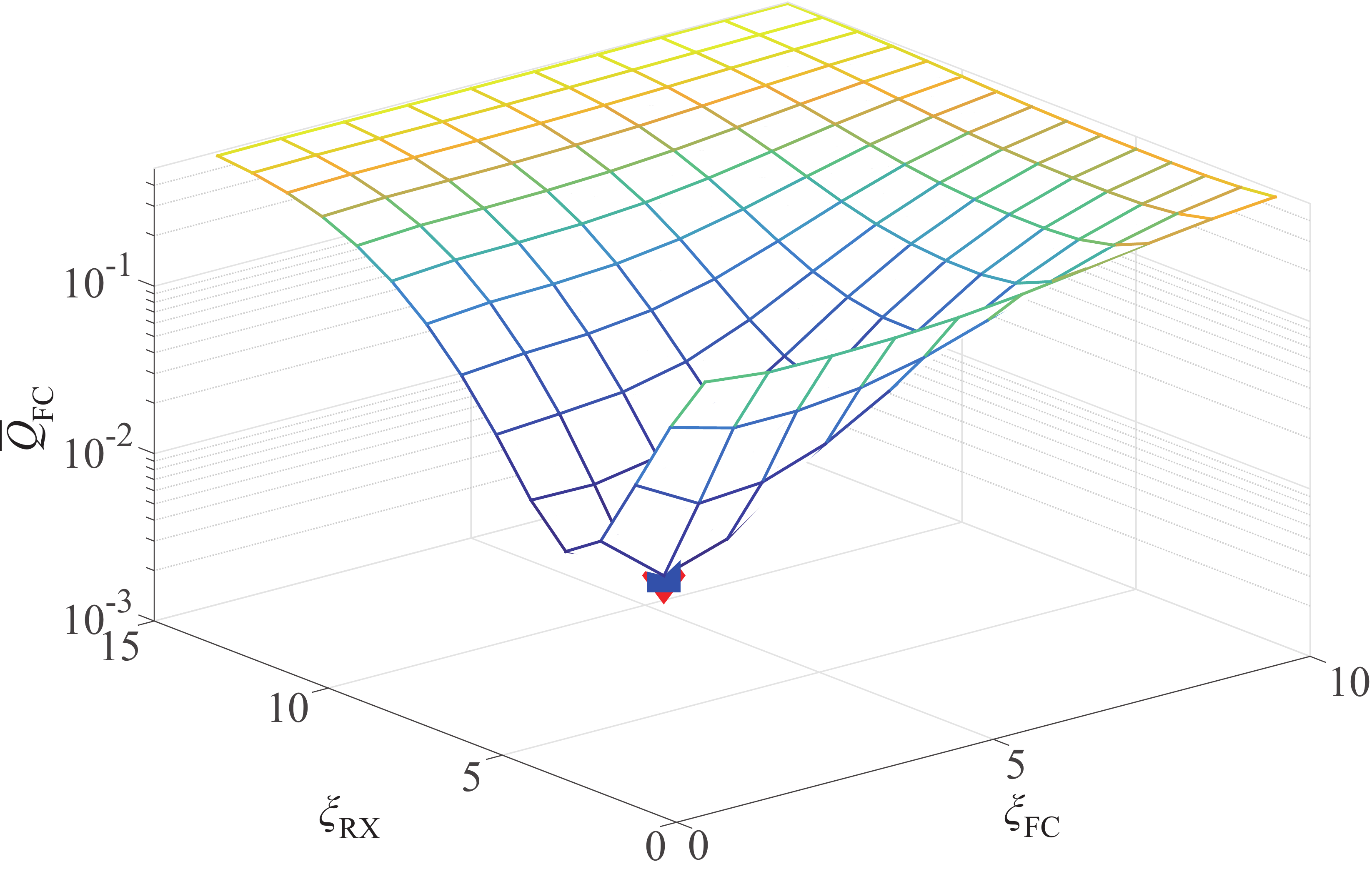}}
\subfigure[Majority Rule]{\label{fig:c}\includegraphics[height=2in]{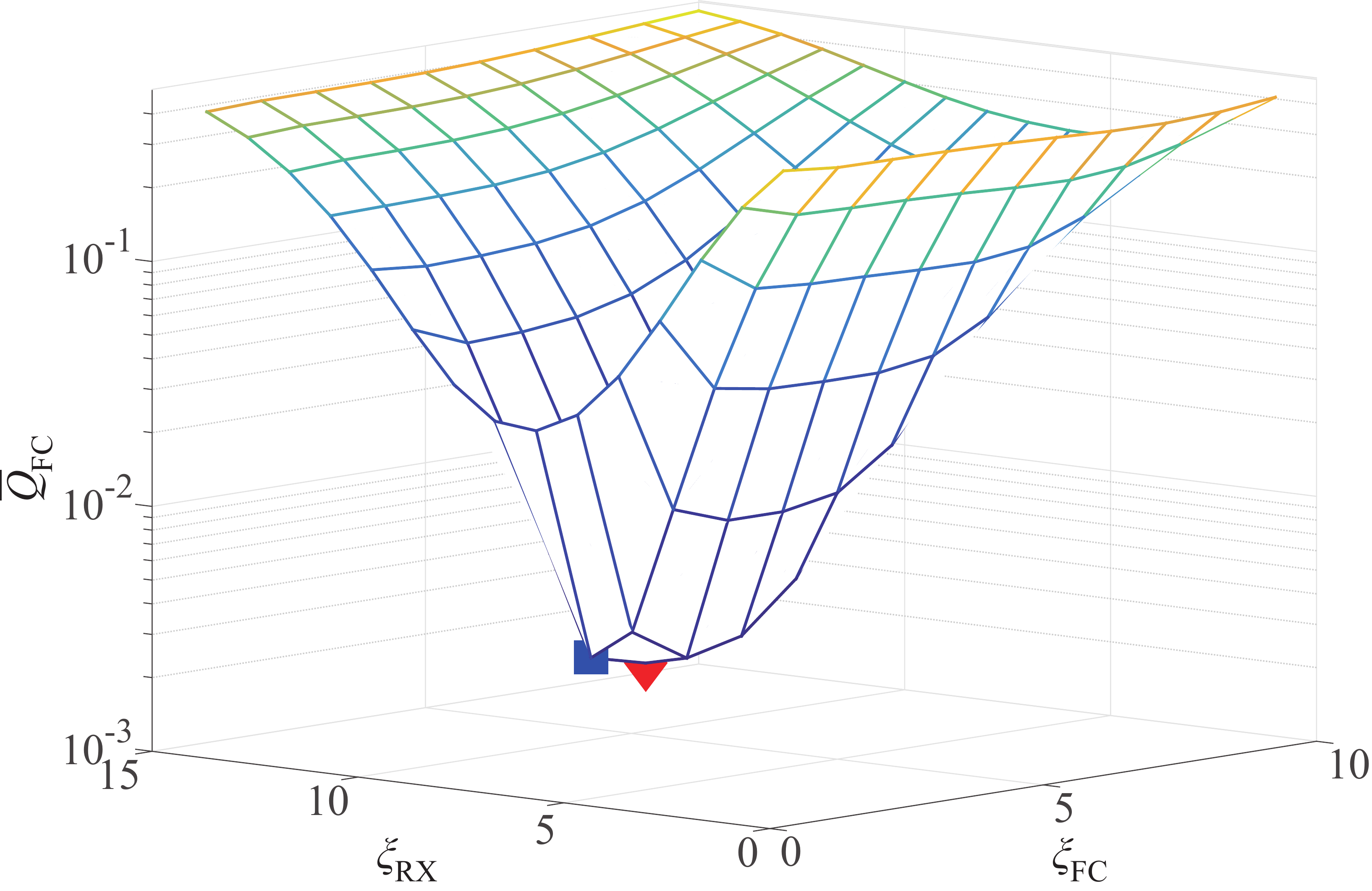}}
\caption{Expected average global error probability $\overline{Q}_{\ss\FC}$ versus the decision threshold at RXs $\xi_{\ss\R}$ and the decision threshold at the FC $\xi_{\ss\FC}$ with $K = 3$ in the noisy reporting scenario for \subref{fig:a} OR rule, \subref{fig:b} AND rule, and \subref{fig:c} majority rule. 
In \subref{fig:a}--\subref{fig:c}, `\textcolor{red}{$\blacklozenge$}' is the optimal $\overline{Q}_{\ss\FC}$ achieved by $\xi_{\ss\R}^{\ast}$ and $\xi_{\ss\FC}^{\ast}$, obtained by exhaustive search, and `\textcolor{blue}{$\blacksquare$}' is the approximated $\overline{Q}_{\ss\FC}$ achieved by $\xi_{\ss\R}^{\circ}$ and $\xi_{\ss\FC}^{\circ}$.
}\label{joint_noisy}
\end{figure}

\begin{table}[!t]
\renewcommand{\arraystretch}{1.2}
\centering
\caption{Coordinates and Values of `\textcolor{red}{$\blacklozenge$}' and `\textcolor{blue}{$\blacksquare$}' in Fig.~\ref{joint_noisy}}\label{tab:optimal points}\vspace{-2mm}
\begin{tabular}{c||c|c|c}
\hline
\bfseries Variable &  \bfseries OR Rule & \bfseries AND rule &  \bfseries Majority Rule \\\hline\hline
$\xi_{\ss\FC}^{\ast}$ of `\textcolor{red}{$\blacklozenge$}' & $4$ & $2$ & $3$\\\hline
$\xi_{\ss\R}^{\ast}$ of `\textcolor{red}{$\blacklozenge$}' & $9$ & $4$ & $7$\\\hline
$\xi_{\ss\FC}^{\circ}$ of `\textcolor{blue}{$\blacksquare$}' & $3$ & $2$ & $2$\\\hline
$\xi_{\ss\R}^{\circ}$ of `\textcolor{blue}{$\blacksquare$}' & $9$ & $4$ & $7$\\\hline
Value of `\textcolor{red}{$\blacklozenge$}' & $2.78\times10^{-3}$  & $8.99\times10^{-3}$ & $2.64\times10^{-3}$\\\hline
Value of `\textcolor{blue}{$\blacksquare$}' & $3.22\times10^{-3}$ & $8.99\times10^{-3}$ & $3.01\times10^{-3}$\\
\hline
\end{tabular}\vspace{-4mm}
\end{table}

In Fig.~\ref{joint_noisy}, we consider a three-RX cooperative system and plot the expected average global error probability versus the decision thresholds at the RXs and FC for the OR rule, AND rule, and majority rule in Fig.~\ref{fig:a}, Fig.~\ref{fig:b}, and Fig.~\ref{fig:c}, respectively.
The expected surfaces for the three fusion rules are obtained from \eqref{Qm_KN}--\eqref{Qf_AND} with \eqref{Pem2} and \eqref{Pef2}. The values of $\xi_{\ss\R}^{\circ}$ and $\xi_{\ss\FC}^{\circ}$, associated with `\textcolor{blue}{$\blacksquare$}', for the OR rule, AND rule, and majority rule are obtained by solving \eqref{noisy OR}, \eqref{noisy AND}, and \eqref{noisy NK}, respectively.
The coordinates and values of `\textcolor{red}{$\blacklozenge$}' and `\textcolor{blue}{$\blacksquare$}' in Figs.~\ref{fig:a}, \ref{fig:b}, and \ref{fig:c} are summarized in Table~\ref{tab:optimal points}. Based on Table~\ref{tab:optimal points}, we quantify the accuracy loss caused by the suboptimal convex optimization for the OR rule, AND rule, and majority rule as 15.7\%, 0\%, and 14\%, respectively. These small losses reveal that the joint $\xi_{\ss\R}^{\circ}$ and $\xi_{\ss\FC}^{\circ}$ we find can achieve near-optimal error performance.

\ifOneCol	
\begin{figure}[!t]
\centering
\includegraphics[height=3in]{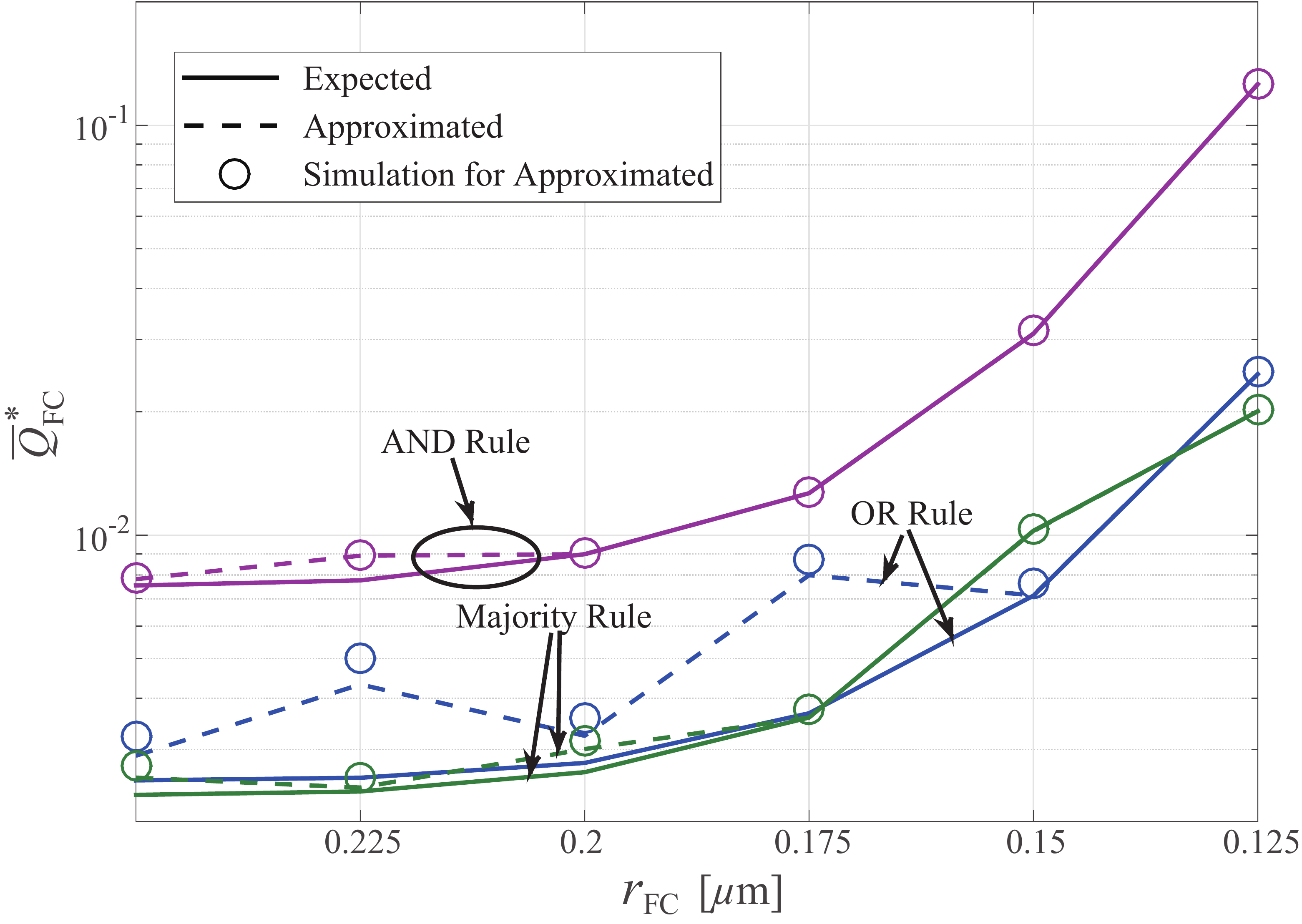}
\caption{Optimal average global error probability $\overline{Q}_{\ss\FC}$ of different fusion rules versus the radius of the FC $r_{\ss\FC}$ with $K = 3$ in the noisy reporting scenario.}
\label{Pe_rFC}
\end{figure}
\else
\begin{figure}[!t]
\centering
\includegraphics[height=2.5in,width=3.25in]{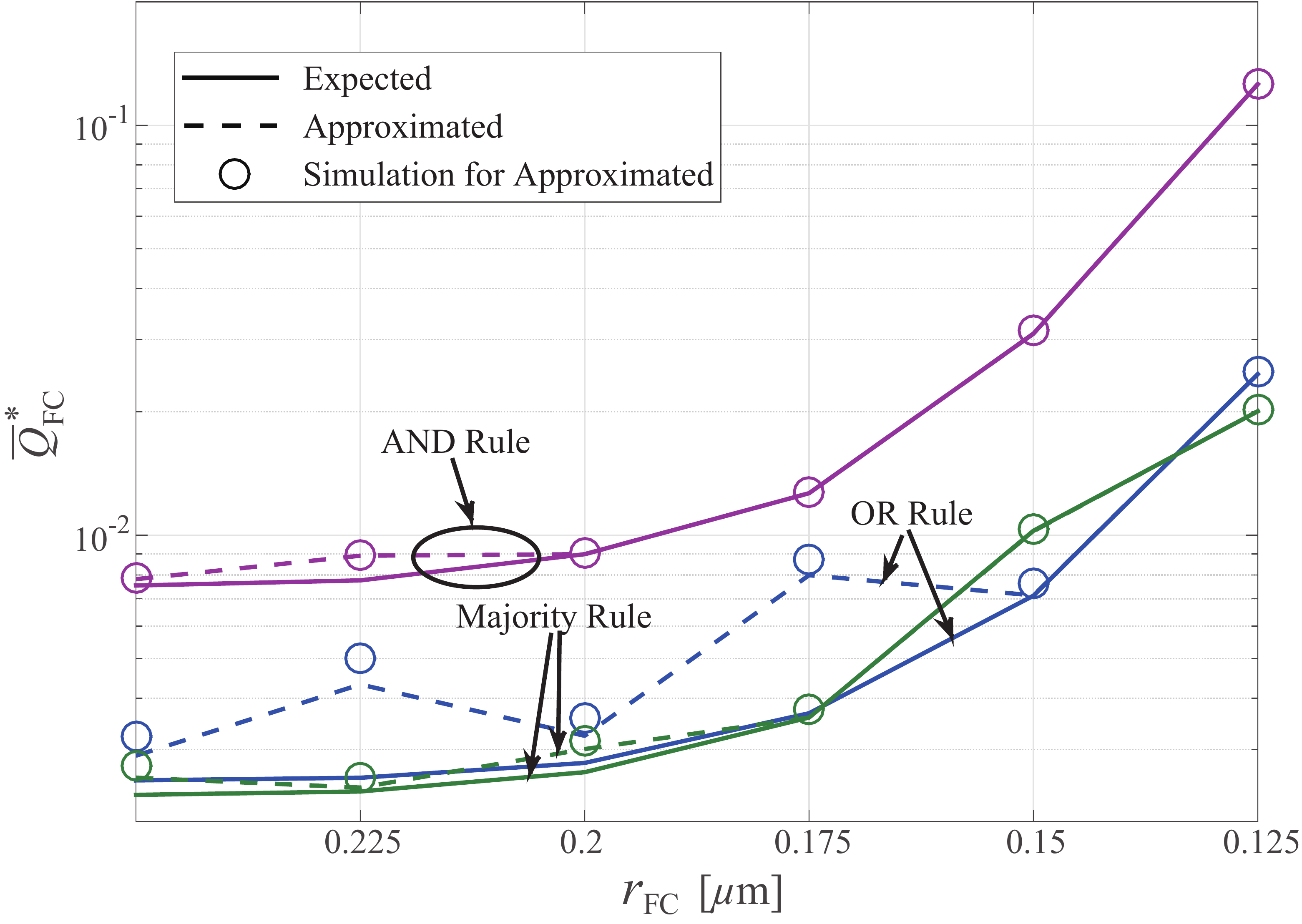}
\caption{Optimal average global error probability $\overline{Q}_{\ss\FC}$ of different fusion rules versus the radius of the FC $r_{\ss\FC}$ with $K = 3$ in the noisy reporting scenario.}
\label{Pe_rFC}
\end{figure}
\fi

In Fig.~\ref{Pe_rFC}, we consider a three-RX cooperative system and plot the average global error probability versus the radius of the FC for the AND rule, OR rule, and majority rule. The value of $\overline{Q}_{\ss\FC}^{\ast}$ for each $r_{\ss\FC}$ in the expected curves for the three fusion rules are obtained via the exhaustive search of \eqref{Qm_KN}--\eqref{Qf_AND} with \eqref{Pem2} and \eqref{Pef2}. The value of $\overline{Q}_{\ss\FC}^{\ast}$ for each $r_{\ss\FC}$ in the approximated curves for the three fusion rules are obtained by first solving \eqref{noisy OR}, \eqref{noisy AND}, and \eqref{noisy NK}, respectively, and then searching the actual values of $\overline{Q}_{\ss\FC}$ achieved by $\xi_{\ss\R}^{\circ}$ and $\xi_{\ss\FC}^{\circ}$. The simulation for approximated curves are obtained by considering $\xi_{\ss\R}^{\circ}$ and $\xi_{\ss\FC}^{\circ}$ for each $r_{\ss\FC}$. We observe that for the AND rule and majority rule, the approximated curves well approximate the expected curves, which confirms the accuracy of jointly optimizing $\xi_{\ss\R}^{\circ}$ and $\xi_{\ss\FC}^{\circ}$. We also observe that for the OR rule, the approximated curve deviates from the expected curve when $r_{\ss\FC} = 0.225$ and $r_{\ss\FC} = 0.175$. This is due to the fact that the global error probability is very sensitive to both thresholds in the region of $\xi_{\ss\FC}^{\ast}$.
Furthermore, we observe that the approximated curves match the expected curves when $r_{\ss\FC}\leq0.2{\mu}\metre$ for the AND rule, $r_{\ss\FC}\leq0.15{\mu}\metre$ for the OR rule, and $r_{\ss\FC}\leq0.175{\mu}\metre$ for the majority rule. Additionally, we observe that the expected error performance degrades as $r_{\ss\FC}$ decreases for all the fusion rules. This can be explained by the fact that the reporting from the RXs to the FC becomes less reliable when $r_{\ss\FC}$ decreases.


\section{Conclusions}\label{sec:con}

In this paper, we optimized the error performance achieved by cooperative detection among distributed RXs in a diffusion-based MC system. For the perfect and noisy reporting scenarios, we derived closed-form expressions for the expected global error probability of the system having a symmetric topology. We also derived approximated expressions for the expected error probability in both reporting scenarios. We then found the convex constraints under which the approximated expressions are jointly convex with respect to the decision thresholds at the RXs and the FC. Based on the derived convex approximations and constraints, we formulated suboptimal convex optimization problems for the system in both reporting scenarios. Furthermore, we extended the suboptimal convex optimization problem for the instantaneous error performance to that for the average error performance over all transmitter symbol sequences. Using numerical and simulation results, we showed that the system error performance can be significantly improved by combining the detection information among distributed RXs, even when the total number of transmitted molecules is limited. We also showed that the suboptimal decision thresholds, obtained by solving our formulated convex optimization problems, achieve near-optimal global error performance. In our future work, we will explore the error performance analysis and optimization of soft fusion schemes at the fusion centre and the cooperative MC system with an asymmetric topology.

\appendix

\section*{Proof of Theorem \ref{theorem 3} and Theorem \ref{theorem 3_1}}\label{app}

The convexity of $\tilde{P}_{{\mdb}}[j]^K$ can be proven by showing that its Hessian is positive semidefinite (PSD)~\cite{Convex Optimization}. Although the Hessian of $\tilde{P}_{{\mdb}}[j]^K$ is not always PSD, we can show that the Hessian of $\tilde{P}_{{\mdb}}[j]^K$ is PSD over a convex region if we impose a set of additional constraints. Recall that a matrix is PSD if and only if all of its principal minors are {nonnegative}~\cite{PSD}. Thus, we prove the joint convexity of $\tilde{P}_{{\mdb}}[j]^K$ with respective to $\xi_{\ss\R}$ and $\xi_{\ss\FC}$ by finding when $\frac{\partial^2 \tilde{P}_{\mdb}[j]^K}{\partial {\xi_{\ss\R}}^2}\geq0$, $\frac{\partial^2 \tilde{P}_{\mdb}[j]^K}{\partial {\xi_{\ss\FC}}^2}\geq0$, and $\left(\frac{\partial^2 \tilde{P}_{\mdb}[j]^K}{\partial {\xi_{\ss\R}}^2}\right)\left(\frac{\partial^2 \tilde{P}_{\mdb}[j]^K}{\partial {\xi_{\ss\FC}}^2}\right)-\left(\frac{\partial^2 \tilde{P}_{\mdb}[j]^K}{\partial {\xi_{\ss\R}}{\xi_{\ss\FC}}}\right)^{2}\geq0$.

We derive the second partial derivatives of $\tilde{P}_{\mdb}[j]^K$ with respect to $\xi_{\ss\R}$ and $\xi_{\ss\FC}$ as
\ifOneCol
\begin{align}\label{Pem2 upp De2}
\frac{\partial^2 {\tilde{P}_{{\mdb}}[j]^K}}{\partial {\xi_{\ss\R}}^2}=&\;\Gamma\left(\xi_{\ss\R},\xi_{\ss\FC},U_{1}[j],\overline{V}_{1}[j]\right)\nonumber\\
=&\;\left(-0.5-U_{1}[j]+\xi_{\ss\R}\right)
\hat{\Upsilon}\left(K-1,1,3/2\right)
+(K-1)\hat{\Upsilon}\left(K-2,2,1\right)
\end{align}	
\else
\begin{align}\label{Pem2 upp De2}
\frac{\partial^2 {\tilde{P}_{{\mdb}}[j]^K}}{\partial {\xi_{\ss\R}}^2}=&\;\Gamma\left(\xi_{\ss\R},\xi_{\ss\FC},U_{1}[j],\overline{V}_{1}[j]\right)\nonumber\\
=&\;\left(-0.5-U_{1}[j]+\xi_{\ss\R}\right)
\hat{\Upsilon}\left(K-1,1,3/2\right)\nonumber\\
&+(K-1)\hat{\Upsilon}\left(K-2,2,1\right)
\end{align}
\fi
and
\begin{align}\label{Pem2 upp De}
\frac{\partial^2{\tilde{P}_{{\mdb}}[j]^K}}{\partial {\xi_{\ss\FC}}^2}=\Gamma\left(\xi_{\ss\FC},\xi_{\ss\R},\overline{V}_{1}[j],U_{1}[j]\right),
\end{align}
respectively, where
\ifOneCol	
\begin{align}\label{Upsilon1}
\hat{\Upsilon}\left(\alpha,\beta,\gamma\right) = &\;\Big(\left(1-\Lambda\left(\xi_{\ss\R},U_{1}[j]\right)\right)\left(1+\Lambda\left(\xi_{\ss\FC},\overline{V}_{1}[j]\right)\right)\nonumber\\
&+2\left(1+\Lambda\left(\xi_{\ss\R},U_{1}[j]\right)\right)\Big)^\alpha
\left({\Lambda}\left(\xi_{\ss\FC},\overline{V}_{1}[j]\right)-1\right)^\beta\frac{K\Theta\left(\xi_{\ss\R},U_{1}[j]\right)^\beta}{U_{1}[j]^\gamma4^\alpha(2\sqrt{2\pi})^\beta}.
\end{align}
\else
\begin{align}\label{Upsilon1}
\hat{\Upsilon}\left(\alpha,\beta,\gamma\right) = &\;\Big(\left(1-\Lambda\left(\xi_{\ss\R},U_{1}[j]\right)\right)\left(1+\Lambda\left(\xi_{\ss\FC},\overline{V}_{1}[j]\right)\right)\nonumber\\
&+2\left(1+\Lambda\left(\xi_{\ss\R},U_{1}[j]\right)\right)\Big)^\alpha
\nonumber\\
&\times\left({\Lambda}\left(\xi_{\ss\FC},\overline{V}_{1}[j]\right)-1\right)^\beta\frac{K\Theta\left(\xi_{\ss\R},U_{1}[j]\right)^\beta}{U_{1}[j]^\gamma4^\alpha(2\sqrt{2\pi})^\beta}.
\end{align}
\fi
Since ${\Lambda}\left(x,\lambda\right)$ is between $-1$ and $1$ and ${\Theta}\left(x,\lambda\right)$ is greater than zero, \eqref{Pem2 upp De2} and \eqref{Pem2 upp De} are always {nonnegative} if we impose the convex constraints \eqref{Constraint 1} and \eqref{Constraint 3}, respectively.


Finally, we show how the third condition of the joint convexity is satisfied. To this end, we derive the second mixed derivative of $\tilde{P}_{\mdb}[j]^K$ with respect to $\xi_{\ss\R}$ and ${\xi_{\ss\FC}}$ as
\ifOneCol
\begin{align}\label{Pem2 upp De12}
\frac{\partial^2\tilde{P}_{\mdb}[j]^K}{\partial{\xi_{\ss\R}}{\xi_{\ss\FC}}}
=&\;\frac{2^{1-2K}K}{\pi\sqrt{U_{1}[j]\overline{V}_{1}[j]}}\Theta\left(\xi_{\ss\R},U_{1}[j]\right)\Theta\left(\xi_{\ss\FC},\overline{V}_{1}[j]\right)
\Big(3-\Lambda\left(\xi_{\ss\R},U_{1}[j]\right)\left(1+\Lambda\left(\xi_{\ss\FC},\overline{V}_{1}[j]\right)\right)\nonumber\\
&-\Lambda\left(\xi_{\ss\FC},\overline{V}_{1}[j]\right)\Big)^{-2+K}\Big(-4+K
\left.+K\Lambda\left(\xi_{\ss\FC},\overline{V}_{1}[j]\right)+K\Lambda\left(\xi_{\ss\R},U_{1}[j]\right)\right.\nonumber\\
&\times\left(1+\Lambda\left(\xi_{\ss\FC},\overline{V}_{1}[j]\right)\right)\Big).
\end{align}
\else
\begin{align}\label{Pem2 upp De12}
\frac{\partial^2\tilde{P}_{\mdb}[j]^K}{\partial{\xi_{\ss\R}}{\xi_{\ss\FC}}}
=&\;\frac{2^{1-2K}K}{\pi\sqrt{U_{1}[j]\overline{V}_{1}[j]}}\Theta\left(\xi_{\ss\R},U_{1}[j]\right)\Theta\left(\xi_{\ss\FC},\overline{V}_{1}[j]\right)\nonumber\\
&\times\Big(3-\Lambda\left(\xi_{\ss\R},U_{1}[j]\right)\left(1+\Lambda\left(\xi_{\ss\FC},\overline{V}_{1}[j]\right)\right)\nonumber\\
&-\Lambda\left(\xi_{\ss\FC},\overline{V}_{1}[j]\right)\Big)^{-2+K}\Big(-4+K\nonumber\\
&\left.+K\Lambda\left(\xi_{\ss\FC},\overline{V}_{1}[j]\right)+K\Lambda\left(\xi_{\ss\R},U_{1}[j]\right)\right.\nonumber\\
&\times\left(1+\Lambda\left(\xi_{\ss\FC},\overline{V}_{1}[j]\right)\right)\Big).
\end{align}
\fi

Combining \eqref{Pem2 upp De2}, \eqref{Pem2 upp De}, and \eqref{Pem2 upp De12}, and performing some algebraic manipulations, we have
\ifOneCol
\begin{align}\label{Pem2 upp De+Pem2 upp De2-Pem2 upp De12}
&\hspace{-5mm}\left(\frac{\partial^2\tilde{P}_{\mdb}[j]^K}{\partial {\xi_{\ss\R}}^2}\right)\left(\frac{\partial^2 \tilde{P}_{\mdb}[j]^K}{\partial {\xi_{\ss\FC}}^2}\right)-\left(\frac{\partial^2 \tilde{P}_{\mdb}[j]^K}{\partial {\xi_{\ss\R}}{\xi_{\ss\FC}}}\right)^2 \nonumber\\
=&\;\Omega\left(\xi_{\ss\R},\xi_{\ss\FC}\right)\frac{K^2 2^{-4K}}{\pi^2 U_{1}[j]\overline{V}_{1}[j]}\Theta\left(\xi_{\ss\FC},\overline{V}_{1}[j]\right)\Theta\left(\xi_{\ss\R},U_{1}[j]\right)\nonumber\\
&\times\Bigl(3-\Lambda\left(\xi_{\ss\FC},\overline{V}_{1}[j]\right)-\Lambda\left(\xi_{\ss\R},U_{1}[j]\right)
\left(1+\Lambda\left(\xi_{\ss\FC},\overline{V}_{1}[j]\right)\right)\Bigr)^{\left(-4+2K\right)},
\end{align}
\else
\begin{align}\label{Pem2 upp De+Pem2 upp De2-Pem2 upp De12}
&\hspace{-5mm}\left(\frac{\partial^2\tilde{P}_{\mdb}[j]^K}{\partial {\xi_{\ss\R}}^2}\right)\left(\frac{\partial^2 \tilde{P}_{\mdb}[j]^K}{\partial {\xi_{\ss\FC}}^2}\right)-\left(\frac{\partial^2 \tilde{P}_{\mdb}[j]^K}{\partial {\xi_{\ss\R}}{\xi_{\ss\FC}}}\right)^2 \nonumber\\
=&\;\Omega\left(\xi_{\ss\R},\xi_{\ss\FC}\right)\frac{K^2 2^{-4K}}{\pi^2 U_{1}[j]\overline{V}_{1}[j]}\Theta\left(\xi_{\ss\FC},\overline{V}_{1}[j]\right)\Theta\left(\xi_{\ss\R},U_{1}[j]\right)\nonumber\\
&\times\Bigl(3-\Lambda\left(\xi_{\ss\FC},\overline{V}_{1}[j]\right)-\Lambda\left(\xi_{\ss\R},U_{1}[j]\right)\nonumber\\
&\times\left(1+\Lambda\left(\xi_{\ss\FC},\overline{V}_{1}[j]\right)\right)\Bigr)^{\left(-4+2K\right)},
\end{align}
\fi
where $\Omega\left(\xi_{\ss\R},\xi_{\ss\FC}\right)$ is
\ifOneCol	
given by
\else
shown in {\eqref{Omega}} at the top of page~{\pageref{Omega}}.
\fi
\ifOneCol	
\begin{align}\label{Omega}
\Omega\left(\xi_{\ss\R},\xi_{\ss\FC}\right)=&\;
-4\Theta\left(\xi_{\ss\R},U_{1}[j]\right)\left(-4+K+K\Lambda\left(\xi_{\ss\FC},\overline{V}_{1}[j]\right)+K\Lambda\left(\xi_{\ss\R},U_{1}[j]\right)\left(1+\Lambda\left(\xi_{\ss\FC},\overline{V}_{1}[j]\right)\right)\right)^2\nonumber\\
&\hspace{-15mm}+\frac{\left(1+\Lambda\left(\xi_{\ss\R},U_{1}[j]\right)\right)}{\sqrt{U_{1}[j]\overline{V}_{1}[j]}}\left(1+\Lambda\left(\xi_{\ss\FC},\overline{V}_{1}[j]\right)\right)\Bigl(2\left(-1+K\right)\sqrt{\overline{V}_{1}[j]}\left(1+\Lambda\left(\xi_{\ss\R},U_{1}[j]\right)\right)\nonumber\\
&\hspace{-15mm}-\frac{\sqrt{2\pi}}{\Theta\left(\xi_{\ss\FC},\overline{V}_{1}[j]\right)}\left(0.5+\overline{V}_{1}[j]-\xi_{\ss\FC}\right)\left(-3+\Lambda\left(\xi_{\ss\FC},\overline{V}_{1}[j]\right)\right.+\Lambda\left(\xi_{\ss\R},U_{1}[j]\right)\left.\left(1+\Lambda\left(\xi_{\ss\FC},\overline{V}_{1}[j]\right)\right)\right)\Bigr)\nonumber\\
&\hspace{-15mm}\times\Bigl(\Theta\left(\xi_{\ss\R},U_{1}[j]\right)\left(-1+K\right)\left(1+\Lambda\left(\xi_{\ss\FC},\overline{V}_{1}[j]\right)\right)
2\sqrt{U_{1}[j]}-\sqrt{2\pi}\left(0.5+U_{1}[j]-\xi_{\ss\R}\right)\nonumber\\
&\hspace{-15mm}\times\left(-3+\Lambda\left(\xi_{\ss\FC},\overline{V}_{1}[j]\right)+\Lambda\left(\xi_{\ss\R},U_{1}[j]\right)\left(1+\Lambda\left(\xi_{\ss\FC},\overline{V}_{1}[j]\right)\right)\right)\Bigr). \end{align}
\else
\begin{figure*}
\begin{align}\label{Omega}
\Omega\left(\xi_{\ss\R},\xi_{\ss\FC}\right)=&\;
-4\Theta\left(\xi_{\ss\R},U_{1}[j]\right)\left(-4+K+K\Lambda\left(\xi_{\ss\FC},\overline{V}_{1}[j]\right)+K\Lambda\left(\xi_{\ss\R},U_{1}[j]\right)\left(1+\Lambda\left(\xi_{\ss\FC},\overline{V}_{1}[j]\right)\right)\right)^2+\frac{\left(1+\Lambda\left(\xi_{\ss\R},U_{1}[j]\right)\right)}{\sqrt{U_{1}[j]\overline{V}_{1}[j]}}\nonumber\\
&\times\left(1+\Lambda\left(\xi_{\ss\FC},\overline{V}_{1}[j]\right)\right)\Bigl(2\left(-1+K\right)\sqrt{\overline{V}_{1}[j]}\left(1+\Lambda\left(\xi_{\ss\R},U_{1}[j]\right)\right)-\frac{\sqrt{2\pi}}{\Theta\left(\xi_{\ss\FC},\overline{V}_{1}[j]\right)}\left(0.5+\overline{V}_{1}[j]-\xi_{\ss\FC}\right)\nonumber\\
&\times\left(-3+\Lambda\left(\xi_{\ss\FC},\overline{V}_{1}[j]\right)\right.+\Lambda\left(\xi_{\ss\R},U_{1}[j]\right)\left.\left(1+\Lambda\left(\xi_{\ss\FC},\overline{V}_{1}[j]\right)\right)\right)\Bigr)\Bigl(\Theta\left(\xi_{\ss\R},U_{1}[j]\right)\left(-1+K\right)\left(1+\Lambda\left(\xi_{\ss\FC},\overline{V}_{1}[j]\right)\right)\nonumber\\
&\times2\sqrt{U_{1}[j]}-\sqrt{2\pi}\left(0.5+U_{1}[j]-\xi_{\ss\R}\right)\left(-3+\Lambda\left(\xi_{\ss\FC},\overline{V}_{1}[j]\right)+\Lambda\left(\xi_{\ss\R},U_{1}[j]\right)\left(1+\Lambda\left(\xi_{\ss\FC},\overline{V}_{1}[j]\right)\right)\right)\Bigr)
\end{align}
\hrulefill
\end{figure*}
\fi

We note that \eqref{Pem2 upp De+Pem2 upp De2-Pem2 upp De12} is always {nonnegative} if the following constraint is satisfied:
\begin{align}\label{Cons 5}
\Omega\left(\xi_{\ss\R},\xi_{\ss\FC}\right)\geq0.
\end{align}
The constraint \eqref{Cons 5} is not convex, and $\xi_{\ss\R}$ and $\xi_{\ss\FC}$ in the exponential and error functions make joint convexity analysis with respect to $\xi_{\ss\R}$ and $\xi_{\ss\FC}$ cumbersome. To tackle this cumbersomeness, we can bound $\xi_{\ss\R}$ with $\xi_{\ss\R}^-$ or $\xi_{\ss\R}^+$, and bound $\xi_{\ss\FC}$ with $\xi_{\ss\FC}^-$ or $\xi_{\ss\FC}^+$ to lower the value of the left-hand side of \eqref{Cons 5}. Thus, we obtain \eqref{cooperative joint miss} to ensure that \eqref{Pem2 upp De+Pem2 upp De2-Pem2 upp De12} is always {nonnegative}. Under the constraints \eqref{Constraint 1}, \eqref{Constraint 3}, and \eqref{cooperative joint miss}, we define a convex region where $\tilde{P}_{\mdb}[j]^K$ is jointly convex with respect to $\xi_{\ss\FC}$ and $\xi_{\ss\R}$.

Similar to the proof of the joint convexity of $\tilde{P}_{\mdb}[j]^K$, it can be proven that $\tilde{P}_{\fab}[j]^K$ is also jointly convex with respect to $\xi_{\ss\R}$ and $\xi_{\ss\FC}$ under the constraints \eqref{Constraint 2}, \eqref{Constraint 4}, and \eqref{cooperative joint false}.


\clearpage

\begin{center}
\textsc{\normalsize{Addition Information}}\\
\end{center}

\bigskip

In this part, we provide the following figure to confirm our clarification that the system error performance degrades as $K$ increases when the \emph{total volume} of all RXs is fixed.

\begin{figure}[!h]
\centering
\includegraphics[height=2.5in,width=3.25in]{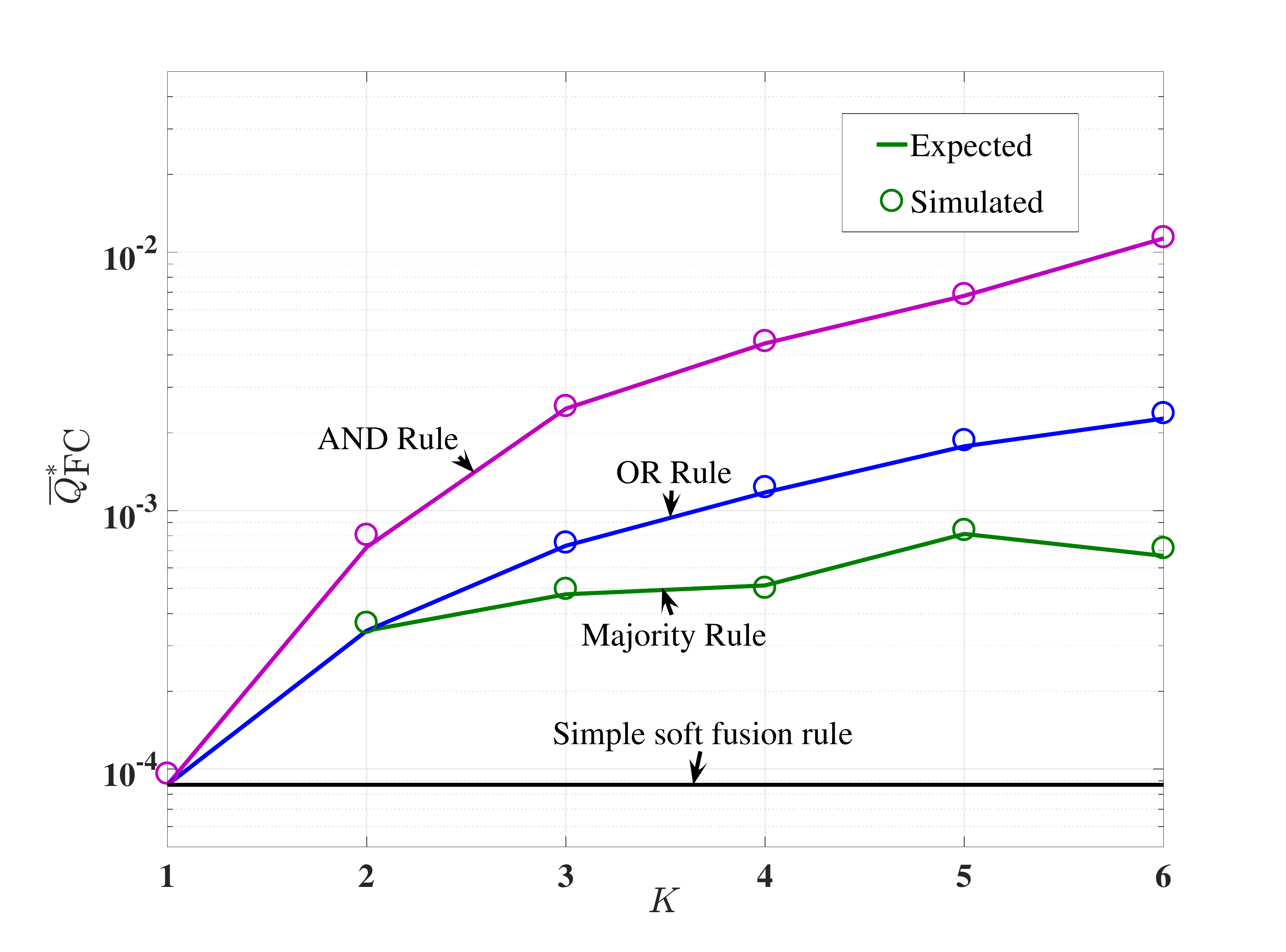}
\caption{Optimal average global error probability $\overline{Q}_{\ss\FC}^{\ast}$ of different hard fusion rules versus the number of cooperative RXs $K$ with fixed total volume of RXs in the perfect reporting scenario. Please note that we use the same parameters as those used for Fig.~\ref{Pe_k}, except that we fix the total volume of RXs as $\sum V_{\ss\R_k}=6\times\frac{4}{3}\pi\times0.2^3\,{\mu}\metre^3$.}
\label{Pe_k_fixedV}
\end{figure}


\end{document}